\documentclass[a4paper,11pt]{article}
\pdfoutput=1 % if your are submitting a pdflatex (i.e. if you have
\date{}
\usepackage{jheppub}
\usepackage[utf8]{inputenc}
\usepackage[english]{babel}
\usepackage{amsmath,amssymb,amsfonts,amsxtra, mathrsfs,graphics,graphicx,amsthm,epsfig, youngtab,bm,longtable,float,tikz,empheq}
\usepackage[margin=0.5cm]{caption}
\usepackage{thmtools}

\allowdisplaybreaks[1]
\usetikzlibrary{positioning}
\usetikzlibrary{automata}
\usetikzlibrary{arrows}
\usetikzlibrary{calc}
\usetikzlibrary{decorations.markings}
\usetikzlibrary{decorations.pathreplacing}
\usetikzlibrary{intersections}
\usetikzlibrary{positioning}
\usetikzlibrary{topaths}
\usetikzlibrary{shapes.geometric}
\usetikzlibrary{shapes.misc}
\tikzset{cf-group/.style = {
    shape = rounded rectangle, minimum size=1.0cm,
    rotate=90,
    rounded rectangle right arc = none,
    draw}}
\tikzset{cross/.style={path picture={ 
  \draw[black]
(path picture bounding box.south east) -- (path picture bounding box.north west) (path picture bounding box.south west) -- (path picture bounding box.north east);
}}}

\tikzset{unode/.style=
{black, circle,draw,thick,fill=black!100 ,minimum size=1mm}}

\tikzset{sunode/.style=
{black, circle,draw,thick,fill=yellow!100 ,minimum size=1mm}}

\tikzset{fnode/.style=
{black, rectangle,draw,thick,minimum size=1mm}}

\tikzset{afnode/.style=
{blue,rectangle,draw,thick,minimum size=1mm}}

\newcommand{\be}{\begin{equation}}
\newcommand{\ee}{\end{equation}}
\newcommand{\ba}{\begin{array}}
\newcommand{\ea}{\end{array}} 
\newcommand{\bi}{\begin{itemize}}
\newcommand{\ei}{\end{itemize}}
\def\vec#1{\bm{#1}}
\def\bea#1\eea{\allowdisplaybreaMs \begin{align}#1\end{align}}
 \newcommand{\ben}{\begin{enumerate}}
\newcommand{\een}{\end{enumerate}}
\newcommand{\bean}{\begin{eqnarray*}}
\newcommand{\eean}{\end{eqnarray*}}
\newcommand{\eref}[1]{(\ref{#1})}

\newcommand{\nn}{\nonumber}

\newcommand{\tr}{\mathrm{Tr}}

\newcommand{\BZ}{\mathbb{Z}}

\newcommand{\comment}[1]{}

\newcommand{\CS}{{\cal S}}

\newcommand{\CT}{{\cal T}}
\newcommand{\CD}{{\cal D}}
\newcommand{\CM}{{\cal M}}
\newcommand{\CO}{{\cal O}}

\newcommand{\CN}{{\cal N}}

\newcommand{\CR}{{\cal R}}
\newcommand{\CI}{{\cal I}}

\newcommand{\frsu}{\mathfrak{su}}
\newcommand{\fru}{\mathfrak{u}}

\newcommand{\frso}{\mathfrak{so}}
\newcommand{\frg}{\mathfrak{g}}
\newcommand{\frh}{\mathfrak{h}}

\newcommand{\hk}{hyperk\"ahler }
\newcommand{\wt}{\widetilde}

\newcommand{\sh}{\sinh \pi}
\newcommand{\ch}{\cosh \pi}
\newcommand{\s}{\sigma}

\newcommand{\Secref}[1]{Section~\ref{#1}}

\newcommand{\Appref}[1]{Appendix~\ref{#1}}
\newcommand{\appref}[1]{App.~\ref{#1}}
\newcommand{\Tabref}[1]{Table~\ref{#1}}

\newcommand{\figref}[1]{Fig.~\ref{#1}}
\renewcommand{\eqref}[1]{(\ref{#1})}

\title{Exploring Seiberg-like Dualities with Eight Supercharges}
\author{Anindya Dey}
\affiliation{Department of Physics and Astronomy, Johns Hopkins University, 3400 North Charles Street,
Baltimore, MD 21218, USA}
\emailAdd{anindya.hepth@gmail.com}
\abstract{We propose a family of IR dualities for 3d $\CN=4$ $U(N)$ SQCD with $N_f$ fundamental flavors and 
$P$ \textit{Abelian hypermultiplets} i.e. $P$ hypermultiplets in the determinant representation of the gauge group. 
These theories are good in the Gaiotto-Witten sense if the number of fundamental flavors obeys the constraint 
$N_f \geq 2N-1$ with generic $P \geq 1$, and in contrast to the standard $U(N)$ SQCD, they do not admit an ugly regime. 
The IR dualities in question arise in the window $N_f=2N+1,2N,2N-1,$ with $P=1$ in the first case and generic $P \geq 1$ for the others. 
The dualities involving $N_f=2N \pm 1$ are characterized by an IR enhancement of the Coulomb branch global symmetry 
on one side of the duality, such that the rank of the emergent global symmetry group is greater than the rank of the UV
global symmetry. The dual description makes the rank of this emergent global symmetry 
manifest in the UV. In addition, one can read off the emergent global symmetry itself from the dual quiver. 
We show that these dualities are related by certain field theory operations and assemble themselves into 
a duality web. Finally, we show that the $U(N)$ SQCDs with $N_f \geq 2N-1$
and $P$ Abelian hypers have Lagrangian 3d mirrors, and this allows one to explicitly write down the 3d mirror associated with a 
given IR dual pair. This paper is the first in a series of four papers on 3d $\CN=4$ Seiberg-like dualities.}

\begin{document}
\maketitle

\clearpage

\section{Introduction and Summary}

%\subsection{Background and the bas}\label{BG}

Supersymmetric gauge theories in $d \leq 6$ space-time dimensions provide a theoretical laboratory for studying 
non-perturbative phenomena in quantum field theories. In particular, theories in $d=3$ space-time dimensions
are important because they include the holographic duals of quantum gravity in four dimensions 
on the one hand, and they turn out to be interesting toy models for a large class of condensed matter 
systems on the other. 

A significant tool for probing non-perturbative/strongly-coupled physics in quantum field theories is an IR duality, 
where two theories which are manifestly different in the UV flow to the same physical theory in the IR. 
Supersymmetric gauge theories in three dimensions have an extremely rich structure of these IR dualities. 
A very well-known class is the $\CN=4$ mirror symmetry \cite{Intriligator:1996ex} -- an IR duality that exchanges the  
Coulomb and the Higgs branches in the deep IR. An $\CN=2$ 
version of the duality can be obtained by deforming the $\CN=4$ duality with an appropriate superpotential.\\

There exists another broad class of IR dualities in three dimensions for $\CN \geq 2$ theories where the 
dual gauge group depends on the number of flavors in the original theory in a fashion similar to the 4d 
$\CN=1$ Seiberg duality \cite{Seiberg:1994pq}. These include the Aharony duality \cite{Aharony:1997gp} for $\CN=2$ 
theories as well as Giveon-Kutasov duality \cite{Giveon:2008zn} for $\CN=2,3$ Chern Simons-Yang Mills-matter theories. Both types of 
dualities may be realized by appropriate Type IIB brane systems. 
The Aharony duality in particular can be a realized by a Type IIB brane system that closely mimics 
that \cite{Elitzur:1997fh} for the 4d Seiberg duality. These 3d dualities are therefore collectively referred to as 
\textit{Seiberg-like dualities}. 

In contrast to mirror symmetry, the 3d $\CN=4$ avatar of the Seiberg-like duality maps a Coulomb branch to a 
Coulomb branch across the duality and a Higgs branch to a Higgs branch. 
This duality has not been studied in much detail, partly because 
very few examples are known in the literature. One can attempt to modify the Elitzur-Giveon-Kutasov brane 
system \cite{Elitzur:1997fh} such that it preserves eight supercharges, as was done in \cite{Hanany:1996ie}, and try to read off an IR 
duality. However, as is now well-known, the naive dualities read off from the set-up are generically incorrect, 
and there is no obvious way to correct them from the brane picture. The simplest example of such a naive duality involves a $U(N)$ 
SQCD with $N_f$ fundamental flavors and a $U(N_f-N_c)$ SQCD with the same number of flavors. The 
duality demonstrably fails for generic $N$ and $N_f$. In the special case of $N_f=2N-1$, a modified version of the 
duality can be shown to hold \cite{Kapustin:2010mh}. In this case, the dual involves a $U(N-1)$ gauge theory with $N_f=2N-1$ and 
a decoupled free twisted hypermultiplet. The $U(N)$ theory with $N_f=2N-1$ is an ugly theory in the Gaiotto -Witten 
sense \cite{Gaiotto:2008ak} for which the IR SCFT is expected to factorize into a free sector and an interacting SCFT. 
In the dual description, this interacting SCFT can be identified as the IR SCFT of the $U(N-1)$ gauge theory 
with $N_f=2N-1$ fundamental flavors, which is a good theory in the Gaiotto -Witten sense. 
Similar dualities were also proposed for the bad theories with $N_f < 2N-1$ \cite{Yaakov:2013fza}, where the putative dual should be understood as the 
low energy effective field theory associated with a certain singular locus on the moduli space of the original theory \cite{Assel:2017jgo}.\\

%For the $N_f < 2N-1$ theories, the IR and the UV R-symmetries do not coincide and these are therefore bad theories 
%in the Gaiotto-Witten sense. \\
%As far as we are aware, all known examples of 3d $\CN=4$ Seiberg-like dualities 
%involve either an ugly or a bad theory on one side of the duality.

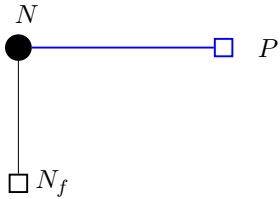
\begin{figure}
\begin{center}
\scalebox{.9}{\begin{tikzpicture}
\node[unode] (1) at (0,0){};
\node[fnode] (2) at (0,-2){};
\node[afnode] (3) at (3,0){};
\draw[-] (1) -- (2);
\draw[-, thick, blue] (1)-- (3);
\node[text width=0.1cm](10) at (0, 0.5){$N$};
\node[text width=1.5cm](11) at (1, -2){$N_f$};
\node[text width=1cm](12) at (4, 0){$P$};
%\node[text width=1cm](11) at (0, 1.1){$\eta=0$};
%\node[text width=0.1cm](20)[below=0.5 cm of 2]{$(\CT)$};
\end{tikzpicture}}
\end{center}
\caption{\footnotesize{$U(N)$ SQCD with $N_f$ fundamental flavors and $P$ abelian hypermultiplets.}}
\label{UNSQCDab}
\end{figure}

In this paper, we propose a family of 3d $\CN=4$ Seiberg-like IR dualities where one has a good theory on either side of a given 
duality. On one side of the duality, we always have a $U(N)$ SQCD with $N_f$ fundamental hypermultiplets and $P$ 
hypermultiplets in the determinant representation of $U(N)$. We will call the latter \textit{Abelian hypermultiplets} 
since they are only charged under the central $U(1)$ subgroup of the $U(N)$ gauge group, with the $U(1)$ charge 
being $N$.
We will denote these gauge theories as $\CT^N_{N_f, P}$ and represent them by the quiver in \figref{UNSQCDab}. 
A detailed description of the quiver notation involving Abelian hypermultiplets can be found in \Secref{Q-notation}.
In this notation, $\CT^N_{N_f, 0}$ denotes the standard $U(N)$ SQCD with $N_f$ flavors \footnote{We will denote an Abelian 
theory with $N_f$ hypers of charge 1 as $\CT^1_{N_f,0}$, although there is no distinction between a fundamental hyper 
and an Abelian hyper in this case.}. 
We will denote the IR duality involving the theory $\CT^N_{N_f, P}$ on one side 
as $\CD^N_{N_f,P}$. In this notation, the duality associated with the ugly SQCD -- $U(N)$ theory with $2N-1$ fundamental flavors -- 
will be denoted as $\CD^N_{2N-1,0}$. We will explicitly determine the dualities $\CD^N_{N_f,P}$ for a certain range of $N_f$ 
and $P$ for any given $N$, and study various aspects of these dualities.\\

In the rest of this section, we summarize the main results of this work followed by a brief description of the upcoming papers which build 
on the results of this paper. In \Secref{SQCD-T}, we discuss different aspects of the theories $\CT^N_{N_f, P}$ -- their classification 
in terms of the good-bad-ugly criterion of Gaiotto -Witten, their global symmetries including emergent IR symmetries, and their 
three-dimensional mirrors. We discuss the proposed dualities $\CD^N_{N_f,P}$ and perform various checks on them in \Secref{Dualities}.  
In \Secref{RG-duality}, we discuss how these dualities are related to each other 
by certain QFT operations and form a duality web. The appendices contain various computational details of the results that appear in 
the main text.

%A primary motivation for studying quiver gauge theories with Abelian hypermultiplets arises from the study of 
%Argyres-Douglas SCFTs \cite{Argyres:1995jj, Argyres:1995xn} in 4d. The 4d Higgs branches of a large class of these SCFTs can be realized as 
%the Higgs branches of certain 3d quiver gauge theories with unitary gauge nodes and Abelian hypermultiplets 
%charged under one or more of those gauge nodes \cite{Dey:2020hfe, Dey:2021rxw}. One of the dualities discussed here ($\CD^N_{2N+1,1}$) 
%has already appeared in an earlier work of the author \cite{Dey:2021rxw}. We discuss this duality in more detail in this paper and also show 
%how it is connected to the other dualities presented here. 

\subsection{Summary of the main results}\label{sum}

\subsection*{IR dualities for $U(N)$ SQCDs with Abelian hypermultiplets}

The main result of this work is to show that there exists a Seiberg-like IR duality for the theory $\CT^N_{N_f, P}$ if the  
parameters $N_f$ and $P$ are in the following regimes for a given $N$ : $(N_f=2N+1, P=1)$, $(N_f=2N, P\geq 1)$ \
and $(N_f=2N-1, P\geq 1)$. The dual pairs associated with the duality $\CD^N_{N_f,P}$ for the different ranges 
of $N_f$ and $P$ are listed in Table \ref{Tab: Result-1}, where we have used the quiver notation discussed 
in \Secref{Q-notation}.

\begin{table}[htbp]
\begin{center}
{%
\begin{tabular}{|c|c|c|}
\hline
Duality &Theory $\CT$ & IR dual  $\CT^\vee$ \\
\hline \hline 
$\CD^N_{2N+1,1}$
& \scalebox{.7}{\begin{tikzpicture}
\node[unode] (1) at (0,0){};
\node[fnode] (2) at (0,-2){};
\node[afnode] (3) at (3,0){};
\draw[-] (1) -- (2);
\draw[-, thick, blue] (1)-- (3);
\node[text width=0.1cm](10) at (0, 0.5){$N$};
\node[text width=1.5cm](11) at (1, -2){$2N+1$};
\node[text width=1cm](12) at (4, 0){$1$};
%\node[text width=1cm](11) at (0, 1.1){$\eta=0$};
%\node[text width=0.1cm](20)[below=0.5 cm of 2]{$(\CT)$};
\end{tikzpicture}}
&\scalebox{.7}{\begin{tikzpicture}
\node[sunode] (1) at (0,0){};
\node[fnode] (2) at (0,-2){};
\draw[-] (1) -- (2);
\node[text width=1 cm](10) at (0, 0.5){$N+1$};
\node[text width=1.5cm](11) at (1, -2){$2N+1$};
%\node[text width=0.1cm](20)[below=0.5 cm of 2]{$(\CT^\vee)$};
\end{tikzpicture}}\\
\hline
$\CD^N_{2N,P}$
&\scalebox{.6}{\begin{tikzpicture}
\node[unode] (1) at (0,0){};
\node[fnode] (2) at (0,-2){};
\node[afnode] (3) at (3,0){};
\draw[-] (1) -- (2);
\draw[-, thick, blue] (1)-- (3);
\node[text width=0.1cm](10) at (0, 0.5){$N$};
\node[text width=1.5cm](11) at (1, -2){$2N$};
\node[text width=1cm](12) at (4, 0){$P$};
%\node[text width=1cm](11) at (0, 1.1){$\eta=0$};
%\node[text width=0.1cm](20)[below=0.5 cm of 2]{$(\CT)$};
\end{tikzpicture}}  
& \scalebox{.6}{\begin{tikzpicture}
\node[unode] (1) at (0,0){};
\node[fnode] (2) at (0,-2){};
\node[afnode] (3) at (3,0){};
\draw[-] (1) -- (2);
\draw[-, thick, blue] (1)-- (3);
\node[text width=0.1cm](10) at (0, 0.5){$N$};
\node[text width=1.5cm](11) at (1, -2){$2N$};
\node[text width=1cm](12) at (4, 0){$P$};
%\node[text width=1cm](11) at (0, 1.1){$\eta=0$};
%\node[text width=0.1cm](20)[below=0.5 cm of 2]{$(\CT)$};
\end{tikzpicture}} \\
\hline
$\CD^N_{2N-1,P}$
& \scalebox{.6}{\begin{tikzpicture}
\node[unode] (1) at (0,0){};
\node[fnode] (2) at (0,-2){};
\node[afnode] (3) at (3,0){};
\draw[-] (1) -- (2);
\draw[-, thick, blue] (1)-- (3);
\node[text width=0.1cm](10) at (0, 0.5){$N$};
\node[text width=1.5cm](11) at (1, -2){$2N-1$};
\node[text width=1cm](12) at (4, 0){$P$};
\end{tikzpicture}}
 &\scalebox{.6}{\begin{tikzpicture}
\node[unode] (1) at (0,0){};
\node[fnode] (2) at (0,-2){};
\node[unode] (3) at (4,0){};
\node[fnode] (4) at (4,-2){};
\draw[-] (1) -- (2);
\draw[line width=0.75mm, blue] (1)-- (3);
\draw[-] (3)-- (4);
\node[text width=2.3cm](10) at (2.0, 0.2){$(1, -(N-1))$};
\node[text width=2cm](11) at (3.0, -0.3){$P$};
\node[text width=0.1 cm](20) at (0,0.5){$1$};
\node[text width=0.1 cm](21) at (0.5,-2){$1$};
\node[text width=1 cm](22) at (4, 0.5){$N-1$};
\node[text width=1.5 cm](23) at (5, -2){$2N-1$};
%\node[text width=1cm](11) at (0, 1.1){$\eta=0$};
%\node[text width=0.1cm](20)[below=0.5 cm of 2]{$(\CT^\vee)$};
%\node[text width=0.1cm](21)[above=0.1 cm of 1]{$\eta$};
%\node[text width=0.25cm](22)[above=0.1 cm of 3]{$-\eta$};
\end{tikzpicture}} \\
\hline
\end{tabular}}
\end{center}
\caption{\footnotesize{Summary of the IR dualities for the $\CT^N_{N_f,P}$ theories.}}
\label{Tab: Result-1}
\end{table}

The gauge group and matter content of the dual pairs in each case can be summarized as follows\footnote{The $\CD^N_{2N+1,1}$ has already appeared in an earlier work of the author \cite{Dey:2021rxw}. We discuss this duality in more detail in this paper and also show 
how it is connected to the other dualities presented here.}:

\begin{itemize}

\item \textbf{Duality $\CD^N_{2N+1,1}$:} The dual pair $(\CT, \CT^\vee)$ involves the theory $\CT=\CT^N_{2N+1,1}$ -- a $U(N)$ SQCD 
with $N_f=2N+1$ fundamental hypers plus a single Abelian hypermultiplet, and the theory $\CT^\vee$ -- an $SU(N+1)$ theory with $N_f=2N+1$ 
fundamental flavors. 

\item \textbf{Duality $\CD^N_{2N,P}$:} This is the self-duality of the theory $\CT=\CT^N_{2N,P}$ -- a $U(N)$ SQCD 
with $N_f=2N$ fundamental hypers plus $P$ Abelian hypermultiplets.

\item \textbf{Duality $\CD^N_{2N-1,P}$:} The dual pair $(\CT, \CT^\vee)$ involves the theory $\CT=\CT^N_{2N-1,P}$ -- a $U(N)$ SQCD 
with $N_f=2N-1$ fundamental hypers plus $P$ Abelian hypermultiplets, and the theory $\CT^\vee$ -- a $U(1) \times U(N-1)$ gauge theory with $1$ and $2N-1$ fundamental hypers respectively, and $P$ Abelian hypermultiplet with charges $(1, -(N-1))$ under the $U(1) \times U(N-1)$ gauge group.

\end{itemize}

The details of these dualities, including various checks, are discussed in \Secref{BD-main}, \Secref{SD-main} and \Secref{D2-main} 
respectively. As a straightforward consequence of the $\CD^N_{2N,P}$ duality, we obtain another interesting duality -- the self-duality of the $SU(N)$ 
gauge theory with $N_f=2N$ fundamental hypers which we will denote as $\CD^{SU(N)}_{N_f=2N}$. The details of this duality is discussed in \Secref{SD-main}.

\subsection*{Matching of Coulomb branch global symmetries and hidden FI parameters}

The matching of Coulomb branch global symmetries across the dualities $\CD^N_{2N+1,1}$ and $\CD^N_{2N-1,P}$ is non-trivial. 
The Coulomb branch symmetry on one side of the duality is partially or completely emergent in the IR. The IR enhancement in question 
is distinct from that of the more familiar case of a standard $U(N)$ SQCD with $N_f=2N$ (or more generally a linear quiver 
with balanced unitary gauge nodes), where the rank of the UV global symmetry always matches 
the rank of the emergent symmetry in the IR. In contrast, the emergent IR symmetry in our case will have a higher rank compared to 
the rank manifest in the UV. This is reminiscent of the \textit{hidden FI parameters} in orthosymplectic quiver gauge theories \cite{Gaiotto:2008ak}. 
On the other side of the duality, however, the rank of the IR global symmetry is manifest from the UV Lagrangian. For cases where the 
associated Coulomb branch global symmetry is non-Abelian, one can correctly predict 
the enhanced IR global symmetry from the UV Lagrangian using the standard notion of balanced/overbalanced gauge nodes in linear quivers \cite{Gaiotto:2008ak}.\\

For the duality $\CD^N_{2N+1,1}$, the $SU(N+1)$ gauge theory does not have any topological symmetry in the UV.
However, the Coulomb branch has an emergent $\fru(1)$ symmetry algebra, which is mapped across the duality to the topological 
symmetry algebra associated with the $U(N)$ gauge node.\\

For the duality $\CD^N_{2N-1,P}$ with $P>1$, the 
theory $\CT^N_{2N-1, P}$ has a UV-manifest $\fru(1)$ topological symmetry which is enhanced to $\fru(1) \oplus \fru(1)$
in the IR. On the dual side, the latter symmetry is manifest in the UV as topological symmetries of two unitary gauge 
nodes -- $U(1)$ and $U(N-1)$. For the special case of $P=1$, the theory $\CT^N_{2N-1, 1}$ has an enhanced 
$\frsu(2) \oplus \fru(1)$ symmetry. On the dual side, the rank of the enhanced symmetry is visible in the UV as before. 
In addition, one observes that the dual quiver $\CT^\vee$ has two gauge nodes -- 
a $U(1)$ gauge node which is balanced and a $U(N-1)$ node which is overbalanced. 
Using the standard intuition that $k$ balanced unitary nodes in a linear quiver give an enhanced $\frsu(k+1)$ Coulomb branch 
global symmetry \cite{Gaiotto:2008ak}, one might expect that the $\fru(1) \oplus \fru(1)$ symmetry in $\CT^\vee$ is enhanced to 
$\frsu(2) \oplus \fru(1)$ in the IR. This expectation indeed turns out to be correct, and can be checked explicitly by 
computing the Coulomb branch Hilbert Series of the dual theory.

\subsection*{The Duality Web}

\begin{figure}
\begin{center}
\scalebox{0.8}{\begin{tikzpicture}
  \node (D00) at (0,0) {$\CD^{N+1}_{2N+1,0}$};
  \node (D10) at (4,0) {$\CD^N_{2N-1,0}$};
  \node (D20) at (6,0) {};
  \node (D30) at (8,0) {};
  \node (D40) at (11,0) {$\CD^N_{2N-1,P-2}$};
   \node (D50) at (14,0) {$\CD^N_{2N-1,P-1}$};
  \node (D01) at (0,3) {$\CD^N_{2N+1,1}$};
  \node (D11) at (4,3) {$\CD^N_{2N,1}$};
  \node (D21) at (10,3) {$\CD^N_{2N,P}$};
  \node (D31) at (14,3) {$\CD^N_{2N-1,P}$};
 % \node (E11) at (4,5) {$\CD^N_{2N,0}$};
   \node (E21) at (7,5) {$\CD^{SU(N)}_{N_f=2N}$};
  \draw[->] (D10) -- (D00) node [midway, above] {\footnotesize $N+1 \leftarrow N$};
   \draw[->] (D20) -- (D10) node [midway, above] {\footnotesize RG};
   \draw[dotted] (D30) -- (D20) node [] {};
    \draw[->] (D40) -- (D30) node [midway, above] {\footnotesize RG};
     \draw[->] (D50) -- (D40) node [midway, above] {\footnotesize RG};
  \draw[->] (D00) -- (D01) node [midway, right=+3pt] {\footnotesize $U(1)$ gauging};
  \draw[->] (D01) -- (D11) node [midway, above] {\footnotesize RG};
   \draw[->] (D11) -- (D21) node [midway, above] {\footnotesize $\circ \CT^1_{P-1,0}$};
    \draw[->] (D11) -- (D21) node [midway, below] {\footnotesize $U(1)_{\rm diag}$ gauging};
   \draw[->] (D21) -- (D31) node [midway, above] {\footnotesize RG};
    \draw[->] (D31) -- (D50) node [midway, right=+3pt] {\footnotesize RG};
    \draw[->] (D10.north east) -- (D31) node [midway, above] {\footnotesize $\circ \CT^1_{P,0}$};
     \draw[->] (D10.north east) -- (D31) node [midway, below=+5pt] {\footnotesize $U(1)_{\rm diag}$ gauging};
     %\draw[->] (D11) -- (E11) node [midway, right=+3pt] {\footnotesize RG};
     % \draw[->] (E11) -- (E21) node [midway, above] {\footnotesize $U(1)$};
      %\draw[->] (E11) -- (E21) node [midway, below] {\footnotesize gauging};
     \draw[->] (D11) -- (E21) node [midway, right=10 pt] {\footnotesize $U(1)$ gauging};
      \draw[->] (D21) -- (E21);
\end{tikzpicture}}
\end{center}
\caption{\footnotesize{The duality web.} }
\label{DualityWeb0}
\end{figure}
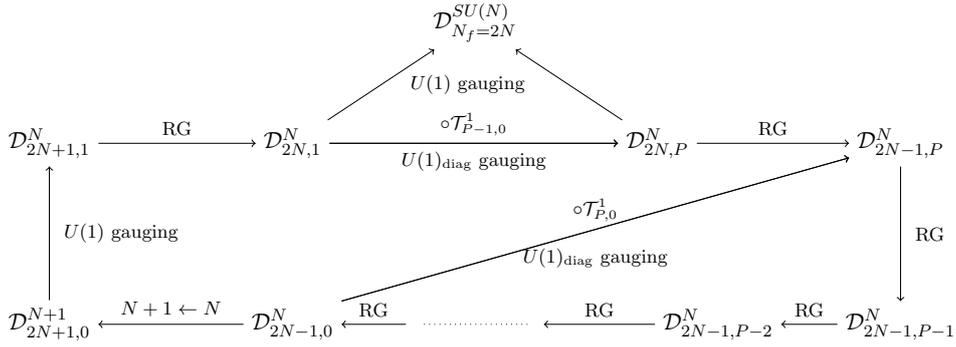

The Seiberg-like dualities proposed in this paper are related among themselves and to the Seiberg-like duality of an ugly 
$U(N)$ SQCD by various QFT operations. These dualities assemble themselves into a duality web, the explicit form of 
which is given in \figref{DualityWeb0}.

A good starting point for reading this diagram is the duality $\CD^{N+1}_{2N+1,0}$ shown in the 
bottom left corner and then moving clock-wise. Note that the duality $\CD^{N+1}_{2N+1,0}$ is the Seiberg-like duality of an ugly 
$U(N+1)$ theory with $2N+1$ fundamental hypers. There are three types of QFT operations in the diagram. Firstly, 
for a dual pair involving unitary gauge nodes on both sides, one can implement a gauging operation of the topological $U(1)$ 
symmetry which is referred to as ``\textit{$U(1)$ gauging}" in the diagram. 
For example, the duality $\CD^{N}_{2N+1,1}$ can be obtained from the duality $\CD^{N+1}_{2N+1,0}$ by such an 
operation. The second operation involves adding a decoupled SQED theory on both sides of a given duality and 
gauging a diagonal subgroup of the $U(1) \times U(1)$ topological symmetry. This is shown in the diagram 
by specifying the coupled SQED above the arrow and the text ``{$U(1)_{\rm diag}$ gauging}" below.
For example, the duality $\CD^{N}_{2N, P}$ can be obtained from the duality $\CD^{N}_{2N, 1}$ by adding a $U(1)$ theory 
with $P-1$ hypers of charge 1 (denoted by $\CT^1_{P-1,0}$ in our notation) on both sides and performing the aforementioned 
gauging operation. The third operation involves RG flows triggered by 3d $\CN=4$-preserving mass terms in the UV theory for the fundamental 
hypers as well as for the Abelian hypers. These are represented in the diagram by arrows with the text ``RG" above them.
Moving clock-wise from $\CD^{N+1}_{2N+1,0}$ in the bottom left corner, the chain of 
dualities ends at $\CD^{N}_{2N-1,0}$ -- the duality of an ugly $U(N)$ theory. Further details of the duality web are summarized 
in \Secref{RG-summary}. Discussions on the gauging operations can be found in \Secref{BD-main}-\Secref{D2-main}, 
while the RG flows are discussed in \Secref{RG-duality-1}-\Secref{RG-duality-3}.

Analogous to the 4d $\CN=1$ Seiberg duality, these IR dualities allow one to construct RG flows between families of 
3d $\CN=4$ SCFTs and therefore lead to exact dualities. This is briefly discussed in \Secref{RG-exact}.

\subsection*{3d mirrors of $U(N)$ SQCD with Abelian hypermultiplets}

\begin{table}
\begin{center}
{%
\begin{tabular}{|c|c|}
\hline
$\CT^N_{N_f,P}$ theory & 3d mirror \\
\hline \hline 
$N_f > 2N$ and $P\geq 1$
& \scalebox{0.7}{\begin{tikzpicture}
\node[unode] (1) {};
\node[unode] (2) [right=.5cm  of 1]{};
\node[unode] (3) [right=.5cm of 2]{};
\node[unode] (4) [right=1cm of 3]{};
\node[] (5) [right=0.5cm of 4]{};
%\node[cnode] (9) [right=1cm of 6]{$\frac{p-v}{2}$};
\node[unode] (6) [right=1 cm of 4]{};
\node[unode] (7) [right=1cm of 6]{};
\node[unode] (8) [right=0.5cm of 7]{};
\node[unode] (9) [right=0.5cm of 8]{};
\node[unode] (13) [above=0.5cm of 4]{};
\node[unode] (14) [left=0.5cm of 13]{};
\node[fnode] (15) [above=0.5cm of 6]{};
\node[unode] (16) [left=0.5cm of 14]{};
\node[fnode] (17) [left=0.5cm of 16]{};
%\node[snode] (18) [above=1 cm of 13]{1};
\node[] (30) [right=0.2cm of 4]{};
\node[] (31) [left=0.2cm of 6]{};
\node[text width=0.1cm](41)[below=0.2 cm of 1]{1};
\node[text width=0.1cm](42)[below=0.2 cm of 2]{2};
\node[text width=0.1cm](43)[below=0.2 cm of 3]{3};
\node[text width=0.1cm](44)[below=0.2 cm of 4]{$N$};
\node[text width=0.1cm](45)[below=0.2 cm of 6]{$N$};
\node[text width=0.1cm](46)[below=0.2 cm of 7]{3};
\node[text width=0.1cm](47)[below=0.2 cm of 8]{2};
\node[text width=0.1cm](48)[below=0.2 cm of 9]{1};
\node[text width=0.1cm](49)[right=0.1 cm of 15]{1};
\node[text width=0.1cm](50)[left=0.2 cm of 17]{1};
\draw[-] (1) -- (2);
\draw[-] (2)-- (3);
\draw[dashed] (3) -- (4);
\draw[-] (4) --(30);
\draw[-] (31) --(6);
\draw[dashed] (30) -- (31);
\draw[dashed] (6) -- (7);
\draw[-] (7) -- (8);
\draw[-] (8) --(9);
%\draw[-] (5) -- (10);
\draw[-] (4) -- (13);
\draw[-] (13) -- (14);
\draw[-] (6) -- (15);
\draw[dashed] (14) -- (16);
\draw[-] (16) -- (17);
%\draw[-] (18) -- (13);
%\node[text width=0.1cm](20)[below=1 cm of 5]{$(\wt{\CT}^N_{N_f,P})$};
\node[text width=0.1cm](25)[above=0.2 cm of 13]{1};
\node[text width=0.1cm](26)[above=0.2 cm of 14]{1};
\node[text width=0.1cm](27)[above=0.2 cm of 16]{$1$};
\end{tikzpicture}}\\
\hline
$N_f = 2N$ and $P\geq 1$
&  \scalebox{0.7}{\begin{tikzpicture}
\node[unode] (1) {};
\node[unode] (2) [right=.5cm  of 1]{};
\node[unode] (3) [right=.5cm of 2]{};
\node[unode] (4) [right=1cm of 3]{};
\node[unode] (5) [right=1cm of 4]{};
\node[unode] (6) [right=1 cm of 5]{};
\node[unode] (7) [right=1cm of 6]{};
\node[unode] (8) [right=0.5cm of 7]{};
\node[unode] (9) [right=0.5cm of 8]{};
\node[unode] (13) [above= 1cm of 5]{};
\node[fnode] (14) [below=1 cm of 5]{};
\node[unode] (15) [left=1 cm of 13]{};
\node[unode] (16) [left=1 cm of 15]{};
\node[fnode] (17) [left=1 cm of 16]{};
%\node[snode] (11) [below=0.5cm of 9]{1};
\node[text width=0.1cm](41)[below=0.2 cm of 1]{1};
\node[text width=0.1cm](42)[below=0.2 cm of 2]{2};
\node[text width=0.1cm](43)[below=0.2 cm of 3]{3};
\node[text width=1cm](44)[below=0.2 cm of 4]{$N-1$};
\node[text width=0.1cm](45)[below=0.2 cm of 5]{$N$};
\node[text width=1cm](46)[below=0.2 cm of 6]{$N-1$};
\node[text width=0.1cm](47)[below=0.2 cm of 7]{3};
\node[text width=0.1cm](48)[below=0.2 cm of 8]{2};
\node[text width=0.1cm](49)[below=0.1 cm of 9]{1};
\node[text width=0.1cm](50)[right=0.1 cm of 14]{1};
\node[text width=0.1cm](51)[left=0.2 cm of 17]{1};
\draw[-] (1) -- (2);
\draw[-] (2)-- (3);
\draw[dashed] (3) -- (4);
\draw[-] (4) --(5);
\draw[-] (5) --(6);
\draw[dashed] (6) -- (7);
\draw[-] (7) -- (8);
\draw[-] (8) --(9);
\draw[-] (5) -- (13);
%\draw[-] (4) -- (13);
\draw[-] (13) -- (15);
\draw[dashed] (16) -- (15);
\draw[-] (16) -- (17);
\draw[-] (5) -- (14);
\node[text width=0.1cm](25)[above=0.2 cm of 13]{1};
\node[text width=0.1cm](26)[above=0.2 cm of 15]{1};
\node[text width=0.1cm](27)[above=0.2 cm of 16]{$1$};
%\node[text width=0.1cm](20)[below=2 cm of 5]{$(\wt{\CT}^N_{2N,P})$};
\end{tikzpicture}}\\
\hline
$N_f = 2N-1$ and $P > 1$
&\scalebox{0.7}{\begin{tikzpicture}
\node[unode] (1) {};
\node[unode] (2) [right=.5cm  of 1]{};
\node[unode] (3) [right=.5cm of 2]{};
\node[unode] (4) [right=1cm of 3]{};
\node[] (5) [right=0.5cm of 4]{};
%\node[cnode] (9) [right=1cm of 6]{$\frac{p-v}{2}$};
\node[unode] (6) [right=1 cm of 4]{};
\node[unode] (7) [right=1cm of 6]{};
\node[unode] (8) [right=0.5cm of 7]{};
\node[unode] (9) [right=0.5cm of 8]{};
\node[unode] (13) [above=0.5cm of 4]{};
\node[unode] (14) [left=0.5cm of 13]{};
\node[fnode] (15) [above=0.5cm of 6]{};
\node[unode] (16) [left=0.5cm of 14]{};
\node[fnode] (17) [left=0.5cm of 16]{};
\node[fnode] (18) [above=1 cm of 13]{};
\node[text width=0.1cm](41)[below=0.2 cm of 1]{1};
\node[text width=0.1cm](42)[below=0.2 cm of 2]{2};
\node[text width=0.1cm](43)[below=0.2 cm of 3]{3};
\node[text width=1cm](44)[below=0.2 cm of 4]{$N-1$};
\node[text width=1cm](45)[below=0.2 cm of 6]{$N-1$};
\node[text width=0.1cm](46)[below=0.2 cm of 7]{3};
\node[text width=0.1cm](47)[below=0.2 cm of 8]{2};
\node[text width=0.1cm](48)[below=0.2 cm of 9]{1};
\node[text width=0.1cm](49)[right=0.1 cm of 15]{1};
\node[text width=0.1cm](50)[left=0.2 cm of 17]{1};
\node[text width=0.1cm](51)[right=0.2 cm of 18]{1};
\draw[-] (1) -- (2);
\draw[-] (2)-- (3);
\draw[dashed] (3) -- (4);
\draw[-] (4) --(6);
%\draw[-] (5) --(6);
\draw[dashed] (6) -- (7);
\draw[-] (7) -- (8);
\draw[-] (8) --(9);
%\draw[-] (5) -- (10);
\draw[-] (4) -- (13);
\draw[-] (13) -- (14);
\draw[-] (6) -- (15);
\draw[dashed] (14) -- (16);
\draw[-] (16) -- (17);
\draw[-] (18) -- (13);
%\node[text width=0.1cm](20)[below=1 cm of 5]{$(\wt{\CT}^{N}_{2N-1,P})$};
\node[text width=0.1cm](25)[above=0.2 cm of 13]{1};
\node[text width=0.1cm](26)[above=0.2 cm of 14]{1};
\node[text width=0.1cm](27)[above=0.2 cm of 16]{1};
\end{tikzpicture}}\\
\hline
$N_f = 2N-1$ and $P =1$
& \scalebox{0.7}{\begin{tikzpicture}
\node[unode] (1) {};
\node[unode] (2) [right=.5cm  of 1]{};
\node[unode] (3) [right=.5cm of 2]{};
\node[unode] (4) [right=1cm of 3]{};
\node[] (5) [right=0.5cm of 4]{};
%\node[cnode] (9) [right=1cm of 6]{$\frac{p-v}{2}$};
\node[unode] (6) [right=1 cm of 4]{};
\node[unode] (7) [right=1cm of 6]{};
\node[unode] (8) [right=0.5cm of 7]{};
\node[unode] (9) [right=0.5cm of 8]{};
\node[unode] (13) [above=0.5cm of 4]{};
\node[fnode] (14) [left=0.5cm of 13]{};
\node[fnode] (15) [above=0.5cm of 6]{};
%\node[snode] (11) [below=0.5cm of 9]{1};
\node[text width=0.1cm](41)[below=0.2 cm of 1]{1};
\node[text width=0.1cm](42)[below=0.2 cm of 2]{2};
\node[text width=0.1cm](43)[below=0.2 cm of 3]{3};
\node[text width=1cm](44)[below=0.2 cm of 4]{$N-1$};
\node[text width=1cm](45)[below=0.2 cm of 6]{$N-1$};
\node[text width=0.1cm](46)[below=0.2 cm of 7]{3};
\node[text width=0.1cm](47)[below=0.2 cm of 8]{2};
\node[text width=0.1cm](48)[below=0.2 cm of 9]{1};
\node[text width=0.1cm](49)[above=0.1 cm of 13]{1};
\node[text width=0.1cm](50)[left=0.2 cm of 14]{2};
\node[text width=0.1cm](51)[right=0.2 cm of 15]{1};
\draw[-] (1) -- (2);
\draw[-] (2)-- (3);
\draw[dashed] (3) -- (4);
\draw[-] (4) --(6);
%\draw[-] (5) --(6);
\draw[dashed] (6) -- (7);
\draw[-] (7) -- (8);
\draw[-] (8) --(9);
%\draw[-] (5) -- (10);
\draw[-] (4) -- (13);
\draw[-] (13) -- (14);
\draw[-] (6) -- (15);
%\node[text width=0.1cm](20)[below=1 cm of 5]{$(\wt{\CT}^N_{2N-1,1})$};
\end{tikzpicture}}\\
\hline
\end{tabular}}
\end{center}
\caption{\footnotesize{Summary of the 3d mirrors for the $\CT^N_{N_f,P}$ theories, with $N_f \geq 2N-1$ and $P\geq 1$. 
The chain of $U(1)$ gauge nodes contains $P \geq 1$ nodes.}}
\label{Tab: Result-2}
\end{table}

The theories $\CT^N_{N_f, P}$ have Lagrangian 3d mirrors for $N_f \geq 2N-1$ and $P \geq 1$, and are summarized in 
Table \ref{Tab: Result-2}. The quivers are qualitatively different in the three regimes $N_f > 2N$, $N_f=2N$ and $N_f=2N-1$ 
as shown. The construction of these 3d mirrors are discussed in \Secref{3dmirr-gen}. Given these 3d mirrors, one 
can readily write down the 3d mirrors associated with the IR dual pairs listed in Table \ref{Tab: Result-1}. These are summarized in 
Table \ref{Tab:D4} of \Secref{Mirr-main}. We discussed earlier that some of the dualities $\CD^N_{N_f, P}$ are 
characterized by an emergent Coulomb branch symmetry on one side of the duality. 
The 3d mirror associated with such a duality $\CD^N_{N_f, P}$ realizes this emergent 
Coulomb branch symmetry as a Higgs branch global symmetry which is manifest in the UV 
Lagrangian.

\subsection{Preview of the upcoming papers}\label{future-dir}

This paper is the first in a series of papers on 3d $\CN=4$ Seiberg-like IR dualities. Below, we present a brief outline of the 
contents of these papers.

\begin{itemize}

\item \textbf{3d $\CN=4$ IR N-ality:} In the second paper  \cite{Dey:2022abc}, we show that a large class of 3d $\CN=4$ quiver gauge theories consisting of unitary and special unitary gauge nodes with fundamental/bifundamental matter can have multiple Seiberg-like IR duals. We will refer to this phenomenon as \textit{N-ality}. The dualities discussed in this paper play a crucial role in the construction of these N-alities. 
A quiver from the aforementioned class will generically have a large emergent Coulomb branch symmetry in the IR. Similar to what we discussed above, the rank of the IR symmetry is greater than that of the UV symmetry and the former becomes UV-manifest in one or more of the dual theories. For certain special families of quivers, one can read off the emergent symmetry algebra itself from one or more of the dual quivers.

\item \textbf{3d Lagrangians of Argyres-Douglas theories:} The phenomenon of IR N-ality has interesting consequences for 4d Argyres-Douglas theories 
of a very large class. 
In the third paper \cite{Dey:2022xyz}, we show that the 3d SCFT, obtained by circle-reducing a given Argyres-Douglas theory of this type, can be associated to multiple 3d $\CN=4$ quiver gauge theories which are related by IR N-ality. This allows one to realize the 4d Higgs branch of the Argyres-Douglas theory in terms of a \hk quotient construction in more than one way. This construction also gives a way of arriving at the 3d mirror associated with the 4d SCFT, which is different from the standard class $\CS$ construction of these mirrors. 

\item \textbf{Mapping extended and local operators:} For the IR dualities studied in this paper as well as for the N-alities discussed in upcoming papers, 
mapping the extended BPS operators as well as the local BPS operators across a given duality provide a much more refined understanding of the IR equivalence. In particular, these are closely related to understanding the map of higher-form symmetries across a duality (or N-ality) for cases where the dual pair does have such a symmetry. This is the subject of the fourth paper \cite{Dey:2022lmn}.

\end{itemize}

% These hypermultiplets arise in quiver gauge theories associated with the 3d SCFT one obtains by putting 
%certain Argyres-Douglas theory on a circle and flowing to the IR, as was shown in \cite{}. 
%
%We will show in this work that there exists an IR duality for the theories $\CT^N_{N_f, P}$ for certain special values of $N_f$ and $P$, 
%i.e. $N_f=2N+1, 2N, 2N-1$, and $P=1$. For $N_f= 2N, 2N-1$, we can extend the duality to a generic $P \geq 1$. We will then present 
%the systematics of constructing the N-ality sequence using the aforementioned dualities as quiver mutations. Finally, we will explicitly 
%present the construction for two infinite families of simple quiver gauge theories...

\section{$U(N)$ SQCD with abelian hypermultiplets}\label{SQCD-T}

In this section, we will study aspects of the IR physics of the theories $\CT^N_{N_f,P}$, i.e. $U(N)$ SQCD 
with $P \geq 1$ abelian hypermultiplets of charge $N$, as shown in \figref{UNSQCDab}. We will begin by discussing the classification of these 
theories into good, bad and ugly theories, following the analysis of Gaiotto-Witten \cite{Gaiotto:2008ak}. We will then study 
the Coulomb branch and the Higgs branch Hilbert Series (HS) of the theories and discuss the global symmetries 
of the respective branches of moduli spaces. Finally, we will discuss the construction of 3d mirrors for these theories.

\subsection{Quiver notation}\label{Q-notation}

In this work, we will be concerned with 3d $\CN=4$ quivers with $U(N)$ and $SU(N)$ gauge nodes, 
fundamental/bifundamental matter and a given number of Abelian hypermultiplets i.e. hypermultiplets 
charged under the determinant/anti-determinant representations of the unitary gauge nodes. A representative 
quiver diagram that will be of interest to us is given as follows:

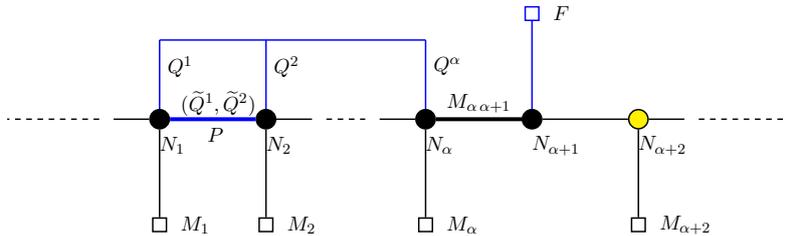
\begin{center}
\scalebox{0.7}{\begin{tikzpicture}
\node[] (100) at (-3,0) {};
\node[] (1) at (-1,0) {};
\node[unode] (2) at (0,0) {};
\node[text width=.2cm](31) at (0.1,-0.5){$N_1$};
\node[unode] (3) at (2,0) {};
\node[text width=.2cm](32) at (2.1,-0.5){$N_2$};
\node[] (4) at (3,0) {};
\node[] (5) at (4,0) {};
\node[unode] (6) at (5,0) {};
\node[text width=.2cm](33) at (5.1,-0.5){$N_{\alpha}$};
\node[unode] (7) at (7,0) {};
\node[text width=.2cm](34) at (7.1,-0.5){$N_{\alpha+1}$};
\node[text width=.2cm](35) at (5.5,0.3){$M_{\alpha\,\alpha+1}$};
\node[sunode] (8) at (9,0) {};
\node[text width=.2cm] (40) at (9.1,-0.5) {$N_{\alpha+2}$};
\node[fnode] (9) at (9,-2) {};
\node[text width=.2cm] (40) at (9.5,-2) {$M_{\alpha+2}$};
\node[] (10) at (10,0) {};
\node[] (11) at (12,0) {};
\node[afnode] (12) at (7,2) {};
\node[text width=.2cm](36) at (7.5,2){$F$};
\node[fnode] (20) at (0,-2) {};
\node[fnode] (21) at (2,-2) {};
\node[fnode] (22) at (5,-2) {};
\node[text width=.2cm](23) at (0.5,-2){$M_1$};
\node[text width=.2cm](24) at (2.5,-2){$M_2$};
\node[text width=.2cm](25) at (5.5,-2){$M_\alpha$};
\draw[thick] (1) -- (2);
\draw[line width=0.75mm, blue] (2) -- (3);
\node[text width=.2cm](50) at (0.5,0.3){$(\wt{Q}^1,\wt{Q}^2)$};
\node[text width=.2cm](51) at (1,-0.3){$P$};
\draw[thick] (3) -- (4);
\draw[thick,dashed] (4) -- (5);
\draw[thick] (5) -- (6);
\draw[line width=0.75mm] (6) -- (7);
\draw[thick] (7) -- (8);
\draw[thick,blue] (7) -- (12);
\draw[thick] (8) -- (9);
\draw[thick] (8) -- (10);
\draw[thick,dashed] (10) -- (11);
\draw[thick,dashed] (1) -- (100);
\draw[thick, blue] (2) -- (0,1.5);
\draw[thick, blue] (0,1.5) -- (5,1.5);
\draw[thick, blue] (2,1.5) -- (3);
\draw[thick, blue] (5,1.5) -- (6);
\draw[thick] (2) -- (20);
\draw[thick] (3) -- (21);
\draw[thick] (6) -- (22);
\node[text width=.2cm](15) at (0.25,1){$Q^1$};
\node[text width=.2cm](16) at (2.25,1){$Q^2$};
\node[text width=.2cm](17) at (5.25,1){$Q^\alpha$};
%\node[text width=.2cm](18) at (7.25,1){$Q^{\alpha+1}$};
\end{tikzpicture}}
\end{center}

The different constituents of the quiver may be 
summarized as follows:

\begin{enumerate}

\item A black node \scalebox{0.7}{\begin{tikzpicture} \node[unode] (1) at (0,0){};\end{tikzpicture}} with label $N$ represents a $U(N)$ gauge node.

\item A yellow node \scalebox{0.7}{ \begin{tikzpicture} \node[sunode] (1) at (0,0){};\end{tikzpicture}} with label $N$ represents an $SU(N)$ gauge node.

\item A black square box \scalebox{0.7}{\begin{tikzpicture} \node[fnode] (1) at (0,0){};\end{tikzpicture}} with label $F$ represents $F$ hypermultiplets in the fundamental representation of the gauge node it is attached to.

\item A thin black line connecting two gauge nodes is a bifundamental hypermultiplet,
while a thick black line with a label $M$ denotes $M$ such bifundamental hypermultiplets.

\item A blue square box \scalebox{0.7}{\begin{tikzpicture} \node[afnode] (1) at (0,0){};\end{tikzpicture}} with label $F$ 
represents $F$ hypermultiplets in the determinant representation of the unitary gauge node it is attached to.

\item A thin blue line connecting two or more unitary gauge nodes is an Abelian hypermultiplet 
associated with those gauge nodes. We show the respective charges $\{Q^i\}$ explicitly to show whether the 
hypermultiplet transforms in the determinant or the anti-determinant  representation of the 
unitary gauge node $i$, i.e. $Q^i= \pm N_i$ where $N_i$ is the label of the unitary gauge node $i$.
In the quiver diagram above, one has a single Abelian hypermultiplet charged under the $U(N_1)$, 
the $U(N_2)$ and the $U(N_\alpha)$ gauge nodes with charges $(Q^1, Q^2, Q^\alpha)$ respectively.
A thick blue line with a label $P$ denotes a collection of $P$ such Abelian hypermultiplets. In the quiver 
diagram, we have $P$ Abelian hypermultiplets charged under the $U(N_1)$ and the $U(N_2)$ gauge 
nodes with respective charges $(\wt{Q}^1, \wt{Q}^2 )$.

\end{enumerate}

\subsection{Monopole operators: The good, the bad and the ugly} \label{GBU}

A monopole operator in a 3d $\CN=4$ gauge theory can be defined by introducing a Dirac monopole singularity 
(labelled by a cocharacter $\vec a$) for the gauge field and a singular configuration for a single real adjoint scalar field in the 
vector multiplet at a given point on the space-time manifold, while the other two real scalars remain regular. This configuration
preserves an $\CN=2$ subalgebra of the full $\CN=4$ supersymmetry algebra. The choice of the real 
adjoint scalar picks a subalgebra $\fru(1)_{\rm C} \cong \frso(2)_{\rm C} \subset \frso(3)_{\rm C} \cong \frsu(2)_{\rm C}$, where 
$\frsu(2)_{\rm C}$ is the Lie algebra of the R-symmetry group $SU(2)_{\rm C}$ acting on the Coulomb branch. 
The Lie algebra of the R-symmetry group $U(1)_{R}$ for the preserved $\CN=2$ subalgebra is then given by 
$\fru(1)_{R}$ -- a Cartan subalgebra of $\frsu(2)_{\rm C} \oplus \frsu(2)_{\rm H}$, 
where $\frsu(2)_{\rm H}$ is the Lie algebra of the R-symmetry group $SU(2)_{\rm H}$ acting on the Higgs branch.
A standard choice of $\fru(1)_R$ corresponds to the assignment of R-charge 1/2 to the complex scalars of the hypermultiplet 
and 1 to the regular complex adjoint scalar in the vector multiplet. This will also be our choice for the rest of this paper. 
 
A given monopole operator breaks the gauge group $G$ to a subgroup $H(\vec a)$. One can turn on a constant background 
for the regular adjoint scalars in the Lie algebra $\frh(\vec a)$ of the subgroup $H(\vec a)$ without breaking the $\CN=2$ subalgebra.
The resultant configuration is referred to as a ``dressed monopole operator" while the configuration where the adjoint scalar 
background is turned off is  referred to as a ``bare monopole operator". The bare monopole operator is uncharged under the standard 
$U(1)_{R}$ classically, but its R-charge receives quantum-mechanical corrections of the following form:
\be \label{MO-gen}
q_R(\vec a) = \Delta(\vec a)= - \sum_{\alpha \in \Delta_+}|\alpha(\vec a)| + \frac{1}{2}\,\sum^n_{i=1}\, \sum_{\rho_i \in \CR_i} |\rho_i(\vec a)|, 
\ee
where the first term is a contribution of the vector multiplets, and the second term is a contribution of $n$ hypermultiplets with 
the $i$-th hypermultiplet transforming in the representation $\CR_i$ of the gauge group. In the deep IR, the 3d $\CN=4$ gauge theory generically 
flows to an interacting SCFT. The monopole operators are chiral operators with conformal dimensions determined by the charges 
under a $U(1)$ superconformal R-symmetry. If the $U(1)_{R}$ defined above is the same as the superconformal $U(1)$ symmetry, then the 
charge $q_R$ in \eref{MO-gen} gives the correct conformal dimension of the chiral operator in the IR SCFT. Such a theory is referred 
to as a good theory. If the $U(1)_{R}$ is not the same as the superconformal $U(1)$, i.e. the $U(1)_{R}$ 
mixes with flavor symmetries in the IR, then the charge $q_R$ does not give the correct IR conformal dimension. Such a theory is 
referred to as a bad theory. Finally, a non-bad theory in the IR can be a product of a good theory and decoupled free hypermultiplets whose complex scalar components have conformal dimension $1/2$-- such a theory is referred to as an ugly theory.

Gaiotto-Witten gave a diagnostic for classifying 3d $\CN=4$ gauge theories into good, bad and ugly categories using the quantum-corrected 
$U(1)_{R}$ charges for the bare monopole operators. If $q_R < \frac{1}{2}$ for one or more monopole operators, the identification 
of $U(1)_{R}$ with the superconformal $U(1)$ will imply the existence of unitarity-violating chiral operators in the SCFT. This can only 
be avoided if $U(1)_{R}$ mixes with flavor symmetries in the IR, which implies that we have a bad theory. If $q_R \geq \frac{1}{2}$ 
for all monopole operators, with one or more saturating the bound, then we have an ugly theory. If $q_R > \frac{1}{2}$ 
for all monopole operators, the theory in question is good. 
For a good theory, this analysis also allows us to determine the 
enhanced global symmetry of the Coulomb branch if it exists. For unitary gauge groups, a subgroup of the global symmetry 
is classically visible as topological $U(1)$ symmetries. However, in the IR, additional conserved currents may appear leading 
to an enhancement of the global symmetry. In the SCFT, conserved currents appear in a superconformal multiplet whose 
lowest component is a chiral operator with R-charge 1. If the UV and the IR R-symmetries are identical, which is 
indeed the case for good theories, a global symmetry generator will be present in the IR SCFT for every monopole operator 
with $q_R=1$. Finding the complete set of $q_R=1$ monopole operators, therefore, allows one to compute the Lie algebra of the 
enhanced global symmetry group of the Coulomb branch.\\

Let us now consider the classification of the $\CT^N_{N_f,P}$ theories into good, bad and ugly theories, following our discussion 
above. The R-charge and the conformal dimension of a monopole operator labelled by a cocharacter $\vec a$ in the theory 
$\CT^N_{N_f,P}$ is given as :
\begin{align}
q_R(\vec a)= \Delta( \vec a)=& \frac{N_f}{2}\,\sum^{N}_i |a_i| +\frac{P}{2}\, |\sum^{N}_{i=1} \,a_i|- \sum_{1 \leq i<j \leq N}|a_i-a_j| \\
=& \frac{N_f -2N +2}{2}\,\sum^{N}_i |a_i| +\frac{P}{2}\, |\sum^{N}_{i=1} \,a_i| + \sum_{1 \leq i<j \leq N}( |a_i| + |a_j| - |a_i-a_j|),
\end{align}
where in the second step, we have written the RHS as a sum of three terms, the last two of which are positive semi-definite. 
To begin with, note that for $N_f \geq 2N$, $\Delta(\vec a)|_{\rm min} > 1$. This implies that there are no unitarity-violating operators with 
$\Delta < \frac{1}{2}$, or operators with $\Delta =\frac{1}{2}$ which decouple as free sectors in the IR, i.e. these theories 
are good in the Gaiotto-Witten sense. In addition, there are no monopole operators with $\Delta =1$, which implies that the 
Coulomb branch global symmetry algebra is simply $\fru(1)$ associated with the topological $U(1)$ symmetry. \\

For $N_f=2N$, the monopole operators $(\pm 1, 0,\ldots,0)$ have conformal dimension $\Delta = 1 + \frac{P}{2}$, 
while the operator $(1,-1,0,\ldots,0)$ has conformal dimension $\Delta = 2$. The monopole operator with the minimal dimension 
is therefore determined by $P$. In any case, there are no operators with $\Delta =1$, which implies that the Coulomb branch 
global symmetry is again $\fru(1)$. \\

For $N_f=2N-1$, the conformal dimensions of the monopole operators satisfy $\Delta(\vec a)|_{\rm min} > \frac{1}{2}$. The conformal 
dimension of the operators $(\pm 1, 0, \ldots,0)$ is $\Delta = \frac{1}{2} + \frac{P}{2}$, while that of $(1,-1,0,\ldots,0)$ is $\Delta = 1$. 
In contrast to the standard $U(N)$ SQCD with $N_f=2N-1$, this is a good theory for all $P >1$. For $P=1$, there are three generators (in addition 
to the generator of the topological $U(1)$ symmetry) with $\Delta = 1$ which lead to an enhanced $\frsu(2) \oplus \fru(1)$ algebra in the IR. 
For $P>1$, there is only one additional generator with $\Delta = 1$, resulting in an $\fru(1) \oplus \fru(1)$ global symmetry in the IR. Note that this symmetry enhancement is  different from the more familiar global symmetry enhancements in $U(N)$ SQCD with $N_f=2N$ flavors (or a linear quiver 
with a set of balanced unitary gauge nodes), where the Cartan subalgebra of the full global symmetry is visible classically, and the IR enhancement gives a larger symmetry group with the same rank as visible in the UV. In our case, only a subalgebra of the full Cartan is manifest in the UV. Symmetry enhancements of this type are common in orthosymplectic quiver gauge theories and are associated with the so-called \textit{hidden FI parameters}. Here, we encounter 
a case where hidden FI parameters are present in a theory with a unitary gauge group, albeit with these Abelian hypermultiplets. \\

For $N_f=2N-2$, the operators $(\pm 1, 0, \ldots,0)$ have dimension $\Delta = \frac{P}{2}$, while the operator $(1,-1, 0, \ldots,0)$ has 
 $\Delta =0$. For $N_f <2N-2$, one has operators with $\Delta < 0$. The theories with $N_f \leq 2N-2$ are therefore bad theories in 
 the Gaiotto-Witten sense.\\
 
 To summarize, the $\CT^N_{N_f, P}$ theories can be classified in terms of their IR properties as follows:
 
 \begin{enumerate}
 
 \item For $N_f \geq 2N-1$, the theories $\CT^N_{N_f, P}$ are good for all $P \geq 1$. 
 
 \item For $N_f > 2N-1$, the Coulomb branch global symmetry algebra in the IR coincides with the $\fru(1)$ algebra associated with the topological $U(1)$ symmetry visible in the UV.
 
 \item For $N_f =2N-1$, the Coulomb branch global symmetry algebra in the IR is enhanced to $\frsu(2) \oplus \fru(1)$  
 and $\fru(1) \oplus \fru(1)$ for the cases $P=1$ and $P>1$ respectively. We will discuss the global structure of the symmetry 
 in the next section and show that the correct global symmetry for $P=1$ is in fact $SO(3) \times U(1)$.
 
 \item There are no ugly theories for any range of $N_f$ and $P$, i.e. there are no $\Delta = \frac{1}{2}$ operators in theories which 
 do not have unitary-violating operators as well. 
 
 \item For $N_f < 2N-1$, the $\CT^N_{N_f, P}$  theories are bad, for any $P$.
 
 \end{enumerate}

In the rest of the paper, we will only focus on the good theories, i.e. $N_f \geq 2N-1$ with different choices of the integer $P$.

\subsection{Hilbert Series: Coulomb/Higgs branch global symmetries}\label{HS-1}

In this section, we will compute the Coulomb branch and the Higgs branch Hilbert Series (HS) of the $\CT^N_{N_f,P}$ theories 
(see \appref{HS-rev} for a brief review of these observables), and discuss the global symmetries of the respective 
branches including global structures. In particular, we will check our predictions on Coulomb branch global symmetries in \Secref{GBU}.\\

Let us first consider the Coulomb branch HS for the theory $\CT^N_{N_f,P}$, given as
\begin{align}
& \CI^C_{\CT^N_{N_f,P}}(t,w) = \sum_{a_1 \geq a_2 \geq \ldots \geq a_{N} > - \infty }\, w^{2\sum^{N}_{i=1} \,a_i}\,\,t^{\Delta( \vec a)}\, P_{U(N)}(t ; \vec a), \label{MonForm-2}\\
& \Delta( \vec a)=\frac{N_f}{2}\,\sum^{N}_{i=1} |a_i| +\frac{P}{2}\, |\sum^{N}_{i=1} \,a_i|- \sum_{i<j}|a_i-a_j| , \label{RchForm-2}
\end{align}
where the factor $P_{U(N)}$, explicitly given by the formula in \eref{PUN-form}, accounts for the dressing of the bare monopole operator by gauge invariant combinations of the adjoint scalar for the residual gauge group $H(\vec a)$ left unbroken by the flux $\vec a$. In addition, we have turned 
on a fugacity $w$ for the topological $U(1)$ symmetry. For concreteness, let us focus on the case of $N=2$ and $P=1$ : the unrefined HS (i.e. setting 
the fugacity $w=1$) and the associated plethystic logarithms are summarized in \Tabref{Tab: CBHS1} for $N_f=2N+2, 2N+1, 2N, 2N-1$, i.e $N_f=6,5,4,3$.

\begin{table}[htbp]
\begin{center}
{%
\begin{tabular}{|c|c|c|}
\hline
Theory & $\CI^C(t)$ & ${\rm PL}[ \CI^C(t)]$ \\
\hline \hline 
$\CT^2_{6,1}$ & \footnotesize{$1+ t + 2 t^2 + 2 t^{5/2}+ 2t^3 + 4t^{7/2} + 4t^4 + O(t^{9/2})$}  & \footnotesize{$t+t^2 +2 t^{5/2}+ 2t^{7/2} +t^4-2t^6 -t^7-2t^{15/2} + O(t^{19/2})$}\\
\hline
$\CT^2_{5,1}$ & \footnotesize{$1+t + 4t^2 + 7t^3 + 13t^4 + 20t^5 + 33t^6 + O(t^7)$} & \footnotesize{$t+3t^2 + 3t^3 - 2t^5 -3t^6 +O(t^8)$} \\
\hline
$\CT^2_{4,1}$ & \footnotesize{$1+ t + 2t^{3/2}+ 3 t^2 + 4 t^{5/2}+ 6t^3 + 8t^{7/2} + O(t^{4})$}  & \footnotesize{$t+2t^{3/2} +2 t^{2}+ 2t^{5/2} - 2t^{4} - 2t^{9/2} - t^5+O(t^6)$}\\
\hline
$\CT^2_{3,1}$ & \footnotesize{$ 1+4t + 13t^2 + 28t^3 + 55t^4 + 92t^5 + 147t^6 + O(t^7)$} & \footnotesize{$4t + 3t^2 - 4t^3 + 4t^5 -6t^6 + O(t^8)$} \\
\hline
\end{tabular}}
\end{center}
\caption{\footnotesize{Coulomb branch HS and the associated plethystic logarithm for $\CT^2_{N_f,1}$ for $N_f=6,5,4,3$.}}
\label{Tab: CBHS1}
\end{table}

Note that for the cases $N_f=6,5,4$, the HS has a $O(t)$ term with coefficient 1, which corresponds to the classically manifest $U(1)$ 
global symmetry of the Coulomb branch. Also note that, in contrast to the $U(N)$ SQCD, the Coulomb branch of the theory 
$\CT^N_{N_f,P}$ is generically not a complete intersection -- the plethystic logarithm of the Hilbert Series does not terminate after a finite 
number of terms unless $N=1$.

Let us now focus on the case $N_f=3$ for which we expect an enhancement in the global 
symmetry. The HS in this case, refined by the UV-manifest $U(1)$ fugacity, has the following form :
\begin{align}
\CI^C_{\CT^2_{3,1}}(t,w) = 1+ t\,\Big([2]_w +1 \Big) + t^2\,\Big([4]_w +2[2]_w +2 \Big) + t^3\,\Big([8]_w + 2[6]_w + 4[4]_w +3[2]_w +3 \Big) + \ldots,
\end{align}
where we have written down the coefficients of the series in terms of characters of representations of $\frsu(2) \oplus \fru(1)$. Note that we 
denote the character of a spin-$s$ representation of $\frsu(2)$ as $[2s]_w$. The appearance of integer spins only implies that 
the correct global symmetry is $SU(2)/\BZ_2 \times U(1) \cong SO(3) \times U(1)$. The coefficient of the $O(t)$ term shows how the 4 conserved currents are assembled as representations of $SO(3) \times U(1)$.\\

Next, consider the Higgs branch HS for the theory $\CT^N_{N_f,P}$, which can be written in the following form \cite{Razamat:2014pta}:
\begin{align}
\CI^H_{\CT^N_{N_f,P}}(x, \vec \mu, \vec y) = \frac{(1-x)^N}{N!}\,\oint_{|\vec z|=1}\, \prod^N_{i=1}\frac{dz_i}{z_i}\,
& \prod_{i\neq j}\, (1-\frac{z_i}{z_j})(1-\frac{x\,z_i}{z_j})\,\prod^N_{i=1}\prod^{N_f}_{k=1}\prod_{s=\pm}\,\frac{1}{(1-x^{1/2}\,z^s_i\,\mu^{-s}_k)} \nn \\
& \times \prod^P_{l=1}\,\prod_{s=\pm}\,\frac{1}{(1-x^{1/2}\,(\prod^N_{i=1}\,z_i)^s\,y^{-s}_l)},
\end{align}
with $x$ is the $U(1)_{\rm H}$ fugacity and $\vec \mu, \vec y$ are the flavor fugacities for the fundamental hypers and the abelian hypers 
respectively. We would like to emphasize that the refined HS is a function of $N_f + P-1$ flavor symmetry fugacities only, corresponding 
to an overall $U(1)$ being factored out. 
The integration is performed over a contour given by the union of the unit circles $|z_i|=1$, $\forall i$. 
For concreteness, let us focus on $N=2$ -- the unrefined Higgs branch HS (i.e. setting the fugacities $\vec \mu, \vec y =1$) for $N_f=5,4,3$ and $P=1,2$ are shown in \Tabref{Tab: HBHS1}. 
In particular, note that the coefficient of the $O(x)$ term gives the dimension of the adjoint representation of the Lie algebra
 $\frg_{\rm H}= \frsu(N_f) \oplus \frsu(P) \oplus \fru(1)$ associated with the Higgs branch global symmetry group $G_{\rm H}$.

\begin{table}[htbp]
\begin{center}
{%
\begin{tabular}{|c|c|c|}
\hline
Theory & $\CI^H(x)$ & ${\rm PL}[ \CI^H(x)]$ \\
\hline \hline 
$\CT^2_{5,1}$ & \footnotesize{$1+ 25x + 20 x^{3/2} + 300 x^{2}+ 370x^{5/2} + O(x^{3})$}  & \footnotesize{$25x+20x^{3/2} -25x^2 -130 x^{5/2}-86x^3+ O(x^{7/2})$}\\
\hline
$\CT^2_{5,2}$ & \footnotesize{$1+ 28x + 40 x^{3/2} + 380 x^{2}+ 820x^{5/2} + O(x^{3})$} & \footnotesize{$28x+40x^{3/2} -26x^2 -300 x^{5/2}-496x^3+ O(x^{7/2})$} \\
\hline
$\CT^2_{4,1}$ & \footnotesize{$1+ 16x + 12 x^{3/2} + 120 x^{2}+ 140x^{5/2} + O(x^{3})$}  & \footnotesize{$16x+12x^{3/2} -16x^2 -52 x^{5/2}-3x^3+ O(x^{7/2})$}\\
\hline
$\CT^2_{4,2}$ & \footnotesize{$ 1+ 19x + 24 x^{3/2} + 173 x^{2}+ 328x^{5/2} + O(x^{3})$} & \footnotesize{$19x+24x^{3/2} -17x^2 -128 x^{5/2}-145x^3+ O(x^{7/2})$} \\
\hline
$\CT^2_{3,1}$ & \footnotesize{$1+ 9x + 6 x^{3/2} + 36 x^{2}+ 36x^{5/2} + O(x^{3})$}  & \footnotesize{$9x+6x^{3/2} -9x^2 -18 x^{5/2}+ O(x^{3})$}\\
\hline
$\CT^2_{3,2}$ & \footnotesize{$1+ 12x + 12 x^{3/2} + 68 x^{2}+ 96x^{5/2} + O(x^{3})$} & \footnotesize{$12x+12x^{3/2} -10x^2 -48 x^{5/2} - 26x^3+ O(x^{7/2})$} \\
\hline
\end{tabular}}
\end{center}
\caption{\footnotesize{Higgs branch HS and the associated plethystic logarithm for $\CT^2_{N_f,P}$ for $N_f=5,4,3$ and $P=1,2$.}}
\label{Tab: HBHS1}
\end{table}

%Let us now consider the global structure of the symmetry $G_{\rm H}$. To understand this aspect, we need to study the refined 
%Higgs branch HS, and we will focus on the case $N_f=3$ with $P=1,2$. Let us first consider the case of $P=1$. 
%To the first few orders in $x$, the HS for $\CT^2_{3,1}$ is given as:
%\begin{align}
% \CI^H_{\CT^2_{3,1}}(x, \vec \mu,  y) = 1 + &(\mu_1 + \mu_2 + \mu_3)(\mu^{-1}_1 + \mu^{-1}_2 + \mu^{-1}_3)\,x + 
%\Big[ (\prod^3_{k=1}\mu_k)(\mu^{-1}_1 + \mu^{-1}_2 + \mu^{-1}_3) \nn \\
%& +(\prod^3_{k=1}\mu_k)^{-1}(\mu_1 + \mu_2 +\mu_3) \Big]\,x^{3/2} + \ldots,
%\end{align}
%where we have implemented a rescaling of the fugacities as $\mu_k \to \mu_k\,\sqrt{y}$, $\forall k$. The correct global 
%symmetry in this case is $G_{\rm H}=U(3)$.\\
%
%Now consider the case of $P=2$. It is sufficient to consider a partially refined HS in this case where we set $\mu_k=1$, $\forall k$.
%The resultant HS has the form:
%\begin{align}
%\CI^H_{\CT^2_{3,2}}(x, \vec \mu= \vec 1, \vec y) = 1+ & \Big(10 +\frac{y_1}{y_2} + \frac{y_2}{y_1}\Big)x + 3\Big( \frac{1}{y_2} + \frac{1}{y_1} +y_1 +y_2\Big)x^{3/2} \nn \\
%& +\Big(46 + \frac{y^2_1}{y^2_2} + \frac{y^2_2}{y^2_1} + 10\frac{y_1}{y_2} + 10\frac{y_2}{y_1}\Big)x^{2} + \ldots,
%\end{align}
%where we also set $y_1.y_2=1$. Note that both even and odd spin representations of $\frsu(2)$ appear on the RHS -- this implies that 
%the correct global symmetry is $G_{\rm H}=U(3)\times SU(2)$.

\subsection{Three dimensional mirrors}\label{3dmirr-gen}

\begin{figure}[htbp]
\begin{center}
\begin{tabular}{ccc}
 \scalebox{0.7}{\begin{tikzpicture}
\node[unode] (1) {};
\node[unode] (2) [right=.5cm  of 1]{};
\node[unode] (3) [right=.5cm of 2]{};
\node[unode] (4) [right=1cm of 3]{};
\node[] (5) [right=0.5cm of 4]{};
\node[unode] (6) [right=1 cm of 4]{};
\node[unode] (7) [right=1cm of 6]{};
\node[unode] (8) [right=0.5cm of 7]{};
\node[unode] (9) [right=0.5cm of 8]{};
\node[fnode, red] (13) [above=0.5cm of 4]{};
%\node[snode] (14) [left=0.5cm of 13]{1};
\node[fnode] (15) [above=0.5cm of 6]{};
%\node[snode] (11) [below=0.5cm of 9]{1};
\node[] (20) [right=0.2cm of 4]{};
\node[] (21) [left=0.2cm of 6]{};
\node[text width=0.1cm](41)[below=0.2 cm of 1]{1};
\node[text width=0.1cm](42)[below=0.2 cm of 2]{2};
\node[text width=0.1cm](43)[below=0.2 cm of 3]{3};
\node[text width=0.1cm](44)[below=0.2 cm of 4]{$N$};
\node[text width=0.1cm](45)[below=0.2 cm of 6]{$N$};
\node[text width=0.1cm](46)[below=0.2 cm of 7]{3};
\node[text width=0.1cm](47)[below=0.2 cm of 8]{2};
\node[text width=0.1cm](48)[below=0.2 cm of 9]{1};
\node[text width=0.1cm](49)[right=0.1 cm of 15]{1};
\node[text width=0.1cm](50)[left=0.2 cm of 13]{1};
\draw[-] (1) -- (2);
\draw[-] (2)-- (3);
\draw[dashed] (3) -- (4);
\draw[-] (4) --(20);
\draw[-] (21) --(6);
\draw[dashed] (20) -- (21);
\draw[dashed] (6) -- (7);
\draw[-] (7) -- (8);
\draw[-] (8) --(9);
%\draw[-] (5) -- (10);
\draw[-] (4) -- (13);
%\draw[-] (13) -- (14);
\draw[-] (6) -- (15);
\node[text width=0.1cm](30)[below=1 cm of 5]{$(X)$};
\end{tikzpicture}}
& \qquad  
&\scalebox{.8}{\begin{tikzpicture}
\node[unode] (1) at (0,0){};
\node[fnode] (2) at (0,-2){};
\node[text width=0.1cm](41)[above=0.2 cm of 1]{$N$};
\node[text width=0.1cm](42)[right=0.2 cm of 2]{$N_f$};
%\node[snode,blue] (3) at (3,0){$1$};
\draw[-] (1) -- (2);
\node[text width=0.1cm](30)[below=0.1 cm of 2]{$(Y)$};
%\draw[-, thick, blue] (1)-- (3);
%\node[text width=1cm](10) at (1.2, 0.2){$N-1$};
\end{tikzpicture}}\\
 \scalebox{.7}{\begin{tikzpicture}
\draw[->] (15,-3) -- (15,-5);
\node[text width=0.1cm](20) at (14.5, -4) {$\CO$};
\end{tikzpicture}}
&\qquad \qquad 
& \scalebox{.7}{\begin{tikzpicture}
\draw[->] (15,-3) -- (15,-5);
\node[text width=0.1cm](29) at (15.5, -4) {$\wt{\CO}$};
\end{tikzpicture}}\\
 \scalebox{0.7}{\begin{tikzpicture}
\node[unode] (1) {};
\node[unode] (2) [right=.5cm  of 1]{};
\node[unode] (3) [right=.5cm of 2]{};
\node[unode] (4) [right=1cm of 3]{};
\node[] (5) [right=0.5cm of 4]{};
%\node[cnode] (9) [right=1cm of 6]{$\frac{p-v}{2}$};
\node[unode] (6) [right=1 cm of 4]{};
\node[unode] (7) [right=1cm of 6]{};
\node[unode] (8) [right=0.5cm of 7]{};
\node[unode] (9) [right=0.5cm of 8]{};
\node[unode] (13) [above=0.5cm of 4]{};
\node[unode] (14) [left=0.5cm of 13]{};
\node[fnode] (15) [above=0.5cm of 6]{};
\node[unode] (16) [left=0.5cm of 14]{};
\node[fnode] (17) [left=0.5cm of 16]{};
%\node[snode] (18) [above=1 cm of 13]{1};
\node[] (30) [right=0.2cm of 4]{};
\node[] (31) [left=0.2cm of 6]{};
\node[text width=0.1cm](41)[below=0.2 cm of 1]{1};
\node[text width=0.1cm](42)[below=0.2 cm of 2]{2};
\node[text width=0.1cm](43)[below=0.2 cm of 3]{3};
\node[text width=0.1cm](44)[below=0.2 cm of 4]{$N$};
\node[text width=0.1cm](45)[below=0.2 cm of 6]{$N$};
\node[text width=0.1cm](46)[below=0.2 cm of 7]{3};
\node[text width=0.1cm](47)[below=0.2 cm of 8]{2};
\node[text width=0.1cm](48)[below=0.2 cm of 9]{1};
\node[text width=0.1cm](49)[right=0.1 cm of 15]{1};
\node[text width=0.1cm](50)[left=0.2 cm of 17]{1};
\draw[-] (1) -- (2);
\draw[-] (2)-- (3);
\draw[dashed] (3) -- (4);
\draw[-] (4) --(30);
\draw[-] (31) --(6);
\draw[dashed] (30) -- (31);
\draw[dashed] (6) -- (7);
\draw[-] (7) -- (8);
\draw[-] (8) --(9);
%\draw[-] (5) -- (10);
\draw[-] (4) -- (13);
\draw[-] (13) -- (14);
\draw[-] (6) -- (15);
\draw[dashed] (14) -- (16);
\draw[-] (16) -- (17);
%\draw[-] (18) -- (13);
\node[text width=0.1cm](20)[below=1 cm of 5]{$(\wt{\CT}^N_{N_f,P})$};
\node[text width=0.1cm](25)[above=0.2 cm of 13]{1};
\node[text width=0.1cm](26)[above=0.2 cm of 14]{1};
\node[text width=0.1cm](27)[above=0.2 cm of 16]{$1$};
\end{tikzpicture}}
&\qquad \qquad 
& \scalebox{.8}{\begin{tikzpicture}
\node[unode] (1) at (0,0){};
\node[fnode] (2) at (0,-2){};
\node[afnode] (3) at (2,0){};
\node[](4) at (-2,0){};
\draw[-] (1) -- (2);
\draw[-, thick, blue] (1)-- (3);
\node[text width=0.1cm](10) at (0, 0.5){$N$};
\node[text width=1.5cm](11) at (1, -2){$N_f$};
\node[text width=1cm](12) at (3, 0){$P$};
\node[text width=0.1cm](20)[below=0.1 cm of 2]{$({\CT}^N_{N_f,P})$};
\end{tikzpicture}}
\end{tabular}
\caption{\footnotesize{The construction of the 3d mirror for $\CT^N_{N_f,P}$ for $N_f > 2N$ and $P\geq 1$. 
The number of gauge nodes in the chain of $U(1)$ nodes is $P$. 
The $S$-type operation ${\CO}$ is a composition of $P$ flavoring-gauging operations. The dual operation $\wt{\CO}$ amounts to attaching 
$P$ hypermultiplets charged in the determinant representation of the gauge group $U(N)$.}}
\label{SOp-IRdual-gen1}
\end{center}
\end{figure}
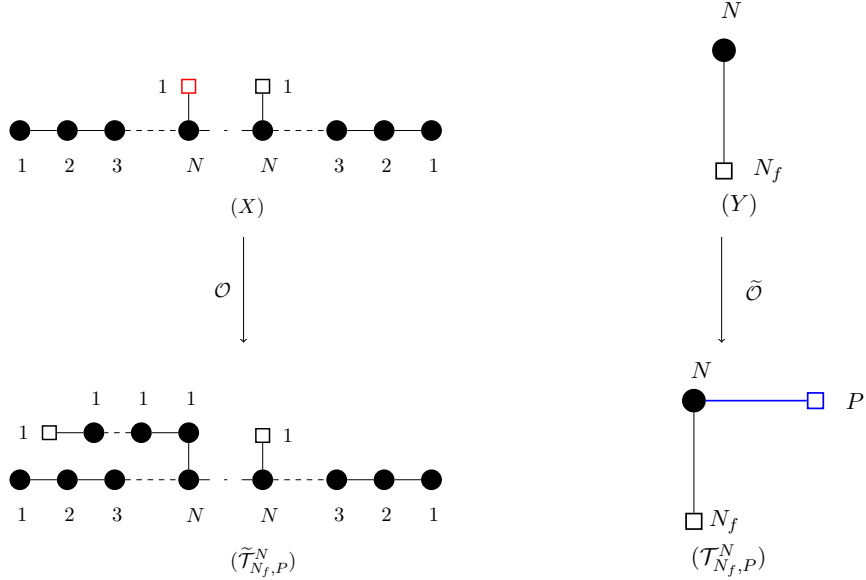

In this section, we will construct the three dimensional mirror of a given theory $\CT^N_{N_f, P}$, for 
$N_f \geq 2N-1$ and $P \geq 1$. 
The 3d mirror can be constructed by implementing an Abelian $S$-type operation \cite{Dey:2020hfe} on a pair $(X,Y)$ of linear quivers 
with unitary gauge groups where the theory $Y$ is a $U(N)$ theory with $N_f$ flavors. The details of the mirror quiver $X$ 
changes with $N_f$, and one should consider the three cases $N_f > 2N$, $N_f=2N$, and $N_f=2N-1$ separately. 
Consider the case $N_f > 2N$ -- the pair $(X,Y)$ are shown in the first row of \figref{SOp-IRdual-gen1}. The quiver $X$ 
has a gauge group $G=\prod^{N-1}_{i=1} U(i) \times U(N)^{N_f-2N+1} \times \prod^{N-1}_{j=1} U(N-j)$ with bifundamental 
hypermultiplets and a single fundamental hypermultiplet each for the gauge nodes $U(N)_1$ and $U(N)_{N_f-2N+1}$. 

Given the quiver $X$, one implements an $S$-type operation of the flavoring-gauging type at the $U(1)$ flavor node (marked in red) 
of $X$ which gives the quiver $\wt{\CT}^N_{N_f,1}$ (see the quiver diagram $\wt{\CT}^N_{N_f,P}$ on the bottom left-hand corner of 
\figref{SOp-IRdual-gen1}). On the dual side, this operation involves attaching a single abelian hypermultiplet to the $U(N)$ gauge 
node of the theory $Y$, which gives the quiver ${\CT}^N_{N_f,1}$. 
In the next step, one implements another flavoring-gauging operation at the $U(1)$ flavor node attached to the $U(1)$ 
gauge node in $\wt{\CT}^N_{N_f,1}$ for which the dual operation involves attaching another abelian hypermultiplet to the $U(N)$ 
gauge node of ${\CT}^N_{N_f,1}$ giving the quiver ${\CT}^N_{N_f,2}$. Repeating the procedure $P$ times, one obtains the 
mirror pairs on the bottom row of \figref{SOp-IRdual-gen1}.

The basic data of the 3d mirror symmetry can be checked as follows. Firstly, one can check that the moduli space 
dimensions agree:
\be
{\rm dim}\,\CM^{(\wt{\CT}^N_{N_f,P})}_{\rm C} = {\rm dim}\,\CM^{({\CT}^N_{N_f,P})}_{\rm H} =NN_f +P -N^2, \qquad 
{\rm dim}\,\CM^{(\wt{\CT}^N_{N_f,P})}_{\rm H} = {\rm dim}\,\CM^{({\CT}^N_{N_f,P})}_{\rm C} =N.
\ee
The theory ${\CT}^N_{N_f,P}$ has a $\fru(1)$ Coulomb branch global symmetry algebra which is realized in the theory $\wt{\CT}^N_{N_f,P}$
as the $\fru(1)$ Higgs branch global symmetry algebra associated with the two fundamental hypers on different gauge nodes.
The $\frsu(N_f) \oplus \frsu(P) \oplus \fru(1)$ algebra of the Higgs branch for ${\CT}^N_{N_f,P}$ arises as the Coulomb branch global 
symmetry algebra of $\wt{\CT}^N_{N_f,P}$. The latter can be read off as follows. The tail of $U(1)$ gauge nodes has $P-1$ balanced nodes and 
a single unbalanced node giving a factor $\frsu(P) \oplus \fru(1)$. In addition, an $\frsu(N_f)$ factor arises from the subquiver 
with $N_f-1$ balanced nodes. Finally, as discussed in \cite{Dey:2020hfe}, one can implement the aforementioned $S$-type operations 
using sphere partition function and superconformal index, thereby checking the 3d mirror symmetry in terms of these 
observables. The partition function computation is summarized in \appref{PF-SOp}, while the superconformal index computation 
works out in an analogous fashion.\\

\begin{figure}[htbp]
\begin{center}
 \scalebox{0.8}{\begin{tikzpicture}
\node[unode] (1) {};
\node[unode] (2) [right=.5cm  of 1]{};
\node[unode] (3) [right=.5cm of 2]{};
\node[unode] (4) [right=1cm of 3]{};
\node[unode] (5) [right=1cm of 4]{};
\node[unode] (6) [right=1 cm of 5]{};
\node[unode] (7) [right=1cm of 6]{};
\node[unode] (8) [right=0.5cm of 7]{};
\node[unode] (9) [right=0.5cm of 8]{};
\node[unode] (13) [above= 1cm of 5]{};
\node[fnode] (14) [below=1 cm of 5]{};
\node[unode] (15) [left=1 cm of 13]{};
\node[unode] (16) [left=1 cm of 15]{};
\node[fnode] (17) [left=1 cm of 16]{};
%\node[snode] (11) [below=0.5cm of 9]{1};
\node[text width=0.1cm](41)[below=0.2 cm of 1]{1};
\node[text width=0.1cm](42)[below=0.2 cm of 2]{2};
\node[text width=0.1cm](43)[below=0.2 cm of 3]{3};
\node[text width=1cm](44)[below=0.2 cm of 4]{$N-1$};
\node[text width=0.1cm](45)[below=0.2 cm of 5]{$N$};
\node[text width=1cm](46)[below=0.2 cm of 6]{$N-1$};
\node[text width=0.1cm](47)[below=0.2 cm of 7]{3};
\node[text width=0.1cm](48)[below=0.2 cm of 8]{2};
\node[text width=0.1cm](49)[below=0.1 cm of 9]{1};
\node[text width=0.1cm](50)[right=0.1 cm of 14]{1};
\node[text width=0.1cm](51)[left=0.2 cm of 17]{1};
\draw[-] (1) -- (2);
\draw[-] (2)-- (3);
\draw[dashed] (3) -- (4);
\draw[-] (4) --(5);
\draw[-] (5) --(6);
\draw[dashed] (6) -- (7);
\draw[-] (7) -- (8);
\draw[-] (8) --(9);
\draw[-] (5) -- (13);
%\draw[-] (4) -- (13);
\draw[-] (13) -- (15);
\draw[dashed] (16) -- (15);
\draw[-] (16) -- (17);
\draw[-] (5) -- (14);
\node[text width=0.1cm](25)[above=0.2 cm of 13]{1};
\node[text width=0.1cm](26)[above=0.2 cm of 15]{1};
\node[text width=0.1cm](27)[above=0.2 cm of 16]{$1$};
\node[text width=0.1cm](20)[below=2 cm of 5]{$(\wt{\CT}^N_{2N,P})$};
\end{tikzpicture}}
\caption{\footnotesize{The 3d mirror of the theory ${\CT}^N_{2N,P}$.}}
\label{SOp-IRdual-gen2}
\end{center}
\end{figure}
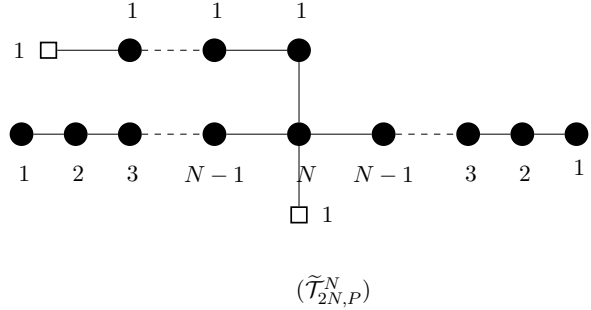

For $N_f=2N$, the 3d mirror $X$ of $Y$ is a quiver gauge theory with gauge group 
$G= \prod^{N-1}_{i=1} U(i) \times U(N)\times \prod^{N-1}_{j=1} U(N-j)$ with bifundamentals 
and two fundamental hypermultiplets for the $U(N)$ gauge node. Following the same construction 
as above, one can engineer the 3d mirror of $\CT^{N}_{N_f, P}$, which is shown in \figref{SOp-IRdual-gen2}.
The matching of moduli space dimensions and global symmetries work out in a similar fashion as above.

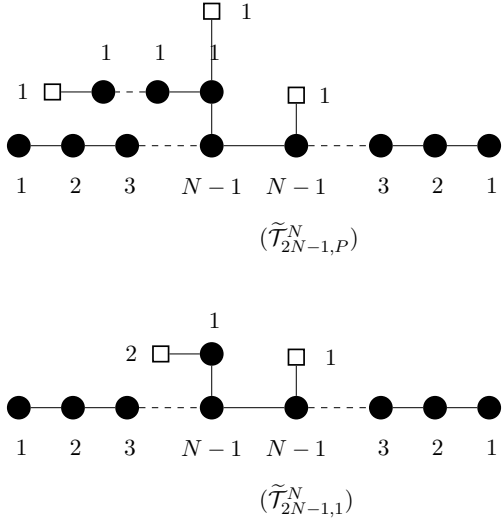
\begin{figure}[htbp]
\begin{center}
 \scalebox{0.8}{\begin{tikzpicture}
\node[unode] (1) {};
\node[unode] (2) [right=.5cm  of 1]{};
\node[unode] (3) [right=.5cm of 2]{};
\node[unode] (4) [right=1cm of 3]{};
\node[] (5) [right=0.5cm of 4]{};
%\node[cnode] (9) [right=1cm of 6]{$\frac{p-v}{2}$};
\node[unode] (6) [right=1 cm of 4]{};
\node[unode] (7) [right=1cm of 6]{};
\node[unode] (8) [right=0.5cm of 7]{};
\node[unode] (9) [right=0.5cm of 8]{};
\node[unode] (13) [above=0.5cm of 4]{};
\node[unode] (14) [left=0.5cm of 13]{};
\node[fnode] (15) [above=0.5cm of 6]{};
\node[unode] (16) [left=0.5cm of 14]{};
\node[fnode] (17) [left=0.5cm of 16]{};
\node[fnode] (18) [above=1 cm of 13]{};
\node[text width=0.1cm](41)[below=0.2 cm of 1]{1};
\node[text width=0.1cm](42)[below=0.2 cm of 2]{2};
\node[text width=0.1cm](43)[below=0.2 cm of 3]{3};
\node[text width=1cm](44)[below=0.2 cm of 4]{$N-1$};
\node[text width=1cm](45)[below=0.2 cm of 6]{$N-1$};
\node[text width=0.1cm](46)[below=0.2 cm of 7]{3};
\node[text width=0.1cm](47)[below=0.2 cm of 8]{2};
\node[text width=0.1cm](48)[below=0.2 cm of 9]{1};
\node[text width=0.1cm](49)[right=0.1 cm of 15]{1};
\node[text width=0.1cm](50)[left=0.2 cm of 17]{1};
\node[text width=0.1cm](51)[right=0.2 cm of 18]{1};
\draw[-] (1) -- (2);
\draw[-] (2)-- (3);
\draw[dashed] (3) -- (4);
\draw[-] (4) --(6);
%\draw[-] (5) --(6);
\draw[dashed] (6) -- (7);
\draw[-] (7) -- (8);
\draw[-] (8) --(9);
%\draw[-] (5) -- (10);
\draw[-] (4) -- (13);
\draw[-] (13) -- (14);
\draw[-] (6) -- (15);
\draw[dashed] (14) -- (16);
\draw[-] (16) -- (17);
\draw[-] (18) -- (13);
\node[text width=0.1cm](25)[above=0.2 cm of 13]{1};
\node[text width=0.1cm](26)[above=0.2 cm of 14]{1};
\node[text width=0.1cm](27)[above=0.2 cm of 16]{1};
\node[text width=0.1cm](20)[below=1 cm of 5]{$(\wt{\CT}^{N}_{2N-1,P})$};
\end{tikzpicture}}\\
\qquad \\
\scalebox{0.8}{\begin{tikzpicture}
\node[unode] (1) {};
\node[unode] (2) [right=.5cm  of 1]{};
\node[unode] (3) [right=.5cm of 2]{};
\node[unode] (4) [right=1cm of 3]{};
\node[] (5) [right=0.5cm of 4]{};
%\node[cnode] (9) [right=1cm of 6]{$\frac{p-v}{2}$};
\node[unode] (6) [right=1 cm of 4]{};
\node[unode] (7) [right=1cm of 6]{};
\node[unode] (8) [right=0.5cm of 7]{};
\node[unode] (9) [right=0.5cm of 8]{};
\node[unode] (13) [above=0.5cm of 4]{};
\node[fnode] (14) [left=0.5cm of 13]{};
\node[fnode] (15) [above=0.5cm of 6]{};
%\node[snode] (11) [below=0.5cm of 9]{1};
\node[text width=0.1cm](41)[below=0.2 cm of 1]{1};
\node[text width=0.1cm](42)[below=0.2 cm of 2]{2};
\node[text width=0.1cm](43)[below=0.2 cm of 3]{3};
\node[text width=1cm](44)[below=0.2 cm of 4]{$N-1$};
\node[text width=1cm](45)[below=0.2 cm of 6]{$N-1$};
\node[text width=0.1cm](46)[below=0.2 cm of 7]{3};
\node[text width=0.1cm](47)[below=0.2 cm of 8]{2};
\node[text width=0.1cm](48)[below=0.2 cm of 9]{1};
\node[text width=0.1cm](49)[above=0.1 cm of 13]{1};
\node[text width=0.1cm](50)[left=0.2 cm of 14]{2};
\node[text width=0.1cm](51)[right=0.2 cm of 15]{1};
\draw[-] (1) -- (2);
\draw[-] (2)-- (3);
\draw[dashed] (3) -- (4);
\draw[-] (4) --(6);
%\draw[-] (5) --(6);
\draw[dashed] (6) -- (7);
\draw[-] (7) -- (8);
\draw[-] (8) --(9);
%\draw[-] (5) -- (10);
\draw[-] (4) -- (13);
\draw[-] (13) -- (14);
\draw[-] (6) -- (15);
\node[text width=0.1cm](20)[below=1 cm of 5]{$(\wt{\CT}^N_{2N-1,1})$};
\end{tikzpicture}}
\caption{\footnotesize{The 3d mirror of the theort $\CT^{N}_{2N-1,P}$ for $P>1$ is shown on top. The mirror for $P=1$ is given on the 
bottom.} }
\label{SOp-IRdual-gen3}
\end{center}
\end{figure}

Finally, for $N_f=2N-1$, the theory $Y$ is ugly and has a Seiberg-like dual involving the theory $\CT^{N-1}_{2N-1,0}$ --
a $U(N-1)$ theory with $2N-1$ fundamental hypers, and a decoupled twisted hyper. 
The theory $X$, which is the 3d mirror of $Y$, is therefore given by the 3d mirror of $\CT^{N-1}_{2N-1,0}$ and a decoupled 
$\CT^{1}_{1,0}$ theory. Using an appropriate $S$-type operation, one can again construct the 3d mirror of $\CT^{N-1}_{2N-1,P}$
-- the resultant quiver is given in \figref{SOp-IRdual-gen3}. For the $P>1$ case, the number of $U(1)$ gauge nodes in the quiver 
tail attached to one of the $U(N-1)$ gauge nodes is $P$.\\

In this case, it is particularly interesting to check how the Coulomb branch global 
symmetry of $\CT^{N-1}_{2N-1,P}$ is realized as the Higgs branch global symmetry of the 3d mirror, since we know that the 
former is enhanced in the IR. From the quiver in \figref{SOp-IRdual-gen3}, one can check that this enhanced symmetry is 
realized as UV-manifest global symmetry. For $P >1$, the Higgs branch global symmetry is $\fru(1) \oplus \fru(1)$, as expected. 
For $P=1$, the global symmetry is $\frsu(2) \oplus \fru(1)$, which is precisely what we found in 
\Secref{GBU} and \Secref{HS-1}. The Higgs branch global symmetry of $\CT^{N-1}_{2N-1,P}$ matches with the Coulomb 
branch of the 3d mirror in a fashion similar to the cases studied above.

\section{Dualities for $U(N)$ SQCD with abelian hypermultiplets}\label{Dualities}

In this section, we state the dualities for the $\CT^N_{N_f,P}$ quiver gauge theories and perform various checks on 
the proposed dualities, which includes matching the sphere partition functions and the Coulomb/Higgs branch Hilbert 
Series. In addition, for a given IR dual, we will determine the associated 3d mirror. 

\subsection{Duality $\CD^N_{2N+1,1}$ : $N_f=2N+1$ and $P = 1$}\label{BD-main}

\begin{figure}[htbp]
\begin{center}
\begin{tabular}{ccc}
\scalebox{.8}{\begin{tikzpicture}
\node[unode] (1) at (0,0){};
\node[fnode] (2) at (0,-2){};
\node[afnode] (3) at (3,0){};
\draw[-] (1) -- (2);
\draw[-, thick, blue] (1)-- (3);
\node[text width=0.1cm](10) at (0, 0.5){$N$};
\node[text width=1.5cm](11) at (1, -2){$2N+1$};
\node[text width=1cm](12) at (4, 0){$1$};
%\node[text width=1cm](11) at (0, 1.1){$\eta=0$};
\node[text width=0.1cm](20)[below=0.5 cm of 2]{$(\CT)$};
\end{tikzpicture}}
& \qquad   \qquad \qquad \qquad
&\scalebox{.8}{\begin{tikzpicture}
\node[sunode] (1) at (0,0){};
\node[fnode] (2) at (0,-2){};
\draw[-] (1) -- (2);
\node[text width=1 cm](10) at (0, 0.5){$N+1$};
\node[text width=1.5cm](11) at (1, -2){$2N+1$};
\node[text width=0.1cm](20)[below=0.5 cm of 2]{$(\CT^\vee)$};
\end{tikzpicture}}
\end{tabular}
\end{center}
\caption{\footnotesize{IR duality involving a $\CT^N_{2N+1,1}$ theory and an $SU(N+1)$ theory with $2N+1$ flavors.}}
\label{IRdual-Ex1}
\end{figure}
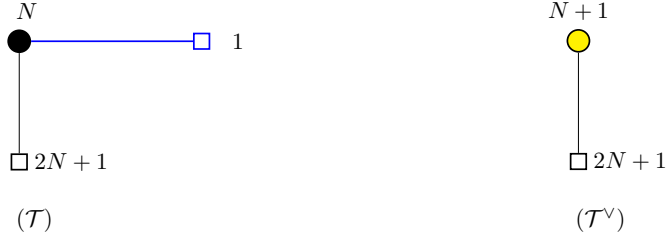

The proposed IR duality $\CD^N_{2N+1,1}$ (the subscript indicating that we have a one-parameter family of dualities for generic 
$N$ and $P=1$), shown in \figref{IRdual-Ex1}, involves the following theories:

\begin{itemize}

\item Theory $\CT$: $U(N)$ gauge theory with $N_f=2N+1$ fundamental hypers and a single Abelian hypermultiplet of 
charge $N$. In the notation of the previous section, this corresponds to the theory $\CT^N_{2N+1, 1}$.

\item Theory $\CT^\vee$:  $SU(N+1)$ gauge theory with $N_f=2N+1$ fundamental hypers.

\end{itemize}

\begin{table}[htbp]
\begin{center}
{%
\begin{tabular}{|c|c|c|}
\hline
Moduli space data & Theory $\CT$ & Theory $\CT^\vee$ \\
\hline \hline 
dim\,$\CM_H$ & $N^2 +N +1$  & $N^2 +N +1$\\
\hline
dim\,$\CM_C$ & $N$ & $N$ \\
\hline
$\frg_H$ & $\frsu(2N+1) \oplus \fru(1)$  & $\frsu(2N+1) \oplus \fru(1)$\\
\hline
$\frg_C$ & $ \fru(1) $ & $\fru(1) $\\
\hline
\end{tabular}}
\end{center}
\caption{\footnotesize{The matching of moduli space data for the proposed dual pair.}}
\label{Tab: D1}
\end{table}

To begin with, note that both theories are good in the Gaiotto-Witten sense. For the theory $\CT$, since $N_f > 2N$, this follows from the 
discussion in \Secref{GBU}, while we will show below that this is true for the $\CT^\vee$ theory. 
The dimensions of the Coulomb and the Higgs branches as well as the respective global symmetry algebras have been summarized in Table \ref{Tab: D1}.
Let us first discuss the matching of global symmetries. The matching of the Higgs branch global symmetry can be readily checked from the 
quivers. The theory $\CT^\vee$ has a $\frsu(2N+1) \oplus \fru(1)$ algebra arising from the fundamental hypermultiplets, since the gauge 
group is special unitary.
The theory $\CT$ has a 
global symmetry $\frsu(2N+1) \oplus \fru(1)$, where the $\frsu(2N+1)$ factor and the $\fru(1)$ factor arise from the fundamental hypers 
and the single Abelian hyper respectively for a unitary gauge group. 
The matching of the Coulomb branch global symmetry, however, is non-trivial. The theory $\CT$ 
does have a topological $\fru(1)$ symmetry manifest in the UV Lagrangian, while the theory $\CT^\vee$ does not. For $\CT^\vee$, the $\fru(1)$ 
symmetry arises as an emergent symmetry in the IR. Recall that the R-charge of a bare monopole operator for the 
$SU(N+1)$ theory, labelled by a GNO charge $\vec p$, are given as:
\begin{align}
\Delta( \vec p)= & \frac{2N+1}{2}\,\sum^{N+1}_i |p_i| - \sum_{i<j}|p_i-p_j| , \qquad \sum_i p_i =0, \nn \\
=& \frac{1}{2}\,\sum^{N+1}_i |p_i|  + \sum_{i<j}(|p_i| + |p_j|-|p_i-p_j|).
\end{align}
The spectrum has no monopole operators with R-charge $\Delta \leq \frac{1}{2}$, and is therefore a good theory.
The spectrum has a single $\Delta =1$ operator labelled by $\vec p=(1,0,\ldots,0,-1)$ -- this monopole operator is 
therefore associated with the generator of the $\fru(1)$ global symmetry algebra in the IR.\\

The duality $\CD^N_{2N+1,1}$ can be derived from the following 3d Seiberg-like duality \cite{Kapustin:2010mh} by a well-defined QFT operation.
A $U(N_c)$ gauge theory with $N_f=2N_c-1$  fundamental hypers,  being an ugly theory in the Gaiotto-Witten sense, 
factorizes in the IR into an interacting SCFT and a single twisted hypermultiplet. The SCFT has a good UV description as 
a $U(N_c-1)$ gauge theory with $N_f=2N_c-1$ fundamental hypers. An $SU(N_c)$ gauge theory with $N_f=2N_c-1$ 
flavors can be obtained from the corresponding $U(N_c)$ theory by gauging the $U(1)$ topological symmetry. On the 
dual side, the gauging operation leads to a $U(N_c-1)$ gauge theory with $N_f=2N_c-1$ fundamental hypers and a 
single Abelian hypermultiplet. For $N_c=N+1$, this reproduces the duality in \figref{IRdual-Ex1}.
Let us demonstrate this operation using the round three-sphere partition function\footnote{The operation can be performed in terms of the 
superconformal index in an analogous fashion.}. The 3d Seiberg-like duality 
translates to the following identity \cite{Kapustin:2010mh} in terms of the sphere partition function (for a slightly 
different derivation see \Appref{UN-2N-1}):

\begin{align} \label{ugly-id}
Z^{\CT^{N_c}_{2N_c-1,0}}(\vec m, \eta) =& \Big(\frac{e^{2\pi i \eta \tr \vec m}}{\ch{\eta}}\Big) \cdot Z^{\CT^{N_c-1}_{2N_c-1,0}}(\vec m, -\eta)
 \nn \\
= & \int\, d\s\,\frac{e^{2\pi i \eta \,\s}}{\ch{(\s - \tr \vec m)}} \cdot Z^{\CT^{N_c-1}_{2N_c-1,0}}(\vec m, -\eta)\nn \\
= & Z^{\CT^1_{1,0}}(\tr \vec m, \eta) \cdot Z^{\CT^{N_c-1}_{2N_c-1,0}}(\vec m, -\eta),
\end{align}
where $\tr \vec m=\sum^{2N_c-1}_{i=1} m_i$. The partition function of the $SU(N_c)$ theory can be obtained from the partition function of the 
unitary theory on the LHS by integrating over the FI parameter $\eta$, i.e.
\be 
Z^{SU(N_c),\, 2N_c-1}(\vec m) = \int \, d\eta \, Z^{\CT^{N_c}_{2N_c-1,0}}(\vec m, \eta) =  \int \, [d \vec s] \, \delta(\tr \vec s)\, \frac{\prod_{j<k}\, \sinh^2{\pi(s_j- s_k)}}{\prod^{N_c}_{j=1} \prod^{2N_c-1}_{i=1}\ch{(s_j- m_i)}}.
\ee
Integrating both sides of the identity \eref{ugly-id} over $\eta$, we get 
\begin{align}\label{Id-1}
Z^{SU(N_c),\,2N_c-1}(\vec m) = & \int \, d\eta \, \Big(\frac{e^{2\pi i \eta \tr \vec m}}{\ch{\eta}}\Big) \, \int \, [d \vec \s] \, e^{-2\pi i \eta \, \tr \s} \,\frac{\prod_{j<k}\, \sinh^2{\pi(\s_j- \s_k)}}{\prod^{N_c-1}_{j=1} \prod^{2N_c-1}_{i=1}\ch{(\s_j- m_i)}} \nn \\
= & \int \, [d \vec \s] \,d\eta \, \Big(\frac{e^{2\pi i \eta (\tr \vec m -\tr \s)}}{\ch{\eta}}\Big) \, \frac{\prod_{j<k}\, \sinh^2{\pi(\s_j- \s_k)}}{\prod^{N_c-1}_{j=1} \prod^{2N_c-1}_{i=1}\ch{(\s_j- m_i)}} \nn \\
= &  \int \, [d \vec \s] \ \frac{1}{\ch{(\tr \vec \s - \tr \vec m)}}\,\frac{\prod_{j<k}\, \sinh^2{\pi(\s_j- \s_k)}}{\prod^{N_c-1}_{j=1} \prod^{2N_c-1}_{i=1}\ch{(\s_j- m_i)}},
\end{align}
where the final form of the matrix integral on the RHS can be identified as the partition function of a $U(N_c-1)$ theory with $2N_c-1$ fundamental 
hypers and a single Abelian hypermultiplet. Finally, setting $N_c=N+1$, we obtain the 
following relation between the partition functions of the theories $(\CT, \CT^\vee)$ in \figref{IRdual-Ex1}:
\be \label{Id-0}
\boxed{Z^{(\CT)} (\vec m, m_{\rm ab}= \tr \vec m, \eta=0) = Z^{(\CT^\vee)}(\vec m),}
\ee
where $m_{\rm ab}$ is the real mass for the Abelian hypermultiplet. Note that the equality of the partition 
functions holds only when the FI parameter of the $U(N)$ vector multiplet is tuned to zero. This 
is expected since the $SU(N+1)$ theory does not have a $U(1)$ topological symmetry for which one can 
turn on a chemical potential in the UV.

In the expression on the RHS of \eref{Id-1}, note that all the masses in the $U(N)$ theory can be shifted by a real parameter without 
changing the partition function relation. 
The $2N+1$ masses for the $SU(N+1)$ gauge theory and the $2N+1$ independent masses for the $U(N)$ gauge theory 
live in the Cartan subalgebra of the respective Higgs branch global symmetry algebra $\frsu(2N+1) \oplus \fru(1)$. 
%The $2N+1$ independent masses for the $U(N)$ gauge theory 
%also live in the Cartan subalgebra of the Higgs branch global symmetry $U(2N+1) \cong \Big(U(2N+1) \times U(1)\Big)/U(1)$.\\

The $SU(N+1)$ theory has a $U(1)_B$ baryonic symmetry, under which the fundamental hypermultiplets are charged.
To see how this symmetry maps across the duality, we need to rewrite partition function identity in a slightly different 
fashion. Let us first reparametrize the real masses of the $SU(N)$ theory in a way that makes the Higgs branch 
global symmetry algebra $\frsu(2N+1) \oplus \fru(1)_B$ manifest, i.e. $m_i := \mu_i + \frac{1}{N+1}\,\mu', \forall i$ such that 
$\tr \vec \mu =0$. The parameter $\mu'=\frac{N+1}{2N+1}\,\tr \vec m$ can then be identified as the real mass for the $\fru(1)_B$ algebra. 
This normalization ensures that the baryons and anti-baryons of the $SU(N+1)$ theory have $\fru(1)_B$ charges $\pm 1$ respectively.
The partition function identity \eref{Id-0} can then be written as
\begin{align}\label{Id-0a}
Z^{SU(N+1),\,2N+1}(\vec \mu, \mu') =& \int \, [d \vec \s] \ \frac{1}{\ch{(\tr \vec \s - \frac{(2N+1)}{N+1}\mu')}}\,\frac{\prod_{j<k}\, \sinh^2{\pi(\s_j- \s_k)}}{\prod^{N}_{j=1} \prod^{2N_c+1}_{i=1}\ch{(\s_j- \mu_i- \frac{\mu'}{N+1})}} \nn \\
=& Z^{\CT^N_{2N+1,1}}(\vec \mu, \mu'; \eta=0),
\end{align}
from which the $\fru(1)_B$ charges of the fundamental hypers and the Abelian hypers in the theory $\CT^N_{2N+1,1}$ can be read off. 
Using the property that the matrix integral is invariant under uniform translations of $\vec \s$, one can check that the $\fru(1)_B$ charges 
can be shifted in the following fashion:
\be
Q^B_{\rm fund} \to Q^B_{\rm fund} - q, \qquad Q^B_{\rm ab} \to Q^B_{\rm ab} - N\,q,
\ee
where $Q^B_{\rm fund}, Q^B_{\rm ab}$ are the $\fru(1)_B$ charges for the fundamental hypers and the Abelian hyper 
respectively, and $q$ is a rational number. Gauging this baryonic symmetry leads us back to the original 
Seiberg-like duality that we started with, as one might expect.\\

A second check of the duality $\CD^N_{2N+1,1}$ can be performed by comparing the Coulomb branch and Higgs branch Hilbert Series of the 
two dual theories. In particular, the matching of the Coulomb branch Hilbert Series is non-trivial given the emergent global 
symmetry for the $SU(N+1)$ theory. Following the general procedure in \cite{Cremonesi:2013lqa} (reviewed in \appref{HS-CB-app}), the agreement 
between the Hilbert Series for the theories $\CT$ and $\CT^\vee$ can be checked for any $N$ -- the specific cases of 
$N=1,2,3$ are worked out in \appref{HS-CB-app}. 
Consider, for example, the case of $N=2$, where one can check that
\begin{align}
& \CI^C_{\CT^2_{5,1}}(t) =\CI^C_{SU(3),5}(t) = 1+t+4 t^2+7 t^3+13 t^4+20 t^5+33 t^6+45 t^7+O\left(t^{8}\right), \\
&  {\rm PL}[ \CI^C_{\CT^2_{5,1}}(t)]= {\rm PL}[ \CI^C_{SU(3),5}(t)] = t+3 t^2+3t^3 -2 t^5 -3t^6+O(t^8),
\end{align}
The existence of the emergent $\fru(1)$ algebra can be read off from the $O(t)$ term of the HS for the $SU(3)$ theory, 
while the $O(t)$ term in the plethystic logarithm corresponds to the associated monopole operator with conformal dimension 1. 
It is instructive to compare this with the HS and the associated plethystic logarithm for an $SU(3)$ theory with 6 flavors 
(i.e. $N_f > 2N_c-1$):
\begin{align}
{\rm PL}[ \CI^C_{SU(3),6}(t)]= 2t^2 + 3t^3 +2t^4 +t^5 -t^6 -2t^7 -3t^8 -2t^9 + O(t^{10}),
\end{align}
where the absence of an $O(t)$ term indicates that the Coulomb branch global symmetry 
of the theory is trivial. 

In contrast to the $U(N)$ theory with $2N+1$ flavors (or any $N_f \geq 2N$), the Coulomb branch of the theory $\CT$ 
is not a complete intersection (except for the case $N=1$), since the PL of the Hilbert Series gives an infinite series. 
The Coulomb branch of the theory $\CT^\vee$ is also not a complete intersection unless $N=1$.\\

Similarly, the agreement of the Higgs branch Hilbert Series (reviewed in \appref{HS-HB-app}) for the dual 
theories $(\CT, \CT^\vee)$ can be checked for any $N$. For example, for the case of $N=2$, one gets:
\begin{align}
& \CI^H_{\CT^2_{5,1}}(x) =\CI^H_{SU(3),5}(x) = 1+ 25x + 20 x^{3/2} + 300 x^{2}+ 370x^{5/2} + O(x^{3}), \\
&  {\rm PL}[ \CI^H_{\CT^2_{5,1}}(x)]= {\rm PL}[ \CI^H_{SU(3),5}(x)] = 25x+20x^{3/2} -25x^2 -130 x^{5/2}-86x^3+ O(x^{7/2}),
\end{align}
where the $O(x)$ term gives the dimension of the adjoint representation of the Lie algebra $\frg_{\rm H} = \frsu(5) \oplus \fru(1)$.

\subsection{Duality $\CD^N_{2N,P}$ : $N_f=2N$ and $P \geq 1$} \label{SD-main}

\begin{figure}[htbp]
\begin{center}
\scalebox{.8}{\begin{tikzpicture}
\node[unode] (1) at (0,0){};
\node[fnode] (2) at (0,-2){};
\node[afnode] (3) at (3,0){};
\node[] (4) at (-2,0){};
\draw[-] (1) -- (2);
\draw[-, thick, blue] (1)-- (3);
\node[text width=0.1cm](10) at (0, 0.5){$N$};
\node[text width=1.5cm](11) at (1, -2){$2N$};
\node[text width=1cm](12) at (4, 0){$P$};
\node[text width=0.1cm](20)[below=0.5 cm of 2]{$(\CT)$};
\end{tikzpicture}}
\end{center}
\caption{\footnotesize{A self-dual theory $\CT^N_{2N,P}$.}}
\label{IRdual-sd}
\end{figure}
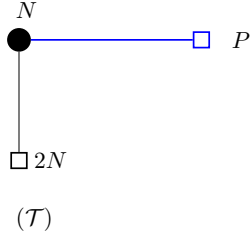

The proposed IR duality $\CD^N_{2N,P}$ (the subscript indicating that we have a two-parameter family of dualities for generic 
$N$ and $P$) is the self-duality of the theory $\CT^N_{2N, P}$ -- a $U(N)$ gauge theory with $N_f=2N$ and $P$ Abelian hypermultiplets, 
as shown in \figref{IRdual-sd}. The Higgs branch global symmetry algebra $\frg_{\rm H} = \frsu(2N) \oplus \frsu(P) \oplus \fru(1)$ can be 
read off from the quiver. The Coulomb branch global symmetry algebra is $\frg_{\rm C} = \fru(1)$ arises from the topological symmetry of the 
unitary gauge group. In contrast to the case of a $U(N)$ SQCD with 2N fundamental flavors, the global symmetry is not enhanced to an $\frsu(2)$ 
for any $P\geq 1$, as we discussed in \Secref{GBU}. 

The self-duality of the theory $\CT^N_{2N,1}$ can be derived from the dual pair in \figref{IRdual-Ex1} 
by giving a large real mass to one of the fundamental hypers in the theory $\CT^N_{2N+1,1}$, and 
reading off the correct low energy effective theory on the dual side. We will perform this exercise in 
\Secref{RG-duality-1}. For now, we simply check this duality using the three-sphere partition function. 
The starting point is the following identity for a $U(N)$ gauge theory with $2N$ fundamental flavors 
(reviewed in \appref{UN-2N}) :
\be \label{UN-2N-pf}
Z^{\CT^N_{2N,0}}(\vec m, \eta) = Z^{\CT^N_{2N,0}}(\vec m, -\eta),
\ee
where $\vec m$ are the fundamental masses satisfying $\sum^{2N}_{a=1}\,m_a=0$, and $\eta$ is the FI parameter. 
The theory $\CT^N_{2N,1}$ can be obtained from $\CT^N_{2N,0} $
by identifying the $U(1)$ gauge group of a $\CT^1_{1,0} $ theory (i.e. $U(1)$ with a single flavor) with the central $U(1)$ subgroup of the $U(N)$ 
gauge group in $\CT^N_{2N,0} $. This amounts to gauging a diagonal combination of the two topological $U(1)$ symmetries in $\CT^1_{1,0} $ 
and $\CT^N_{2N,0} $ respectively. At the level of the partition function, the operation can be implemented as follows. Let 
$\eta=\frac{1}{2}\,(\eta_+ +\eta_-)$, and one can readily check that
\begin{align}
Z^{\CT^N_{2N,1}}(\vec m, m', \eta_+) = & \int\,d\eta_-\, Z^{\CT^1_{1,0}}(m', \frac{(\eta_+ -\eta_-)}{2})\, Z^{\CT^N_{2N,0}}(\vec m, \frac{(\eta_+ +\eta_-)}{2}) \nn \\
=&  \int\,d\eta_-\,\int\,du\,\frac{e^{2\pi i \frac{(\eta_+ - \eta_-)}{2}\,u}}{\ch{(u -m')}}\, \int \,  [d \vec \s] \, e^{2\pi i \frac{(\eta_+ +\eta_-)}{2}\,\tr \vec \s}\,Z^{\CT^N_{2N,0}}_{\rm 1-loop} (\vec \s, \vec m) \nn \\
=&  \int \,  [d \vec \s] \, \frac{e^{2\pi i \eta_+ \tr \vec \s}}{\ch{(\tr \vec \s -m')}}\,Z^{\CT^N_{2N,0}}_{\rm 1-loop} (\vec \s, \vec m),
\end{align}
where $m'$ is the real mass associated with the Abelian hypermultiplet.
Using the identity \eref{UN-2N-pf}, one can then show that:
\be \label{Id-2a}
Z^{\CT^N_{2N,1}}(\vec m, m', \eta) = Z^{\CT^N_{2N,1}}(\vec m, -m', -\eta),
\ee
confirming the self-duality of the theory. Written in this form, it is evident that implementing the duality twice gives back the 
original theory. By a simple change of variables on both sides, the above equation can be rewritten in the following 
form:
\begin{align} \label{Id-2}
\boxed{Z^{\CT^N_{2N,1}}(\vec \mu, m_{\rm ab}=\tr \vec \mu,\eta) =  \int \,  [d \vec \s] \, \frac{e^{2\pi i \eta \tr \vec \s}}{\ch{(\tr \vec \s - \tr \vec \mu')}}\,Z^{\CT^N_{2N,0}}_{\rm 1-loop} (\vec \s, \vec \mu') = Z^{\CT^N_{2N,1}}(\vec \mu',m_{\rm ab}=\tr \vec \mu', -\eta),}
\end{align}
where the masses $\vec \mu$ are unconstrained, i.e. in particular $\tr \vec \mu \neq 0$, and the masses $\vec \mu'$ are 
related to $\vec \mu$ as follows:
\be
\mu'_a = \mu_a - \frac{1}{N}\,\tr \vec \mu, \qquad a=1,\ldots, 2N.
\ee
In the next step, the theory $\CT^N_{2N,2}$ can be obtained by gauging a diagonal combination of the two topological 
$U(1)$ symmetries in $\CT^N_{2N,1} $ and another copy of $\CT^1_{1,0} $. Repeating the operation $P$ times (or simply 
introducing a $\CT^1_{P-1,0}$ theory on both sides of \eref{Id-2a} and gauging the diagonal $U(1)$), we have 
the partition function identity
\be
\boxed{Z^{\CT^N_{2N,P}}(\vec m, \vec{m}^{\rm ab}_l, \eta) = Z^{\CT^N_{2N,P}}(\vec m, -\vec{m}^{\rm ab}_l, -\eta),}
\ee
where $\{ m^{\rm ab}_l\}_{l=1,\ldots,P}$ are the masses of the Abelian hypermultiplets. The above equation implies that 
the theory $\CT^N_{2N, P}$ is self-dual. Alternatively, one could have constructed $\CT^N_{2N, P}$ by gauging a 
diagonal combination of the topological $U(1)$ symmetries in $\CT^N_{2N,0} $ and in $\CT^1_{P,0}$ (a $U(1)$ gauge 
theory with $P$ fundamental hypers). Starting from the identity \eref{UN-2N-pf}, the operation leads to the same result as above.\\

One can obtain another interesting duality from the $P=1$ duality in \figref{IRdual-sd} by gauging the topological $U(1)$ symmetry. 
It is convenient for this purpose to rewrite the partition function identity \eref{Id-2a} after a simple change of variables in the following form:
\be
Z^{\CT^N_{2N,1}}(\vec m - \frac{1}{N}\,m'\vec 1, 0, \eta) = Z^{\CT^N_{2N,1}}(\vec m +\frac{1}{N}\,m'\vec 1, 0, -\eta).
\ee
Gauging the topological $U(1)$ now amounts to integrating over the FI parameter $\eta$ on both sides of the identity. 
This leads to the new identity:
\be
\boxed{Z^{SU(N),2N}(\vec m - \frac{1}{N}\,m'\vec 1) = Z^{SU(N),2N}(\vec m + \frac{1}{N}\,m'\vec 1),}
\ee
which we interpret as the self-duality of an $SU(N)$ gauge theory with $N_f=2N$. From the above expression, $m'$ 
can be identified as the real mass of the baryonic symmetry. The duality acts by changing the sign of the baryonic 
charge of the fundamental hypermultiplet. This duality was proposed in \cite{Kubo:2021ecs} by a dimensional reduction argument 
from the 4d $SU(N)$ theory with $N_f=2N$ which is known to be self-dual. From our perspective, this duality is a 
straightforward consequence of the duality in \figref{IRdual-sd}.

\subsection{Duality $\CD^N_{2N-1,P}$: $N_f=2N-1$ and $P \geq 1$}\label{D2-main}

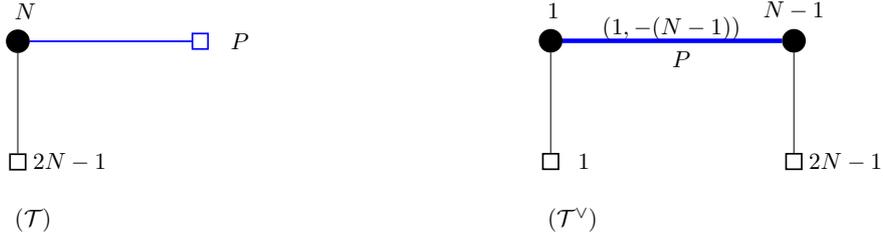
\begin{figure}[htbp]
\begin{center}
\begin{tabular}{ccc}
\scalebox{.8}{\begin{tikzpicture}
\node[unode] (1) at (0,0){};
\node[fnode] (2) at (0,-2){};
\node[afnode] (3) at (3,0){};
\draw[-] (1) -- (2);
\draw[-, thick, blue] (1)-- (3);
\node[text width=0.1cm](10) at (0, 0.5){$N$};
\node[text width=1.5cm](11) at (1, -2){$2N-1$};
\node[text width=1cm](12) at (4, 0){$P$};
\node[text width=0.1cm](20)[below=0.5 cm of 2]{$(\CT)$};
\end{tikzpicture}}
& \qquad   \qquad \qquad \qquad
&\scalebox{.8}{\begin{tikzpicture}
\node[unode] (1) at (0,0){};
\node[fnode] (2) at (0,-2){};
\node[unode] (3) at (4,0){};
\node[fnode] (4) at (4,-2){};
\draw[-] (1) -- (2);
\draw[line width=0.75mm, blue] (1)-- (3);
\draw[-] (3)-- (4);
\node[text width=2.3cm](10) at (2.0, 0.2){$(1, -(N-1))$};
\node[text width=2cm](11) at (3.0, -0.3){$P$};
\node[text width=0.1 cm](20) at (0,0.5){$1$};
\node[text width=0.1 cm](21) at (0.5,-2){$1$};
\node[text width=1 cm](22) at (4, 0.5){$N-1$};
\node[text width=1.5 cm](23) at (5, -2){$2N-1$};
\node[text width=0.1cm](20)[below=0.5 cm of 2]{$(\CT^\vee)$};
\end{tikzpicture}}
\end{tabular}
\end{center}
\caption{\footnotesize{IR duality involving the $\CT^N_{2N-1,P}$ theory and the quiver gauge theory $\CT^\vee$.}}
\label{IRdual-Ex2}
\end{figure}

The proposed IR duality $\CD^N_{2N-1,P}$ (the subscript indicating that we have a two-parameter family of dualities for generic 
$N$ and $P$), shown in \figref{IRdual-Ex2}, involves the following theories:

\begin{itemize}

\item Theory $\CT$: $U(N)$ gauge theory with $2N-1$ fundamental hypers and $P\geq 1$ Abelian hypermultiplets,
 i.e. the theory $\CT^N_{2N-1,P}$.

\item Theory $\CT^\vee$: $U(1) \times U(N-1)$ gauge theory with $1$ and $2N-1$ fundamental hypers respectively, and $P$ Abelian 
hypermultiplet with charge $(1, -(N-1))$ under the $U(1) \times U(N-1)$ gauge group \footnote{The Abelian hypermultiplet is constituted of 
a chiral multiplet with charges $(1, -(N-1))$ and a chiral multiplet in the complex conjugate representation, i.e. with charges $(-1, N-1)$.}. 

\end{itemize}

%\begin{figure}[htbp]
%\begin{center}
%\begin{tabular}{ccc}
%\scalebox{.8}{\begin{tikzpicture}[node distance=2cm, nnode/.style={circle,draw,thick, fill, inner sep=1 pt},cnode/.style={circle,draw,thick,minimum size=1.0 cm},snode/.style={rectangle,draw,thick,minimum size=1.0 cm}]
%\node[cnode] (1) at (0,0){$N$};
%\node[snode] (2) at (0,-2){$2N-1$};
%\node[snode,blue] (3) at (3,0){$1$};
%\draw[-] (1) -- (2);
%\draw[-, thick, blue] (1)-- (3);
%\node[text width=1cm](10) at (1.5, 0.2){$N$};
%%\node[text width=1cm](11) at (0, 1.1){$\eta=0$};
%\node[text width=0.1cm](20)[below=0.5 cm of 2]{$(\CT)$};
%\node[text width=0.1cm](21)[above=0.1 cm of 1]{$\eta$};
%\end{tikzpicture}}
%& \qquad   \qquad \qquad \qquad
%&\scalebox{.8}{\begin{tikzpicture}[node distance=2cm, nnode/.style={circle,draw,thick, fill, inner sep=1 pt},cnode/.style={circle,draw,thick,minimum size=1.0 cm},snode/.style={rectangle,draw,thick,minimum size=1.0 cm}]
%\node[cnode] (1) at (0,0){$1$};
%\node[snode] (2) at (0,-2){$1$};
%\node[cnode] (3) at (4,0){$N-1$};
%\node[snode] (4) at (4,-2){$2N-1$};
%\draw[-] (1) -- (2);
%\draw[-, thick, blue] (1)-- (3);
%\draw[-] (3)-- (4);
%\node[text width=2.3cm](10) at (2.0, 0.2){$(1,-(N-1))$};
%%\node[text width=1cm](11) at (0, 1.1){$\eta=0$};
%\node[text width=0.1cm](20)[below=0.5 cm of 2]{$(\CT^\vee)$};
%\node[text width=0.1cm](21)[above=0.1 cm of 1]{$\eta$};
%\node[text width=0.25cm](22)[above=0.1 cm of 3]{$-\eta$};
%\end{tikzpicture}}
%\end{tabular}
%\end{center}
%\caption{\footnotesize{IR duality involving the $\CT^N_{2N-1,1}$ theory and the quiver gauge theory $\CT^\vee$.}}
%\label{IRdual-Ex2}
%\end{figure}

Let us first discuss the duality for $P=1$. 
Note that both theories are good in the Gaiotto-Witten sense. For the theory $\CT=\CT^N_{2N-1,1}$, we have already 
discussed this fact in the previous section. We will show below that this is true for the $\CT^\vee$ theory. 
The dimensions of the Coulomb and the Higgs branches as well as the respective global symmetries have been summarized in Table \ref{Tab:D2}.
The matching of the Higgs branch global symmetry can be readily checked from the 
quivers. The matching of the Coulomb branch global symmetry, however, is non-trivial. The theory $\CT$ 
has a topological $\fru(1)$ symmetry manifest in the UV Lagrangian, but the global symmetry algebra in the IR is enhanced to 
$\frsu(2) \oplus \fru(1)$ as we have seen in \Secref{GBU}. The theory $\CT^\vee$ has a manifest $\fru(1) \oplus \fru(1)$ symmetry algebra, 
which corresponds to the correct Cartan subalgebra of $\frsu(2) \oplus \fru(1)$. 
In addition, since the $U(1)$ gauge node is balanced and the $U(N-1)$ node is overbalanced, one 
might expect that the symmetry is enhanced to $\frsu(2) \oplus \fru(1)$ in the IR. One can check that this intuition 
is indeed correct from the spectrum of monopole operators in the $\CT^\vee$ theory:
\begin{align}
\Delta( m,\vec p)= & \frac{1}{2}\,|m| + \frac{1}{2}\,|m- \sum^{N-1}_{i=1} p_i | + \frac{2N-1}{2}\,\sum^{N-1}_i |p_i| - \sum_{i<j}|p_i-p_j|,   \\
=& \frac{1}{2}\,|m| + \frac{1}{2}\,|m- \sum^{N-1}_{i=1} p_i | + \frac{3}{2}\,\sum^{N-1}_{i=1} |p_i|  + \sum_{i<j}(|p_i| + |p_j|-|p_i-p_j|),
\end{align}
where $m$ and $\vec p$ are GNO fluxes for the $U(1)$ and $U(N-1)$ gauge groups respectively.
The spectrum has no monopole operators with conformal dimensions $\Delta \leq \frac{1}{2}$, and is therefore a good theory.
The spectrum has two operators with $\Delta =1$ labelled by the fluxes $m=\pm 1, \vec p=0$ which together with the
generator of the topological symmetry associated with the $U(1)$ gauge node gives an $\frsu(2)$ algebra in the IR. 
Together with the generator of the topological symmetry associated with the $U(N-1)$ gauge 
group, these generate the $\frsu(2) \oplus \fru(1)$ global symmetry algebra. The fact that the global symmetry group 
is actually $SO(3) \times U(1)$ can be read off from the refined Coulomb branch Hilbert Series of the theory $\CT^\vee$. \\

\begin{table}[htbp]
\begin{center}
%\resizebox{\textwidth}{!}
{%
\begin{tabular}{|c|c|c|}
\hline
Moduli space data & Theory $\CT$ & Theory $\CT^\vee$ \\
\hline \hline 
dim\,$\CM_H$ & $N^2 -N +1$  & $N^2 -N +1$\\
\hline
dim\,$\CM_C$ & $N$ & $N$ \\
\hline
$\frg_H$ & $\frsu(2N-1) \oplus \fru(1)$ & $\frsu(2N-1) \oplus \fru(1) $\\
\hline
$\frg_C$ & $ \frsu(2) \oplus \fru(1) $ & $\frsu(2) \oplus \fru(1) $\\
\hline
\end{tabular}}
\end{center}
\caption{\footnotesize{The matching of moduli space data for the proposed dual pair in \figref{IRdual-Ex2} for $P=1$.}}
\label{Tab:D2}
\end{table}

For $P>1$, the dual theories are again good in the Gaiotto-Witten sense. We already discussed this in \Secref{GBU} for the 
theory $\CT=\CT^N_{2N-1,P}$, while for the $\CT^\vee$ one can check this directly from the R-charges of monopole operators.
The moduli space data for the duality, including the global symmetries are summarized in Table \ref{Tab:D3}. As before, the 
matching of the Higgs branch global symmetry can be readily checked from the quivers. For the Coulomb branch, the theory $\CT$ 
has a topological $\fru(1)$ symmetry manifest in the UV Lagrangian, but the global symmetry algebra in the IR is enhanced to 
$\fru(1) \oplus \fru(1)$. For the theory $\CT^\vee$, this global symmetry is manifest in the UV as the $\fru(1) \oplus \fru(1)$ 
algebra associated with the topological symmetry of the two unitary gauge nodes.

\begin{table}[htbp]
\begin{center}
%\resizebox{\textwidth}{!}
{%
\begin{tabular}{|c|c|c|}
\hline
Moduli space data & Theory $\CT$ & Theory $\CT^\vee$ \\
\hline \hline 
dim\,$\CM_H$ & $N^2 -N +P$  & $N^2 -N +P$\\
\hline
dim\,$\CM_C$ & $N$ & $N$ \\
\hline
$\frg_H$ & $\frsu(2N-1) \oplus \frsu(P) \oplus \fru(1)$ & $ \frsu(2N-1) \oplus \frsu(P) \oplus \fru(1)$\\
\hline
$\frg_C$ & $ \fru(1) \oplus \fru(1) $ & $ \fru(1) \oplus \fru(1) $\\
\hline
\end{tabular}}
\end{center}
\caption{\footnotesize{The matching of moduli space data for the proposed dual pair in \figref{IRdual-Ex2} for generic $P>1$.}}
\label{Tab:D3}
\end{table}

The duality in \figref{IRdual-Ex2} for $P=1$ can be derived by an RG flow argument starting from the duality $\CD^N_{2N+1,1}$ in 
 \figref{IRdual-Ex1}, which we discuss in \Secref{RG-duality}. In this section, we show that this duality (for generic $P \geq 1$) can 
 be obtained from the Seiberg-like duality of the ugly $U(N)$ theory discussed earlier by a simple QFT operation. 
 Recall that the ugly $U(N)$ theory (i.e. the theory $\CT^N_{2N-1,0}$) is dual to the theory $\CT^{N-1}_{2N-1,0}$
and a decoupled $\CT^1_{1,0}$ theory. The theory $\CT=\CT^N_{2N-1,P}$ can be constructed by gauging a diagonal $U(1)$ 
subgroup of the $U(1)$ topological symmetries of the theory $\CT^N_{2N-1,0}$ and the theory $\CT^1_{P,0}$. In terms of the 
three-sphere partition function, this can be explicitly written as
\begin{align}
Z^{\CT^N_{2N-1,P}}(\vec m, \vec{m}_{\rm{Ab}};\eta_+) = & \int\,d\eta_-\, Z^{\CT^1_{P,0}}(\vec{m}_{\rm{Ab}}, \frac{(\eta_+ -\eta_-)}{2})\, Z^{\CT^N_{2N-1,0}}(\vec m, \frac{(\eta_+ +\eta_-)}{2}) \nn \\
=&  \int\,d\eta_-\,\int\,du\,\frac{e^{2\pi i \frac{(\eta_+ - \eta_-)}{2}\,u}}{\prod^P_{l=1}\,\ch{(u - {m}^l_{\rm{Ab}})}}\, \int \,  [d \vec s] \, e^{2\pi i \frac{(\eta_+ +\eta_-)}{2}\,\tr \vec s}\,Z^{\CT^N_{2N-1,0}}_{\rm 1-loop} (\vec s, \vec m) \nn \\
=&  \int \,  [d \vec s] \, \frac{e^{2\pi i \eta_+ \tr \vec s}}{\prod^P_{l=1}\,\ch{(\tr \vec s -{m}^l_{\rm{Ab}})}}\,Z^{\CT^N_{2N-1,0}}_{\rm 1-loop} (\vec s, \vec m).
\end{align}
The dual theory can be read off by using the identity \eref{ugly-id} to substitute $Z^{\CT^N_{2N-1,0}}$ in the first line of the above equation, 
which gives:
\begin{align}
&Z^{\CT^N_{2N-1,P}}(\vec m, \vec{m}_{\rm{Ab}};\eta_+) =  \int\,d\eta_-\, Z^{\CT^1_{P,0}}(\vec{m}_{\rm{Ab}}, \frac{(\eta_+ -\eta_-)}{2})\, Z^{\CT^N_{2N-1,0}}(\vec m, \frac{(\eta_+ +\eta_-)}{2}) \nn \\
=& \int\,d\eta_-\,Z^{\CT^1_{P,0}}(\vec{m}_{\rm{Ab}}, \frac{(\eta_+ -\eta_-)}{2}) \,Z^{\CT^1_{1,0}}(\tr m, \frac{(\eta_+ +\eta_-)}{2})\,
Z^{\CT^{N-1}_{2N-1,0}}(\vec m, -\frac{(\eta_+ +\eta_-)}{2}).
\end{align}
Integrating over $\eta_-$ and implementing the resultant delta function, the partition function of $\CT^N_{2N-1,P}$ can be written as:
\begin{align}\label{Id-3}
Z^{\CT^N_{2N-1,P}}(\vec m, \vec{m}_{\rm{Ab}};\eta_+) =  &\int \,d{\s'}\,[d \vec \s]\,\frac{e^{2\pi i \eta_+\,(\s' - \tr \vec \s)}\,Z^{\CT^{N-1}_{2N-1,0}}_{\rm 1-loop} (\vec \s, \vec m)}{\ch{(\s'- \tr m)}\,\prod^P_{l=1}\,\ch{(\s'-\tr \vec \s -m^l_{\rm{Ab}})} } \nn \\
=& Z^{\CT^\vee}(m^\vee_{(1)} =\tr m, m^\vee_{(2)}=\vec m, m^\vee_{\rm Ab}=\vec m_{\rm{Ab}};\eta_+, -\eta_+),
\end{align}
where in the final step we have identified the matrix integral on the RHS as the sphere partition function of the theory $\CT^\vee$ as 
given in \figref{IRdual-Ex2}. We therefore have the following relation of the two partition functions: 
\be
\boxed{Z^{\CT}(\vec m, m_{\rm{Ab}};\eta)=  Z^{\CT^\vee}(m^\vee_{(1)}, m^\vee_{(2)}, m^\vee_{\rm Ab};\eta, -\eta).}
\ee

Note that a linear combination of the FI parameters of the theory $\CT^\vee$ has to be set to zero for the two 
partition functions to agree. This is expected since the theory $\CT^\vee$ has a manifest $\fru(1) \oplus \fru(1)$ topological symmetry 
in the UV, while the theory $\CT$ has only a $\fru(1)$. 
The number of mass parameters in the theory $\CT^\vee$ is $(2N-1)+1+P=2N+P$ by naive counting. The independent mass parameters can be obtained by shifting the integration variables $u'$ and $\vec \s$ -- one can check that there are $(2N+P-2)$ of them which live in the Cartan 
subalgebra of the global symmetry algebra $\frg_H= \frsu(2N-1) \oplus \frsu(P) \oplus \fru(1)$. This matches the number of independent mass parameters in the theory $\CT$.\\

A second check of the duality can be performed by comparing the Coulomb branch and Higgs branch Hilbert Series of the 
two dual theories. As before, the matching of the Coulomb branch Hilbert Series is non-trivial given the emergent global 
symmetry for the $\CT$ theory. Following the general procedure reviewed in \appref{HS-rev}, the agreement 
between the Hilbert Series for the theories $\CT$ and $\CT^\vee$ can be checked for any $N$. 
Consider, for example, the case of $N=2$ and $P=1$, where one can check that
\begin{align}
& \CI^C_{\CT^2_{3,1}}(t)=  \CI^C_{\CT^\vee_{N=2,P=1}}(t)=  1+4t + 13t^2 + 28t^3 + 55t^4 + 92t^5 + 147t^6 + O(t^7), \\
& {\rm PL}[ \CI^C_{\CT^2_{3,1}}(t)]= {\rm PL}[ \CI^C_{\CT^\vee_{N=2,P=1}}(t)] = 4t + 3t^2 - 4t^3 + 4t^5 -6t^6 + O(t^8).
\end{align}
The existence of the emergent $\frsu(2) \oplus \fru(1)$ global symmetry can be read off from the coefficient of the $O(t)$ term 
of the HS for the $\CT^2_{3,1}$ theory, while the $O(t)$ term in the plethystic logarithm corresponds to the associated monopole 
operators with conformal dimension 1.
Similarly, the Higgs branch Hilbert Series for the dual theories also match:
\begin{align}
& \CI^H_{\CT^2_{3,1}}(x)=  \CI^H_{\CT^\vee_{N=2,P=1}}(x)= 1+ 9x + 6 x^{3/2} + 36 x^{2}+ 36x^{5/2} + O(x^{3}), \\
& {\rm PL}[ \CI^H_{\CT^3_{5,1}}(x)]= {\rm PL}[ \CI^H_{\CT^\vee_{N=2,P=1}}(x)] = 9x+6x^{3/2} -9x^2 -18 x^{5/2}+ O(x^{3}),
\end{align}
where the $O(x)$ term gives the dimension of the adjoint representation of the Lie algebra $\frg_{\rm H} = \frsu(3)\oplus \fru(1)$ as expected.

\subsection{Three dimensional mirrors for the dual pairs}\label{Mirr-main}

In this section, we comment on the 3d mirror associated with each pair of IR dual theories discussed above. 
Since one of the theories in the IR dual pair is always a $\CT^N_{N_f,P}$ quiver, the 3d mirrors can be read off 
from our analysis in \Secref{3dmirr-gen} -- we summarize the answers in Table \ref{Tab:D4}.\\

\begin{table}[htbp]
\begin{center}
{%
\begin{tabular}{|c|c|}
\hline
Duality &  3d mirror \\
\hline \hline 
$\CD^N_{2N+1,1}$
&  \scalebox{0.6}{\begin{tikzpicture}
\node[unode] (1) {};
\node[unode] (2) [right=.5cm  of 1]{};
\node[unode] (3) [right=.5cm of 2]{};
\node[unode] (4) [right=1cm of 3]{};
\node[] (5) [right=0.5cm of 4]{};
%\node[cnode] (9) [right=1cm of 6]{$\frac{p-v}{2}$};
\node[unode] (6) [right=1 cm of 4]{};
\node[unode] (7) [right=1cm of 6]{};
\node[unode] (8) [right=0.5cm of 7]{};
\node[unode] (9) [right=0.5cm of 8]{};
\node[unode] (13) [above=0.5cm of 4]{};
\node[fnode] (14) [left=0.5cm of 13]{};
\node[fnode] (15) [above=0.5cm of 6]{};
%\node[snode] (11) [below=0.5cm of 9]{1};
\node[text width=0.1cm](41)[below=0.2 cm of 1]{1};
\node[text width=0.1cm](42)[below=0.2 cm of 2]{2};
\node[text width=0.1cm](43)[below=0.2 cm of 3]{3};
\node[text width=0.1cm](44)[below=0.2 cm of 4]{$N$};
\node[text width=0.1cm](45)[below=0.2 cm of 6]{$N$};
\node[text width=0.1cm](46)[below=0.2 cm of 7]{3};
\node[text width=0.1cm](47)[below=0.2 cm of 8]{2};
\node[text width=0.1cm](48)[below=0.2 cm of 9]{1};
\node[text width=0.1cm](49)[above=0.2 cm of 13]{1};
\node[text width=0.1cm](50)[left=0.2 cm of 14]{1};
\node[text width=0.1cm](51)[right=0.2 cm of 15]{1};
\draw[-] (1) -- (2);
\draw[-] (2)-- (3);
\draw[dashed] (3) -- (4);
\draw[-] (4) --(6);
%\draw[-] (5) --(6);
\draw[dashed] (6) -- (7);
\draw[-] (7) -- (8);
\draw[-] (8) --(9);
%\draw[-] (5) -- (10);
\draw[-] (4) -- (13);
\draw[-] (13) -- (14);
\draw[-] (6) -- (15);
\end{tikzpicture}}\\
\hline
$\CD^N_{2N,P}$
 &  \scalebox{0.6}{\begin{tikzpicture}
\node[unode] (1) {};
\node[unode] (2) [right=.5cm  of 1]{};
\node[unode] (3) [right=.5cm of 2]{};
\node[unode] (4) [right=1cm of 3]{};
\node[unode] (5) [right=1cm of 4]{};
\node[unode] (6) [right=1 cm of 5]{};
\node[unode] (7) [right=1cm of 6]{};
\node[unode] (8) [right=0.5cm of 7]{};
\node[unode] (9) [right=0.5cm of 8]{};
\node[unode] (13) [above= 1cm of 5]{};
\node[fnode] (14) [below=1 cm of 5]{};
\node[unode] (15) [left=1 cm of 13]{};
\node[unode] (16) [left=1 cm of 15]{};
\node[fnode] (17) [left=1 cm of 16]{};
%\node[snode] (11) [below=0.5cm of 9]{1};
\node[text width=0.1cm](41)[below=0.2 cm of 1]{1};
\node[text width=0.1cm](42)[below=0.2 cm of 2]{2};
\node[text width=0.1cm](43)[below=0.2 cm of 3]{3};
\node[text width=1cm](44)[below=0.2 cm of 4]{$N-1$};
\node[text width=0.1cm](45)[below=0.2 cm of 5]{$N$};
\node[text width=1cm](46)[below=0.2 cm of 6]{$N-1$};
\node[text width=0.1cm](47)[below=0.2 cm of 7]{3};
\node[text width=0.1cm](48)[below=0.2 cm of 8]{2};
\node[text width=0.1cm](49)[below=0.1 cm of 9]{1};
\node[text width=0.1cm](50)[right=0.1 cm of 14]{1};
\node[text width=0.1cm](51)[left=0.2 cm of 17]{1};
\draw[-] (1) -- (2);
\draw[-] (2)-- (3);
\draw[dashed] (3) -- (4);
\draw[-] (4) --(5);
\draw[-] (5) --(6);
\draw[dashed] (6) -- (7);
\draw[-] (7) -- (8);
\draw[-] (8) --(9);
\draw[-] (5) -- (13);
%\draw[-] (4) -- (13);
\draw[-] (13) -- (15);
\draw[dashed] (16) -- (15);
\draw[-] (16) -- (17);
\draw[-] (5) -- (14);
\node[text width=0.1cm](25)[above=0.2 cm of 13]{1};
\node[text width=0.1cm](26)[above=0.2 cm of 15]{1};
\node[text width=0.1cm](27)[above=0.2 cm of 16]{$1$};
%\node[text width=0.1cm](20)[below=2 cm of 5]{$(\wt{\CT}^N_{2N,P})$};
\end{tikzpicture}} \\
\hline
$\CD^N_{2N-1,P}$
 & \scalebox{0.6}{\begin{tikzpicture}
 \node[unode] (1) {};
\node[unode] (2) [right=.5cm  of 1]{};
\node[unode] (3) [right=.5cm of 2]{};
\node[unode] (4) [right=1cm of 3]{};
\node[] (5) [right=0.5cm of 4]{};
%\node[cnode] (9) [right=1cm of 6]{$\frac{p-v}{2}$};
\node[unode] (6) [right=1 cm of 4]{};
\node[unode] (7) [right=1cm of 6]{};
\node[unode] (8) [right=0.5cm of 7]{};
\node[unode] (9) [right=0.5cm of 8]{};
\node[unode] (13) [above=0.5cm of 4]{};
\node[unode] (14) [left=0.5cm of 13]{};
\node[fnode] (15) [above=0.5cm of 6]{};
\node[unode] (16) [left=0.5cm of 14]{};
\node[fnode] (17) [left=0.5cm of 16]{};
\node[fnode] (18) [above=1 cm of 13]{};
\node[text width=0.1cm](41)[below=0.2 cm of 1]{1};
\node[text width=0.1cm](42)[below=0.2 cm of 2]{2};
\node[text width=0.1cm](43)[below=0.2 cm of 3]{3};
\node[text width=1cm](44)[below=0.2 cm of 4]{$N-1$};
\node[text width=1cm](45)[below=0.2 cm of 6]{$N-1$};
\node[text width=0.1cm](46)[below=0.2 cm of 7]{3};
\node[text width=0.1cm](47)[below=0.2 cm of 8]{2};
\node[text width=0.1cm](48)[below=0.2 cm of 9]{1};
\node[text width=0.1cm](49)[right=0.1 cm of 15]{1};
\node[text width=0.1cm](50)[left=0.2 cm of 17]{1};
\node[text width=0.1cm](51)[right=0.2 cm of 18]{1};
\draw[-] (1) -- (2);
\draw[-] (2)-- (3);
\draw[dashed] (3) -- (4);
\draw[-] (4) --(6);
%\draw[-] (5) --(6);
\draw[dashed] (6) -- (7);
\draw[-] (7) -- (8);
\draw[-] (8) --(9);
%\draw[-] (5) -- (10);
\draw[-] (4) -- (13);
\draw[-] (13) -- (14);
\draw[-] (6) -- (15);
\draw[dashed] (14) -- (16);
\draw[-] (16) -- (17);
\draw[-] (18) -- (13);
\node[text width=0.1cm](25)[above=0.2 cm of 13]{1};
\node[text width=0.1cm](26)[above=0.2 cm of 14]{1};
\node[text width=0.1cm](27)[above=0.2 cm of 16]{1};
\end{tikzpicture}} \\
\hline
\end{tabular}}
\end{center}
\caption{\footnotesize{Summary of the 3d mirrors associated with the dual IR pairs.}}
\label{Tab:D4}
\end{table}

To begin with, consider the duality $\CD^N_{2N+1,1}$ in \figref{IRdual-Ex1}, where the theory $\CT=\CT^N_{2N+1,1}$.
The 3d mirror in this case can be read off from the quiver gauge theory $\wt{\CT}^N_{N_f,P}$ in \figref{SOp-IRdual-gen1} 
for $N_f=2N+1$ and $P=1$. The 3d mirrors for the self-duality in \figref{IRdual-sd} and the duality in \figref{IRdual-Ex2} 
can be read off from \figref{SOp-IRdual-gen2} and \figref{SOp-IRdual-gen3} respectively. Note that the Coulomb branch 
symmetry which is emergent in the IR for one (or both) of the dual theories is manifest in the 3d mirror as a Higgs branch 
symmetry.

\section{Relating the dualities via RG flows and the duality web}\label{RG-duality}

In this section, we discuss how the dualities proposed in \Secref{Dualities} are related to one another by various 
QFT operations. In \Secref{RG-duality-1}, we show that the duality $\CD^N_{2N,1}$ discussed in \Secref{SD-main} 
can be realized by an appropriate mass deformation of the duality $\CD^N_{2N+1,1}$. Given the duality $\CD^N_{2N+1,1}$, one can turn on a real mass for a single fundamental hypermultiplet in the theory $\CT=\CT^N_{2N+1,1}$ and take the large mass limit, thereby flowing to the theory $\CT^N_{2N,1}$. On the dual side, this turns on a large non-trivial vev for one of the real adjoint scalars, which leads to a partial Higgsing of the gauge group and it turns out that the low energy effective theory around the new vacuum is described by another (dual) $\CT^N_{2N,1}$ theory. This realizes the duality  $\CD^N_{2N,1}$. 
In the next step, one can turn on a real mass for a single fundamental hypermultiplet in the theory $\CT=\CT^N_{2N,1}$ and repeat the above
procedure, which leads to the duality $\CD^N_{2N-1,1}$ discussed in \Secref{D2-main}. We perform this exercise in \Secref{RG-duality-2}. 
Following the procedure in \Secref{RG-duality-2}, one can show that the duality $\CD^N_{2N-1,P}$ arises from the duality $\CD^N_{2N,P}$ in an analogous fashion. In \Secref{RG-duality-3}, we demonstrate how the duality $\CD^N_{2N-1,P-1}$ arises from the duality $\CD^N_{2N-1,P}$ by turning on a real mass for a single Abelian hypermultiplet and taking the large mass limit. In $P$ steps, this allows one to flow to the duality $\CD^N_{2N-1,0}$ which is the well-known Seiberg-like duality for an ugly theory. In \Secref{RG-summary}, we use our findings from this section and the previous one to show how the dualities assemble themselves into a duality web. Finally, we briefly comment how the IR dualities proposed in this paper lead to exact dualities in \Secref{RG-exact}.\\

 We will perform the aforementioned RG flow analysis using the sphere partition function \cite{Aharony:2013dha, Yaakov:2013fza}. 
 Let us briefly review the basic principle 
 underlying the procedure. Recall that given a good theory with generic real masses, the dominant saddle point of the matrix integral 
 comes from the region $\vec \s \sim 0$ of the $\vec \s$-space. The Lagrangian associated with the matrix model can 
 be thought of as the effective field theory in the neighborhood of the Coulomb branch vacuum $\vec \s \sim 0$. 
 One can, however, evaluate subdominant contributions to the matrix integral as well, by implementing a transformation:
 \be \label{sigma-rep}
 \vec \s \to \vec \s + \Lambda_g\,\vec A_g,
 \ee
 where $\vec A_g$ is a diagonal matrix, and taking $\Lambda_g \to \infty$. For a finite $\Lambda_g$, the above transformation is 
 simply a change of variables, but in the large $\Lambda_g$ limit it picks out a subdominant contribution to the matrix integral 
 coming from a vacuum in the neighborhood of the region $\vec \s \sim \Lambda_g\,\vec A_g$. Equivalently, one can think of \eref{sigma-rep} as 
 implementing an RG flow along the Coulomb branch from the $\vec \s \sim 0$ vacuum to the $\vec \s \sim \Lambda_g\,\vec A_g$ vacuum. 
 The original theory will generically be 
 partially Higgsed at the latter vacuum with the pattern of the symmetry breaking being encoded in the matrix $A_g$. 
At the level of the matrix integral, the low energy effective theory at this new vacuum can be read off from the hypermultiplets  
and the vector multiplets whose contributions remain finite, while the remaining ($\Lambda_g$-dependent) terms 
contribute to the FI parameters and a prefactor depending exponentially on $\Lambda_g$. The latter gives a measure 
of the sub-dominance of the aforementioned vacuum with respect to the $\vec \s \sim 0$ vacuum in the original matrix model.

 Now, given a good theory, one can make certain real masses large -- this can be done by a 
 reparametrization of the masses in the matrix integral:
 \be \label{massrep-gen}
 \vec m \to \vec m + \Lambda_m \,\vec A_m,
 \ee 
 with $\vec A_m$ being a diagonal matrix with at least a single non-zero entry, and taking the limit $\Lambda_m \to \infty$. 
 The dominant contribution to the resultant matrix integral will not generically arise from the $\vec \s \sim 0$ vacuum. 
 To determine the dominant saddle point, one combines the reparametrization \eref{massrep-gen} with the transformation 
 \eref{sigma-rep} setting $\Lambda_m = \Lambda_g = \Lambda \to \infty$. For a given choice of the matrix $\vec A_g$, 
 this operation again picks out the vacuum in the neighborhood of $\vec \s \sim \Lambda\,\vec A_g$.
 The low energy effective theory for this vacuum can be read off as described above, along with a $\Lambda$-dependent 
 prefactor of the matrix integral which measures the relative dominance of the vacuum. In this fashion, one can determine 
 the dominant vacuum as well as the various sub-dominant vacua for the massive theory and their respective 
 contributions to the matrix integral. \\
 
 Let us now apply this procedure to our specific goal i.e. to generate a new IR duality by deforming a given IR duality 
 with large real masses.  Let $(\CT, \CT^\vee)$ be a pair of IR dual theories, where $\CT$ is of the form $\CT^N_{N_f,P}$.
We want to introduce large real masses for the fundamental hypers and/or the Abelian hypers such that 
we flow to a theory $\CT'$ of the form $\CT^N_{N'_f,P'}$ with $N'_f < N_f$ and/or $P' < P$. 
As discussed above, this can be implemented 
at the level of the matrix model by the reparametrization \eref{massrep-gen} combined with the transformation 
\eref{sigma-rep} of the integration variables. The form of the theory $\CT'$ fixes the matrices $\vec A_m$ 
and $\vec A_g$. In particular, since we demand that the gauge group remains unbroken, the matrix $\vec A_g$ 
 must be proportional to a unit matrix of $N$, and not a generic diagonal matrix. The $\Lambda$-dependent exponential 
 prefactor can be read off from the matrix integral as discussed above.

Now, consider what happens to the dual theory $\CT^\vee$. The masses are reparametrized as in \eref{massrep-gen}, and 
one should combine this with the transformation of the form \eref{sigma-rep} : $\vec s \to \vec s + \Lambda\,\vec A^\vee_g$
where $\vec \s$ is the integration variable of the dual matrix model, and $\vec A^\vee_g$ is a diagonal matrix (but not necessarily proportional to the unit matrix), before taking the limit $\Lambda \to \infty$. 
A priori there many choices for $\vec A^\vee_g$ each corresponding to a certain vacuum. For each such vacua one can read off 
the low energy effective theory as well as the $\Lambda$-dependent expontential prefactors. We are, however, 
interested in the flow of $\CT^\vee$ to the theory $\CT'^\vee$ such that the latter is IR dual to the theory $\CT'$. This will correspond 
to the choice of $\vec A^\vee_g$ for which $\Lambda$-dependent prefactor is the same as that obtained for the theory $\CT'$ above. 
In this fashion, one arrives at a new IR dual pair $(\CT', \CT'^\vee)$ by deforming the dual pair $(\CT, \CT^\vee)$ by large real masses.

%The choice of 
%$\vec A^\vee_g$ which corresponds to 
%
%a priori there are many choices for shifting the integration variables in the matrix model as $\vec s \to \vec s + \Lambda\,\vec B^\vee$, 
%where $\vec B^\vee$ is a diagonal matrix (but not necessarily proportional to the unit matrix), before taking the limit $\Lambda \to \infty$. For a given choice of $\vec B^\vee$, this implies that the the matrix integral will receive its dominant contribution from the region $\vec s \sim \Lambda\,\vec B^\vee$. Note that the transformation $\vec s \to \vec s + \Lambda\,\vec B^\vee$, for a given $\vec B^\vee$, will not 
%keep the vector multiplet contribution invariant generically, and therefore one expects to have a partial Higgsing of the 
%gauge group. As before, a given choice of $\vec B^\vee$ will correspond to a low energy effective theory that can be read off 
%from the finite vector multiplet and hypermultiplet contributions of the matrix model, while the $\Lambda$-dependent terms will 
%contribute to FI terms as well as prefactors scaling exponentially with $\Lambda$. The correct choice of $\vec B^\vee$ is the one that gives the 
%same scaling factor as the matrix integral obtained above from the theory $\CT$ after the reparametrization \eref{massrep-gen} -- the associated theories read off from the respective matrix models are then dual to each other. 
%In this fashion, one arrives at a new duality by deforming the original one with large real masses.

\subsection{Flowing from the duality $\CD^N_{2N+1,1}$ to the duality $\CD^N_{2N,1}$}\label{RG-duality-1}

Consider the IR duality $\CD^N_{2N+1,1}$ given in \figref{IRdual-Ex1}. The sphere partition function of the theory $\CT=\CT^N_{2N+1, 1}$ 
with a vanishing FI parameter and generic real masses is given as:
\be
Z^{\CT^N_{2N+1, 1}} (\vec m, \eta =0) =  \int \, [d \vec \s] \ \frac{1}{\ch{(\tr \vec \s - \tr \vec m)}}\,\frac{\prod_{1\leq j<k \leq N}\, \sinh^2{\pi(\s_j- \s_k)}}{\prod^{N}_{j=1} \prod^{2N+1}_{i=1}\ch{(\s_j- m_i)}},
\ee
where $\tr \vec m = \sum^{2N+1}_{i=1}\, m_i$. Following the discussion above, let us reparametrize the masses along 
with a shift in the integration variables in the following fashion:
\begin{align}
& m_{2N+1} = -N\,\Lambda, \qquad m_a = \mu_a + \Lambda \, \, (a=1,\ldots,2N), \label{massrep-1}\\
& \s_i \to \s_i + \Lambda \,\, (i=1, \ldots,N). \label{shift-1}
\end{align}
Note that this shift in the integration variables keeps the vector multiplet contribution to the matrix integral invariant 
and therefore preserves the gauge group. With the above parametrization of the masses and the change of integration variables, 
the partition function assumes the form:
\begin{align}
Z^{\CT^N_{2N+1, 1}} (\vec m, 0) =  \int \, [d \vec \s] \,&\frac{\prod_{1\leq j<k \leq N}\, \sinh^2{\pi(\s_j- \s_k)}}{\prod^{N}_{j=1} \prod^{2N}_{a=1}\ch{(\s_j- \mu_a)}\,\ch{(\s_j + (N+1)\Lambda)}} \nn \\
& \times \frac{1}{\ch{(\tr \vec \s - \sum^{2N}_{a=1}\,\mu_a)}}.
\end{align}
In the limit $\Lambda \to \infty$, the fundamental hyper associated with the real mass $m_{2N+1}$ only contributes to the FI term 
and a $\Lambda$-dependent prefactor. The partition function can then be written as:
\be \label{lhs-pf-II}
Z^{\CT^N_{2N+1, 1}} (\vec m, 0) \xrightarrow{\Lambda \to \infty} e^{-\pi \Lambda\,N(N+1)}\, Z^{\CT^N_{2N, 1}} (\vec \mu, \eta = \frac{i}{2}).
\ee

Let us now consider what happens on the dual side. The partition function of the dual theory $\CT^\vee$
-- an $SU(N+1)$ theory with $2N+1$ fundamental hypers -- is given as
\be \label{suN-pfL}
Z^{SU(N+1),\,2N+1}(\vec m) =  \int \, [d \vec s] \, \delta(\tr \vec s)\, \frac{\prod_{1\leq j<k \leq N+1}\, \sinh^2{\pi(s_j- s_k)}}{\prod^{N+1}_{j=1} \prod^{2N+1}_{a=1}\ch{(s_j- m_a)}}.
\ee
Given the parametrization of masses in \eref{massrep-1}, the appropriate change of the integration variables is given as
\be
s_i \to s_i + \Lambda \,\,(i=1,\ldots,N+1,\, i\neq j), \qquad s_{j} \to s_{j} - N\,\Lambda, \label{shift-1d}
\ee
for any of the $N+1$ choices of $j$, which are related by Weyl symmetry.  
In the limit $\Lambda \to \infty$, for $j=N+1$, the vector multiplet and the hypermultiplet contributions in the matrix model 
integrand of \eref{suN-pfL} are respectively given as
\begin{align}
 \prod_{1\leq j<k \leq N+1}\,\sinh^2{\pi(s_j- s_k)} \to  & \prod_{1\leq j<k \leq N}\,\sinh^2{\pi(s_j- s_k)}\,e^{2\pi \Lambda N(N+1)}\, e^{2\pi (\sum^N_{j=1}s_j -Ns_{N+1})}, \\
\prod^{N+1}_{j=1} \prod^{2N+1}_{a=1}\ch{(s_j- m_a)} \to & \prod^{N}_{j=1} \prod^{2N}_{a=1}\ch{(s_j- \mu_a)}\,\ch{s_{N+1}} \nn \\
& \times e^{3\pi \Lambda N(N+1)}\, e^{\pi \sum^{2N}_{a=1}\mu_a}\, e^{\pi (\sum^N_{j=1}s_j - 2Ns_{N+1})}.
\end{align}
The matrix integral \eref{suN-pfL} then assumes the following form in the $\Lambda \to \infty$ limit : 
\begin{align}
Z^{SU(N),\,2N+1}(\vec m)  \xrightarrow{\Lambda \to \infty} & \,(N+1)\, \int \, [d \vec s] \,
\frac{\delta(\tr \vec s)\,\prod_{1\leq j<k \leq N}\,\sinh^2{\pi(s_j- s_k)}\,e^{\pi \sum^N_{j=1}s_j }}{\prod^{N}_{j=1} \prod^{2N}_{a=1}\ch{(s_j- \mu_a)}\,\ch{s_{N+1}}\,e^{\pi \Lambda N(N+1)}\,e^{\pi \sum^{2N}_{a=1}\mu_a}},
\end{align}
where $(N+1)$ is a combinatorial factor that arises from the fact that there are $(N+1)$ possible choices of $j$ in \eref{shift-1d}.
Integrating over $s_{N+1}$ and shifting integration variables $s_j \to s_j + \frac{1}{N}\,\tr \vec \mu$ (with $j=1,\ldots,N$), we have
\begin{align} \label{rhs-pf-II}
Z^{SU(N),\,2N+1}(\vec m)  \xrightarrow{\Lambda \to \infty} &  \int \, [d \vec s] \,
\frac{\prod_{1\leq j<k \leq N}\,\sinh^2{\pi(s_j- s_k)}\,e^{\pi \tr \vec s }}{\prod^{N}_{j=1} \prod^{2N}_{a=1}\ch{(s_j- \mu_a +\frac{1}{N}\,\tr \vec \mu)}\,\ch{(\tr \vec s + \tr \vec \mu)}\,e^{\pi \Lambda N(N+1)}} \nn \\
= & e^{-\pi \Lambda N(N+1)}\, Z^{\CT^N_{2N, 1}} (\vec \mu', \eta = -\frac{i}{2}),
\end{align}
where the masses $\vec \mu'$ are related to the masses $\vec \mu$ in the following fashion:
\be
\mu'_a = \mu_a - \frac{1}{N}\,\tr \vec \mu.
\ee
Comparing \eref{lhs-pf-II} and \eref{rhs-pf-II}, we note that the $\Lambda$-dependent exponential scaling factor 
does match on both sides as expected. Therefore, analytically continuing to real values of $\eta$, we obtain
\be \label{SD-pf}
\boxed{Z^{\CT^N_{2N, 1}} (\vec \mu, \eta ) = Z^{\CT^N_{2N, 1}} (\vec \mu', -\eta),}
\ee
where the masses $\vec \mu$ and $\vec \mu'$ are related as above. This precisely reproduces the duality $\CD^N_{2N,1}$ i.e. 
self-duality of the theory $\CT^N_{2N, 1}$ found in \Secref{SD-main}.

\subsection{Flowing from the duality $\CD^N_{2N,1}$ to the duality $\CD^N_{2N-1,1}$}\label{RG-duality-2}

In the next step, let us introduce a large mass for a fundamental hyper in the theory $\CT^N_{2N, 1}$ to flow to the theory  $\CT^N_{2N-1, 1}$.
Starting from the sphere partition function of $\CT^N_{2N, 1}$ as obtained earlier:
\be
Z^{\CT^N_{2N, 1}} (\vec \mu, \eta =\frac{i}{2}) =  \int \, [d \vec \s] \ \frac{e^{-\pi \tr \vec \s}}{\ch{(\tr \vec \s - \tr \vec \mu)}}\,\frac{\prod_{1\leq j<k \leq N}\, \sinh^2{\pi(\s_j- \s_k)}}{\prod^{N}_{j=1} \prod^{2N}_{a=1}\ch{(\s_j- \mu_a)}},
\ee
where $\tr \vec \mu = \sum^{2N}_{a=1}\, \mu_a$, let us consider a parametrization of masses along with a shift in the integration variables:
\begin{align}
& \mu_{2N} = -(N-1)\,\Lambda, \qquad \mu_a = M_a + \Lambda \, \, (a=1,\ldots,2N-1), \label{massrep-2}\\
& \s_j \to \s_j + \Lambda \,\, (j=1, \ldots,N). \label{shift-2}
\end{align}
The partition function then assumes the form:
\begin{align}
Z^{\CT^N_{2N, 1}} (\vec \mu, \frac{i}{2}) =  \int \, [d \vec \s] \ \frac{e^{-\pi \tr \vec \s}\, e^{-\pi N \Lambda}}{\ch{(\tr \vec \s - \sum^{2N-1}_{a=1}\,M_a)}}\cdot\frac{\prod_{1\leq j<k \leq N}\, \sinh^2{\pi(\s_j- \s_k)}}{\prod^{N}_{j=1} \prod^{2N-1}_{a=1}\ch{(\s_j- M_a)}\,\ch{(\s_j + N\Lambda)}}.
\end{align}
In the limit $\Lambda \to \infty$, the fundamental hyper associated with the real mass $\mu_{2N}$ is integrated out, 
and one observes that 
\be \label{lhs-pf-III}
Z^{\CT^N_{2N, 1}} (\vec \mu, \eta =\frac{i}{2}) \xrightarrow{\Lambda \to \infty} e^{-\pi \Lambda\,N(N+1)}\, Z^{\CT^N_{2N-1, 1}} (\vec M, \eta = i).
\ee

The theory dual to $\CT^N_{2N-1, 1}$ can be worked out in the following fashion. Given the partition function of the theory on the RHS of 
the self-duality relation \eref{SD-pf}:
\be
Z^{\CT^N_{2N, 1}} (\vec \mu', \eta = -\frac{i}{2})= 
 \int \, [d \vec s] \,\frac{\prod_{1\leq j<k \leq N}\,\sinh^2{\pi(s_j- s_k)}\,e^{\pi \tr \vec s }}{\prod^{N}_{j=1} \prod^{2N}_{a=1}\ch{(s_j- \mu_a +\frac{1}{N}\,\tr \vec \mu)}\,\ch{(\tr \vec s + \tr \vec \mu)}},
\ee
and the reparametrization of masses in \eref{massrep-2}, the appropriate shift in the integration variables is
\be
s_k \to s_k - N\Lambda, \qquad s_j \to s_j \quad {\rm for} \quad j=1,\ldots,N-1 \neq k.
\ee
for any of the $N$ choices of $k$, which are related by Weyl symmetry. In terms of the reparametrized masses and shifted integration variables, the partition function can be cast in the form in the limit $\Lambda \to \infty$:
\begin{align}
Z^{\CT^N_{2N, 1}} (\vec \mu', \eta = -\frac{i}{2}) \to  &
N\, \int \, [d \vec s] \,\frac{\prod_{1\leq j<k \leq N-1}\,\sinh^2{\pi(s_j- s_k)}\,e^{\pi \sum^{N-1}_{j=1} s_j + \pi s_N }}{\prod^{N-1}_{j=1} \prod^{2N-1}_{a=1}\ch{(s_j- M_a +\frac{1}{N}\,\tr \vec M)}\, \ch{(s_N + \frac{1}{N}\,\tr \vec M)} } \nn \\
 \times & \frac{1}{\ch{( \sum^{N-1}_{j=1} s_j + s_N+ \tr \vec M)}}\,\Big[ \frac{\prod^{N-1}_{j=1}\,\sinh^2{\pi(s_j- s_N + N\Lambda)}}{\prod^{2N-1}_{a=1}\, \ch{(s_N- N\Lambda -M_a + \frac{1}{N}\,\tr \vec M)}} \nn \\
\times & \frac{e^{-\pi N\Lambda}}{\prod^{N-1}_{j=1}\ch{(s_j + N\Lambda + \frac{1}{N}\,\tr \vec M)}} \Big]_{\Lambda \to \infty},
\end{align}
where only the terms inside the square brackets are $\Lambda$-dependent. Note that the matrix integral has been written corresponding to the choice $k=N$, 
while the prefactor $N$ is a combinatorial factor that arises from the $N$ possible choices of $k$ as mentioned above. 
In the limit $\Lambda \to \infty$, the $\Lambda$-dependent terms in the square brackets simplify as 
\begin{align}
\Big[ \ldots \Big]  \xrightarrow{\Lambda \to \infty} \, e^{\pi \sum^{N-1}_{j=1}\,s_j}\, e^{\pi s_N}\, e^{-\pi \Lambda\, N(N+1)}.
\end{align}
Finally, redefining the integration variables as variables for an $U(N-1) \times U(1)$ matrix integral, i.e.
\be
s_j \to s_j \,\, (j=1,\ldots,2N-1), \qquad s_N \to - u, 
\ee
we get 
\begin{align}\label{rhs-pf-III}
Z^{\CT^N_{2N, 1}} (\vec \mu', \eta = -\frac{i}{2})   \xrightarrow{\Lambda \to \infty} &
\int \,du\, [d \vec s] \,\frac{\prod_{1\leq j<k \leq N-1}\,\sinh^2{\pi(s_j- s_k)}\,e^{2\pi \tr \vec s}\, e^{-2 \pi u }}{\prod^{N-1}_{j=1} \prod^{2N-1}_{a=1}\ch{(s_j- M_a +\frac{1}{N}\,\tr \vec M)}\, \ch{(u - \frac{1}{N}\,\tr \vec M)} } \nn \\
 \times & \frac{1}{\ch{( \sum^{N-1}_{j=1} s_j - u + \tr \vec M)}}  \nn \\
 = & e^{-\pi \Lambda\, N(N+1)}\, Z^{\CT^\vee}( \vec M', \eta_1 = i, \eta_2=-i). 
\end{align}
In the last line we have identified the matrix integral as the partition function of the $U(1) \times U(N-1)$ 
quiver gauge theory $\CT^\vee$ in \figref{IRdual-Ex2}. The masses $\vec M'$ are related to the masses $\vec M$ 
in the following fashion:
\begin{align}
M'^{(1)} =  \frac{1}{N}\,\tr \vec M, \qquad M'^{(2)}_a = M_a - \frac{1}{N}\,\tr \vec M \, (a=1,\ldots, 2N-1),
\qquad M'^{(12)} = \tr \vec M,
\end{align}
where $M'^{(1)}, \vec M'^{(2)}$ denote the masses for the hypers charged under the $U(1)$ and the $U(N-1)$ 
gauge groups respectively, while $M'^{(12)}$ is the mass for the hyper charged under both the gauge groups.
Comparing \eref{lhs-pf-III} and \eref{rhs-pf-III}, and analytically continuing to real values of $\eta$, we obtain
\be
\boxed{Z^{\CT^N_{2N-1, 1}} (\vec M, \eta) = Z^{\CT^\vee}( \vec M', \eta_1=\eta, \eta_2=-\eta).}
\ee
This precisely reproduces the duality $\CD^N_{2N-1,1}$ found in \figref{IRdual-Ex2} of \Secref{D2-main}. 
It was noted in \Secref{SD-main} that the duality $\CD^N_{2N,P}$ can be obtained from the duality $\CD^N_{2N,1}$ 
by introducing a decoupled $\CT^1_{P-1,0}$ theory ($U(1)$ SQED with a $P$ hypers with charge 1) on both 
sides and then gauging the diagonal $U(1)$ subgroup of the $U(1) \times U(1)$ topological symmetry. This 
 Proceeding in an analogous fashion as above, one can also flow from the duality $\CD^N_{2N,P}$ to the duality 
$\CD^N_{2N-1,P}$ for $P>1$.

\subsection{Flowing from the duality $\CD^N_{2N-1,P}$ to the duality $\CD^N_{2N-1,P-1}$}\label{RG-duality-3}

Let us now study the RG flow between dualities triggered by large mass deformations for the Abelian hypers. 
Consider the IR duality $\CD^N_{2N-1,P}$ with $P>1$ given in \figref{IRdual-Ex2} as the starting point.
The sphere partition function of the theory $\CT=\CT^N_{2N-1, P}$ with generic real masses is given as:
\begin{align}
Z^{\CT^N_{2N-1,P}}(\vec m, \vec{m}_{\rm{Ab}};\eta) 
=&  \int \,  [d \vec s] \, \frac{e^{2\pi i \eta \tr \vec s}}{\prod^P_{l=1}\,\ch{(\tr \vec s -{m}^l_{\rm{Ab}})}}\,Z^{\CT^N_{2N-1,0}}_{\rm 1-loop} (\vec s, \vec m).
\end{align}
Let us now parametrize the real mass of the $P$-th abelian hypermultiplet as ${m}^P_{\rm{Ab}}=\Lambda$ 
and take the limit $\Lambda \to \infty$, keeping all the other masses finite. In this limit, the partition function can be written as:
\be \label{lhs-pf-IIIa}
Z^{\CT^N_{2N-1, P}} (\vec m, \vec{m}_{\rm{Ab}};\eta) \xrightarrow{\Lambda \to \infty} e^{-\pi \Lambda}\, Z^{\CT^N_{2N-1, P-1}} (\vec m, \vec{m}_{\rm{Ab}};\eta - \frac{i}{2}).
\ee
Now let us consider what happens on the dual side.
The dual theory $\CT^\vee$ which has the following partition function:
\begin{align}
Z^{\CT^\vee}(\vec m, \vec{m}_{\rm{Ab}};\eta, -\eta) =  &\int \,d{\s'}\,[d \vec \s]\,\frac{e^{2\pi i \eta\,(\s' - \tr \vec \s)}\,Z^{\CT^{N-1}_{2N-1,0}}_{\rm 1-loop} (\vec \s, \vec m)}{\ch{(\s'- \tr m)}\,\prod^P_{l=1}\,\ch{(\s'-\tr \vec \s -m^l_{\rm{Ab}})} },
\end{align}
which in the limit ${m}^P_{\rm{Ab}}=\Lambda \to \infty$ assumes the following form:
\begin{align} \label{rhs-pf-IIIa}
Z^{\CT^\vee} \xrightarrow{\Lambda \to \infty} \, & e^{-\pi \Lambda}\,\int \,d{\s'}\,[d \vec \s]\,\frac{e^{2\pi i (\eta - \frac{i}{2})\,(\s' - \tr \vec \s)}\,Z^{\CT^{N-1}_{2N-1,0}}_{\rm 1-loop} (\vec \s, \vec m)}{\ch{(\s'- \tr m)}\,\prod^{P-1}_{l=1}\,\ch{(\s'-\tr \vec \s -m^l_{\rm{Ab}})} } \nn \\
= &e^{-\pi \Lambda}\,Z^{\CT^{N\, \vee}_{2N-1, P-1}} (\vec m, \vec{m}_{\rm{Ab}};\eta- \frac{i}{2}, -(\eta - \frac{i}{2})),
\end{align}
where $\CT^{N\, \vee}_{2N-1, P-1}$ is the quiver gauge theory on the RHS of \figref{IRdual-Ex2} with $P \to P-1$. 
Comparing \eref{lhs-pf-IIIa} and \eref{rhs-pf-IIIa}, and analytically continuing to real $\eta$, we have
\be
\boxed{Z^{\CT^N_{2N-1, P-1}} (\vec m, \vec{m}_{\rm{Ab}};\eta) = Z^{\CT^{N\, \vee}_{2N-1, P-1}} (\vec m, \vec{m}_{\rm{Ab}};\eta, -\eta),}
%\int \,d{\s'}\,[d \vec \s]\,\frac{e^{2\pi i (\eta - \frac{i}{2})\,(\s' - \tr \vec \s)}\,Z^{\CT^{N-1}_{2N-1,0}}_{\rm 1-loop} (\vec \s, \vec m)}{\ch{(\s'- \tr m)}\,\prod^{P-1}_{l=1}\,\ch{(\s'-\tr \vec \s -m^l_{\rm{Ab}})} },
\ee
which is precisely the IR duality $\CD^N_{2N-1,P-1}$ in terms of the sphere partition function. One can implement this 
procedure sequentially to flow to the dualities with fixed $N$ and decreasing values of $P$, i.e. the dualities $\CD^N_{2N-1,P-2}$, $\CD^N_{2N-1,P-3}$ and so on, down to $\CD^N_{2N-1,1}$. Finally, for the duality $\CD^N_{2N-1,1}$, repeating the procedure for the single Abelian hypermultiplet in 
the $\CT=\CT^N_{2N-1,1}$ theory leads to the Seiberg-like duality for an ugly $U(N)$ SQCD, i.e.
\be
Z^{\CT^N_{2N-1, 0}} (\vec m;\eta)= Z^{\CT^1_{1, 0}} (\tr \vec m;\eta)\cdot Z^{\CT^{N-1}_{2N-1, 0}}(\vec m; -\eta ).
\ee
Therefore, turning on large masses for the Abelian hypermultiplets allows one to flow from the duality $\CD^N_{2N-1,P}$ to 
$\CD^N_{2N-1,0}$ with decreasing $P$ and fixed $N$.

\subsection{Summarizing the duality web} \label{RG-summary}

\begin{figure}[htbp]
\begin{center}
\begin{tikzpicture}
  \node (D00) at (0,0) {$\CD^{N+1}_{2N+1,0}$};
  \node (D10) at (4,0) {$\CD^N_{2N-1,0}$};
  \node (D20) at (6,0) {};
  \node (D30) at (8,0) {};
  \node (D40) at (11,0) {$\CD^N_{2N-1,P-2}$};
   \node (D50) at (14,0) {$\CD^N_{2N-1,P-1}$};
  \node (D01) at (0,3) {$\CD^N_{2N+1,1}$};
  \node (D11) at (4,3) {$\CD^N_{2N,1}$};
  \node (D21) at (10,3) {$\CD^N_{2N,P}$};
  \node (D31) at (14,3) {$\CD^N_{2N-1,P}$};
 % \node (E11) at (4,5) {$\CD^N_{2N,0}$};
   \node (E21) at (7,5) {$\CD^{SU(N)}_{N_f=2N}$};
  \draw[->] (D10) -- (D00) node [midway, above] {\footnotesize $N+1 \leftarrow N$};
   \draw[->] (D20) -- (D10) node [midway, above] {\footnotesize RG};
   \draw[dotted] (D30) -- (D20) node [] {};
    \draw[->] (D40) -- (D30) node [midway, above] {\footnotesize RG};
     \draw[->] (D50) -- (D40) node [midway, above] {\footnotesize RG};
  \draw[->] (D00) -- (D01) node [midway, right=+3pt] {\footnotesize $U(1)$ gauging};
  \draw[->] (D01) -- (D11) node [midway, above] {\footnotesize RG};
   \draw[->] (D11) -- (D21) node [midway, above] {\footnotesize $\circ \CT^1_{P-1,0}$};
    \draw[->] (D11) -- (D21) node [midway, below] {\footnotesize $U(1)_{\rm diag}$ gauging};
   \draw[->] (D21) -- (D31) node [midway, above] {\footnotesize RG};
    \draw[->] (D31) -- (D50) node [midway, right=+3pt] {\footnotesize RG};
    \draw[->] (D10.north east) -- (D31) node [midway, above] {\footnotesize $\circ \CT^1_{P,0}$};
     \draw[->] (D10.north east) -- (D31) node [midway, below=+5pt] {\footnotesize $U(1)_{\rm diag}$ gauging};
     %\draw[->] (D11) -- (E11) node [midway, right=+3pt] {\footnotesize RG};
     % \draw[->] (E11) -- (E21) node [midway, above] {\footnotesize $U(1)$};
      %\draw[->] (E11) -- (E21) node [midway, below] {\footnotesize gauging};
     \draw[->] (D11) -- (E21) node [midway, right=10 pt] {\footnotesize $U(1)$ gauging};
      \draw[->] (D21) -- (E21);
\end{tikzpicture}
\end{center}
\caption{\footnotesize{The duality web.} }
\label{DualityWeb}
\end{figure}
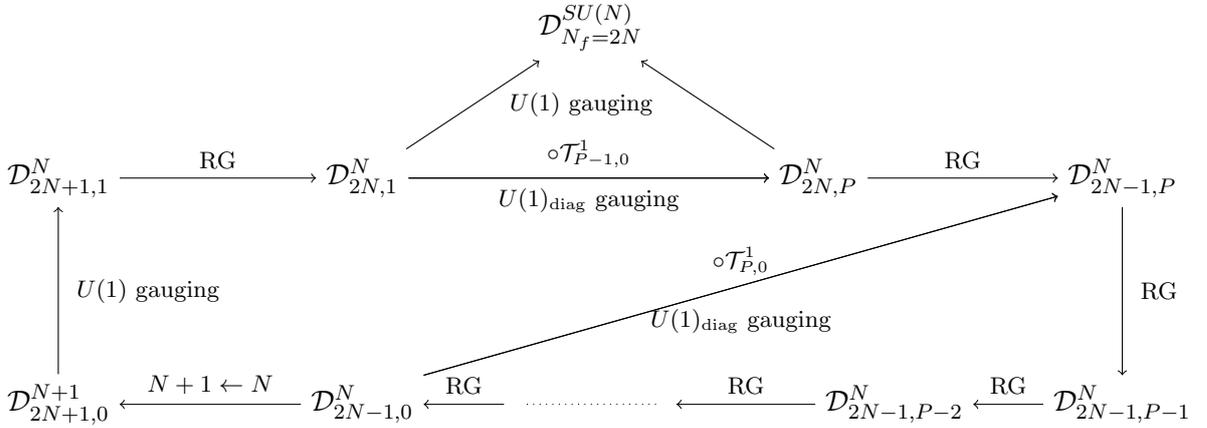

In this section, we discuss how the different dualities discussed in this paper are connected by various QFT operations. 
The duality web is summarized in \figref{DualityWeb}. Let us pick the duality $\CD^{N+1}_{2N+1,0}$ -- the Seiberg-like duality 
for an ugly $U(N+1)$ SQCD -- shown in the bottom left corner of the figure as our starting point. A $U(1)$ gauging operation 
on both sides of the duality $\CD^{N+1}_{2N+1,0}$ gives rise to the duality $\CD^{N}_{2N+1,1}$. In the next step, one 
introduces a large mass for one of the fundamental hypermultiplets and flows to the self-duality $\CD^{N}_{2N,1}$.

Given the self-duality $\CD^{N}_{2N,1}$, gauging the $U(1)$ topological symmetry 
on both sides of the duality $\CD^{N}_{2N,1}$ leads to the self-duality of an $SU(N)$ gauge theory 
with $N_f=2N$ fundamental hypers (denoted as $\CD^{SU(N)}_{N_f=2N}$). 
Next, one can introduce a decoupled $\CT^1_{P-1,0}$ theory -- a $U(1)$ SQED with $P-1$ hypers with charge 1 -- 
on both sides of the duality $\CD^{N}_{2N,1}$ and gauge a diagonal subgroup of the $U(1) \times U(1)$ topological 
symmetry. This leads to the duality $\CD^{N}_{2N,P}$. One can also obtain the 
duality $\CD^{SU(N)}_{N_f=2N}$ by gauging the $U(1)$ topological symmetry on both sides of the duality $\CD^{N}_{2N,P}$.
A mass deformation for a single fundamental hypermultiplet 
in $\CD^{N}_{2N,P}$ leads to the duality $\CD^{N}_{2N-1,P}$. Any further mass deformation for a fundamental hyper will take us into the bad regime 
of the $\CT^N_{N_f,P}$ theories, which we will not explore in this paper. However, given the duality $\CD^{N}_{2N-1,P}$, 
one can introduce large masses for the Abelian hypers and flow sequentially to dualities where $N$ is held fixed and $P$ 
decreases by 1 in every step. The sequence ends at the duality $\CD^{N}_{2N-1,0}$ -- the IR duality for an ugly $U(N)$ 
SQCD. The duality $\CD^{N}_{2N-1,P}$ can also be obtained by introducing a decoupled $\CT^1_{P,0}$ theory on both sides of 
the duality  $\CD^{N}_{2N-1,0}$ and gauging a diagonal subgroup of the $U(1) \times U(1)$ topological 
symmetry. 

We would like to emphasize that there are many additional RG flows relating dualities shown in \figref{DualityWeb}.
For example, one can flow from the duality $\CD^N_{2N,P}$ to the duality $\CD^N_{2N-1,P-1}$ by turning on a 
large mass for a fundamental hyper and a large mass for an Abelian hyper simultaneously.

\subsection{Comments on exact dualities from IR dualities}\label{RG-exact}

Let us briefly comment on how the above analysis leads to RG flows between various families of 3d $\CN=4$ SCFTs. A more detailed 
treatment of the issue will be covered in a future paper. 
In \Secref{RG-duality-1}, we showed that one can flow from the duality $\CD^N_{2N+1,1}$ to the duality $\CD^N_{2N,1}$ by introducing a large real 
mass for one of fundamental hypers, as discussed above. Let $CFT[\CD^N_{2N+1,1}]$ and $CFT[\CD^N_{2N,1}]$ denote the 3d $\CN=4$ 
interacting IR SCFTs associated with the dualities $\CD^N_{2N+1,1}$ and $\CD^N_{2N,1}$ respectively. If the mass scale $\Lambda$ is much larger 
compared to the strong coupling scale $\Lambda_s$, i.e. $\Lambda \gg \Lambda_s$, then the theory $\CT=\CT^N_{2N+1,1}$ 
and its dual $\CT^\vee$ will flow as 
\begin{align}
& \CT=\CT^N_{2N+1,1} \, \to \, \CT^N_{2N,1}\, \to \, CFT[\CD^N_{2N,1}], \\
& \CT^\vee \, \to \, \CT^N_{2N,1}\, \to \, CFT[\CD^N_{2N,1}].
\end{align}

Now consider the opposite limit $\Lambda \ll \Lambda_s$. In this case, the theory $\CT$ and its dual $\CT^\vee$ 
will flow as :
\begin{align}
& \CT=\CT^N_{2N+1,1} \, \to \,CFT[\CD^N_{2N+1,1}]\, \to \, CFT[\CD^N_{2N,1}], \\
& \CT^\vee \, \to \,CFT[\CD^N_{2N+1,1}]\, \to \, CFT[\CD^N_{2N,1}]. 
\end{align}
We therefore have two different descriptions of a single RG flow from $CFT[\CD^N_{2N+1,1}]$ to $CFT[\CD^N_{2N,1}]$. 
This is an example of an ``exact duality" arising out of an IR duality \cite{Strassler:2005qs}. 
In fact, we have a one-parameter family of such flows labelled by the integer $N$.

Similarly, the flow from the duality $\CD^N_{2N,P}$ to the duality $\CD^N_{2N-1,P}$ studied in \Secref{RG-duality-2} also leads 
to a two-parameter family of exact dualities associated with the flow $CFT[\CD^N_{2N,P}] \to CFT[\CD^N_{2N-1,P}]$.
For the special case of $P=1$, the conclusions from \Secref{RG-duality-1} and \Secref{RG-duality-2} can be combined to construct the 
exact duality sequence: 
\be
CFT[\CD^N_{2N+1,1}] \to CFT[\CD^N_{2N,1}] \to CFT[\CD^N_{2N-1,1}].
\ee

In \Secref{RG-duality-3}, we studied a sequence of flows between dualities: $\CD^N_{2N-1,P} \to \CD^N_{2N-1,P-1} 
\to \CD^N_{2N-1,P-2} \to \ldots \to \CD^N_{2N-1,1} \to \CD^N_{2N-1,0}$, where $\CD^N_{2N-1,0}$ is the known Seiberg-like 
duality of an ugly $U(N)$ SQCD. This leads to a one-parameter family of exact duality sequences associated with 
the flow: 
\be
CFT[\CD^N_{2N-1,P}] \to CFT[\CD^N_{2N-1,P-1}] \to CFT[\CD^N_{2N-1,P-2}] \to \ldots \to CFT[\CD^N_{2N-1,1}] \to CFT[\CD^N_{2N-1,0}].
\ee
Obviously, one can construct additional RG flows from a given duality by choosing to introduce large masses for both fundamental and 
Abelian hypermultiplets. These can be readily worked out using the technology of \Secref{RG-duality-1}-\Secref{RG-duality-3}.

\section*{Acknowledgements}
The author would like to thank Ibrahima Bah, Stefano Cremonesi, Amihay Hanany, Zohar Komargodski, Andrew Neitzke, 
Jaewon Song and Itamar Yaakov for discussion at different stages of the project. The author was supported in part at the Johns Hopkins 
University by the NSF grant PHY-2112699, and also by the Simons Collaboration on Global Categorical Symmetries, while 
this work was in preparation. 

\appendix

\section{Partition function identity for $U(N)$ theory with $2N$ flavors}\label{UN-2N}

In this section, we derive the identity \eref{UN-2N-pf} for a $U(N)$ gauge theory with $2N$ fundamental flavors. 
The partition function of the theory can be written in terms of the real masses $\vec m$ and a real FI parameter $\eta$, 
where we parametrize the latter as $\eta=t_1-t_2$, in the following fashion:
\be \label{UN-2N-pf0}
Z^{\CT^N_{2N,0}}(\vec m; \eta) = Z^{\CT^N_{2N,0}}(\vec m; t_1, t_2)=  \int \, [d \vec \s] \, e^{2\pi i (t_1-t_2)\,\tr \vec \s}\, \frac{\prod_{j<k}\, \sinh^2{\pi(\s_j- \s_k)}}{\prod^{N}_{j=1} \prod^{2N}_{i=1}\ch{(\s_j- m_i)}}.
\ee
Note that the sign of the FI parameter is flipped if we interchange $t_1$ and $t_2$, i.e.
\be
Z^{\CT^N_{2N,0}}(\vec m; -\eta) = Z^{\CT^N_{2N,0}}(\vec m; t_2, t_1). 
\ee
To obtain the desired identity, the above matrix integral should be manipulated in the following fashion. 
Recall the Cauchy determinant identity:
\be
\frac{\prod_{i<j} \sh{(x_i - x_j)}\,\sh{(y_i-y_j)}}{\prod_{i,j} \ch{(x_i-y_j)}}= \sum_{\rho} \, (-1)^{\rho}\, \frac{1}{\prod_i\,\ch{(x_i - y_{\rho(i)})}},
\ee
where the indices $i,j=1,\ldots, N$, and $\rho$ is an element of the permutation group of $N$ objects $S_N$.
First, we use the Cauchy determinant identity twice to reduce the matrix integral on the RHS of \eref{UN-2N-pf0} 
to the following form: 
\begin{align}
& Z^{\CT^N_{2N,0}}(\vec m; t_1, t_2)= \sum_{\rho, \rho'} \, (-1)^{\rho+\rho'}\,\int \, [d \vec \s] \, 
\frac{F(\vec m)\, e^{2\pi i (t_1-t_2)\,\tr \vec \s}}{\prod_i \ch{(\s_i - m_{\rho(i)})}\,\ch{(\s_i - m_{\rho'(i)+N})}}, \\
& F(\vec m) = \frac{1}{\prod_{i<j} \sh{(m_i - m_j)}\,\sh{(m_{i+N} - m_{j+N})}}.
\end{align}
Note that we have reduced to the matrix integral to a finite sum where the matrix integral in each summand can be 
treated as a product of $N$ Abelian integrals. A generic Abelian integral in this product can then be performed 
in the following fashion:
\be
\int\, d\s_i\, \frac{e^{2\pi i (t_1-t_2)\,\s_i}}{\ch{(\s_i - m_{\rho(i)})}\,\ch{(\s_i - m_{\rho'(i)+N})}} 
=(-i)\,\frac{e^{2\pi i (t_1-t_2)m_{\rho(i)}} -e^{2\pi i (t_1-t_2)m_{\rho'(i)+N}} }{\sh{(t_1-t_2)}\, \sh{(m_{\rho(i)} - m_{\rho'(i)+N})}}.
\ee
Collecting all the terms and after some straightforward manipulation, we have
\begin{align}
Z^{\CT^N_{2N,0}}= \sum_{\rho, \rho'} \,\frac{(-1)^{\rho+\rho'}\,(-i)^N\, F(\vec m)\, e^{-2\pi i t_2 \tr \vec m}}{N! \, \sinh^N{\pi(t_1-t_2)}}\,
\prod_i \frac{e^{2\pi i (t_1m_{\rho(i)} +t_2m_{\rho'(i)+N})} - e^{2\pi i(t_2m_{\rho(i)} +t_1m_{\rho'(i)+N})}}{\sh{(m_{\rho(i)} - m_{\rho'(i)+N})}}.
\end{align}
Written in this form, and imposing the condition $\tr \vec m=0$, one can readily check that
\be
Z^{\CT^N_{2N,0}}(\vec m; t_1, t_2) = Z^{\CT^N_{2N,0}}(\vec m; t_2, t_1),
\ee
which reproduces the identity \eref{UN-2N-pf}.

\section{Partition function identity for the ugly theory}\label{UN-2N-1}

In this section, we present a derivation of the identity \eref{ugly-id} which is different from the one 
discussed in \cite{Kapustin:2010mh}. Consider the partition function of the $\CT^N_{2N,0}$ 
theory with generic real masses and the real FI parameter set to zero:
\be
Z^{\CT^N_{2N, 0}} (\vec m, \eta =0) =  \int \, [d \vec \s] \,\frac{\prod_{1\leq j<k \leq N}\, \sinh^2{\pi(\s_j- \s_k)}}{\prod^{N}_{j=1} \prod^{2N}_{i=1}\ch{(\s_j- m_i)}}.
\ee
Now, let us set $m_{2N} =\Lambda$ and study the RG flow as $\Lambda \to \infty$ with other masses being finite. In the 
neighborhood of the $\vec \s \sim 0$ vacuum, the partition function takes the following form:
\be \label{ugly-fl1}
Z^{\CT^N_{2N, 0}} (\vec m, \eta =0) \xrightarrow{\Lambda \to \infty} e^{-\pi \Lambda\,N}\, Z^{\CT^N_{2N-1, 0}} (\vec m, \eta = -\frac{i}{2}).
\ee
Now, suppose we want to flow to a different vacuum where the gauge group is partially Higgsed to $U(N-1)$. This is 
achieved by combining the mass reparametrization $m_{2N} =\Lambda$ with the transformation:
\be
\s_i \to \s_i  \,\,(i=1,\ldots,N,\, i\neq j), \qquad \s_{j} \to \s_{j} + \,\Lambda,
\ee
and then taking the limit $\Lambda \to \infty$. The vector multiplet and the hypermultiplet contributions to the partition function 
assume the following form:
\begin{align}
 \prod_{1\leq j<k \leq N}\,\sinh^2{\pi(\s_j- \s_k)} \to  & \prod_{1\leq j<k \leq N-1}\,\sinh^2{\pi(\s_j- \s_k)}\,e^{2\pi \Lambda \,(N-1)}\, e^{-2\pi (\sum^{N-1}_{j=1}\s_j + (N-1)\s_{N})}, \\
\prod^{N}_{j=1} \prod^{2N}_{a=1}\ch{(\s_j- m_a)} \to & \prod^{N-1}_{j=1} \prod^{2N-1}_{a=1}\ch{(\s_j- m_a)}\,\ch{\s_{N}} \nn \\
& \times e^{\pi \Lambda (3N-2)}\, e^{\pi \sum^{2N-1}_{a=1}m_a}\, e^{-\pi \,\sum^{N-1}_{j=1}\s_j}\,e^{ \pi (2N-1)\,\s_{N}}.
\end{align}
Using the above expressions, the sphere partition function in the neighborhood of the second vacuum assumes the form:
\be  \label{ugly-fl2}
Z^{\CT^N_{2N, 0}} (\vec m, \eta =0) \xrightarrow{\Lambda \to \infty} e^{-\pi \Lambda\,N}\,Z^{\CT^{N-1}_{2N-1, 0}} (\vec m, \eta = \frac{i}{2})
\cdot Z^{\CT^1_{1, 0}} (\tr \vec m, \eta = -\frac{i}{2}), 
\ee
where $\tr \vec m=\sum^{2N-1}_{a=1}m_a$. Comparing \eref{ugly-fl1} and \eref{ugly-fl2}, we arrive at the identity:
\be
Z^{\CT^N_{2N-1, 0}} (\vec m, \eta = -\frac{i}{2}) = Z^{\CT^{N-1}_{2N-1, 0}} (\vec m, \eta = \frac{i}{2})
\cdot Z^{\CT^1_{1, 0}} (\tr \vec m, \eta = -\frac{i}{2}),
\ee
which on analytic continuation to real $\eta$ reproduces the identity \eref{ugly-id}.

\section{Hilbert Series and emergent global symmetries} \label{HS-rev}

In this section, we briefly review the Coulomb branch and the Higgs branch Hilbert Series for
the quiver gauge theories discussed in the main text. 

\subsection{Coulomb branch Hilbert Series}\label{HS-CB-app}

The Coulomb branch HS of the $U(N)$ theory with $N_f$ flavors \cite{Cremonesi:2013lqa} is given as
\begin{align}
& \CI^C_{\CT^N_{N_f,P}}(t) = \sum_{p_1 \geq p_2 \geq \ldots \geq p_{N} > - \infty }\, t^{\Delta( \vec p)}\, P_{U(N)}(t ; \vec p), \label{MonForm-2}\\
& \Delta( \vec p)=\frac{N_f}{2}\,\sum^{N-1}_i |p_i|  - \sum_{i<j}|p_i-p_j| , \label{RchForm-2}
\end{align}
where the term $t^{\Delta(\vec p)}$ counts the bare monopole operators while the factor 
$P_{U(N)}(t;\vec p)$ accounts for the dressing of the bare monopole operator by gauge invariant combinations 
of the adjoint scalar for the residual gauge group left unbroken by the flux $\vec p$. 
The explicit form of $P_{U(N)}$ is given as follows. Associate with every magnetic flux $\vec p$ a partition of $N$: 
$(\lambda_j (\vec p))^k_{j=1}$, where $\lambda_j (\vec p)$ counts how many times a distinct flux $p_j$ appears,
with the total number of distinct fluxes being $k$ and taking the convention $\lambda_i (\vec p) \geq \lambda_{i+1} (\vec p)$. 
Obviously, $\sum^k_{j=1} \lambda_j (\vec p) =N$. To this partition, we can associate a Young diagram $\lambda(\vec p)$ 
where the number of boxes in the $j$-th row is given by $(\lambda_j (\vec p))^k_{j=1}$. Then the factor $P_{U(N)}(t ; \vec p)$ 
is given as
\begin{align} \label{PUN-form}
& P_{U(N)}(t ; \vec p) = \prod^N_{i=1}\, Z^U_{\lambda_i(\vec p)}, \\
& Z^U_k = \prod^k_{i=1}\,\frac{1}{(1-t^i)}, \qquad k \geq 1, \nn \\
& Z^U_0 =1, \nn 
\end{align}
where we also set $\lambda_i (\vec p) =0$, for all $i > k$.
The Coulomb branch HS of the $SU(N)$ theory with $N_f$ flavors was also discussed in \cite{Cremonesi:2013lqa} and we state the result here:
\begin{align}
& \CI^C_{SU(N),N_f}(t) = \sum_{\substack{p_1 \geq p_2 \geq \ldots \geq p_N > - \infty \\ \sum_i p_i=0}}\, t^{\Delta( \vec p)}\, P_{SU(N)}(t ; \vec p), \label{MonForm}\\
& \Delta( \vec p)=\frac{N_f}{2}\,\sum^N_i |p_i| - \sum_{i<j}|p_i-p_j| , \label{RchForm}\\
& P_{SU(N)}(t, \vec p) = (1-t)\, P_{U(N)}(t ; \vec p)|_{\sum_i p_i=0}, \label{ClassFact}
\end{align}
where $P_{U(N)}$ is given by the formula \eref{PUN-form}. After performing the sum over fluxes, the final form of the result is 
\be
\CI^C_{SU(N),N_f}(t) = \frac{F_{N,N_f}(t)}{\prod^{N-1}_{i=1}\,(1-t^{i+1})\,(1- t^{N_f-N+1-i})},
\ee
where $F_{N,N_f}(t)$ is a palindromic polynomial of degree $(N-1)(N_f-N+1)$ and for small $N$:
\begin{align}
& F_{2,N_f}(t)=1 + t^{-1+N_f}, \\
& F_{3,N_f}(t)=1 + t^{-3+N_f} + 2t^{-2+N_f} + t^{-1+N_f} + t^{-4+2N_f},\\
& F_{4,N_f}(t)=1 + t^{-5+N_f} + 2t^{-4+N_f} + 3t^{-3+N_f} + 2t^{-2+N_f} + t^{-1+N_f} \nn \\
&+ t^{-8+2N_f} + 2t^{-7+2N_f} + 3t^{-6+2N_f} + 2t^{-5+2N_f} + t^{-4+2N_f} + t^{-9+3N_f}.
\end{align}

For the theory $\CT$, we have $N_f=2N-1$, and for small $N$, the Coulomb branch Hilbert Series is given as
\begin{align}
& \CI^C_{SU(2),3}(t) = 1+t+3 t^2+3 t^3+5 t^4+5 t^5+7 t^6+7 t^7+9 t^8+9 t^9+11 t^{10}+O\left(t^{11}\right), \label{CHS-1}\\
& \CI^C_{SU(3),5}(t) = 1+t+4 t^2+7 t^3+13 t^4+20 t^5+33 t^6+45 t^7+66 t^8+87 t^9+117 t^{10}+O\left(t^{11}\right), \label{CHS-2}\\
&\CI^C_{SU(4),7}(t) = 1+t+4 t^2+8 t^3+17 t^4+29 t^5+54 t^6+86 t^7+141 t^8+213 t^9+322 t^{10}+O\left(t^{11}\right),\label{CHS-3}
\end{align}
Note that in each case there exits a term at the order $t$ with coefficient 1. This indicates that the Coulomb branch has a 
$U(1)$ global symmetry, which is not visible in the UV Lagrangian. Contrast this with the index for $SU(3)$ with $N_f=6$:
\be
 \CI^C_{SU(3),6}(t) = 1+2 t^2+3 t^3+5 t^4+7 t^5+13 t^6+15 t^7+24 t^8+30 t^9+41 t^{10}+O\left(t^{11}\right),
\ee
where the $O(t)$ term is absent, implying there is no emergent $U(1)$ global symmetry in the IR in this case.

The generators of the Coulomb branch chiral ring and the relations governing them can be read off from the plethystic logarithm of 
the Hilbert Series. For the theory $\CT$, we have $N_f=2N-1$, and for small $N$, the plethystic logarithms of the Coulomb branch Hilbert Series are 
given as follows:
\begin{align}
& {\rm PL}[ \CI^C_{SU(2),3}(t)] = t+2 t^2 - t^4, \label{PLCHS-1}\\
& {\rm PL}[ \CI^C_{SU(3),5}(t)] = t+3 t^2+3t^3 -2 t^5 -3t^6+O(t^8), \label{PLCHS-2}\\
&{\rm PL}[ \CI^C_{SU(4),7}(t)] = t +3t^2 +4t^3 + 3t^4 -4t^6 - 4t^7-2t^8  +O\left(t^{9}\right),\label{PLCHS-3}
\end{align}
To contrast this with a case where $N_f > 2N-1$, consider
\be
{\rm PL}[ \CI^C_{SU(3),6}(t)]= 2t^2 + 3t^3 +2t^4 +t^5 -t^6 -2t^7 -3t^8 -2t^9 + O(t^{10}) .
\ee
For $N_f = 2N-1$, there exists an $O(t)$ term on the RHS with coefficient 1 - this 
generator of conformal dimension 1 corresponds to the monopole operator associated with the emergent $U(1)$ global symmetry. 
Also, note that the CB is a complete intersection for the $N=2$ theory, but not for $N > 2$.\\

Let us now consider the Coulomb branch HS for the theory $\CT^N_{N_f,P}$, given as
\begin{align}
& \CI^C_{\CT^N_{N_f,P}}(t) = \sum_{a_1 \geq a_2 \geq \ldots \geq a_{N} > - \infty }\, t^{\Delta( \vec a)}\, P_{U(N)}(t ; \vec a), \label{MonForm-2}\\
& \Delta( \vec a)=\frac{N_f}{2}\,\sum^{N-1}_i |a_i| +\frac{P}{2}\, |\sum^{N-1}_i \,a_i|- \sum_{i<j}|a_i-a_j| , \label{RchForm-2}
\end{align}
where $P_{U(N)}$ is given by the formula in \eref{PUN-form}. The second term in the R-charge 
formula gives the contribution of the $P$ Abelian multiplets.\\

First consider the case of $N_f=2N+1$ and $P=1$. For $N=1$, this is a $U(1)$ gauge theory 
with four flavors. The unrefined Coulomb branch HS is given as \cite{Cremonesi:2013lqa}:
\begin{align}
\CI^C_{U(1),4}(t) = & \Big[\frac{(1-t^{N_f})}{(1-t)(1-z\,t^{N_f/2})(1-z^{-1}\,t^{N_f/2})} \Big]_{z=1,\, N_f=4} =\frac{(1-t^{4})}{(1-t)\,(1-t^{2})^2} \nn \\
= & 1+t+3 t^2+3 t^3+5 t^4+5 t^5+7 t^6+7 t^7+9 t^8+9 t^9+11 t^{10}+O\left(t^{11}\right) \nn \\
=& \CI^C_{SU(2),3}(t).
\end{align}

For $N=2,3$, the unrefined Coulomb branch HS are given as
\begin{align}
\CI^C_{U(2),5,1}(t) = & 1+t+4 t^2+7 t^3+13 t^4+20 t^5+33 t^6+45 t^7+66 t^8+87 t^9+117 t^{10}+O\left(t^{11}\right) \nn \\
=& \CI^C_{SU(3),5}(t), \\
\CI^C_{U(3),7,1}(t) =& 1+t+4 t^2+8 t^3+17 t^4+29 t^5+54 t^6+86 t^7+141 t^8+213 t^9+322 t^{10}+O\left(t^{11}\right) \nn \\
=& \CI^C_{SU(4),7}(t).
\end{align}
The plethystic logarithms for the CB Hilbert Series of these theories are given by \eref{PLCHS-1}-\eref{PLCHS-3}. 
In contrast to the $U(N)$ SQCD, the CB of the theory $\CT^N_{N_f,P}$ is not a complete intersection.

\subsection{Higgs branch Hilbert Series}\label{HS-HB-app}

The Higgs branch HS for a $U(N)$ gauge theory with $N_f$ flavors is given as \cite{Razamat:2014pta}:
\begin{align}
\CI^H_{\CT^N_{N_f,0}}(x, \vec \mu) = &\frac{(1-x)^N}{N!}\,\oint_{|\vec z|=1}\, \prod^N_{i=1}\frac{dz_i}{z_i}\,
 \prod_{i\neq j}\, (1-\frac{z_i}{z_j})(1-\frac{x\,z_i}{z_j})\,\prod^N_{i=1}\prod^{N_f}_{k=1}\prod_{s=\pm}\,\frac{1}{(1-x^{1/2}\,z^s_i\,\mu^{-s}_k)},\\
= :& \frac{(1-x)^{N}}{N!}\,\oint_{|\vec z|=1}\, \prod^{N}_{i=1}\frac{dz_i}{z_i}\, \CI^{\rm int}_{\CT^N_{N_f,0}}(\vec z, x, \vec \mu),
\end{align}
where $x$ is the $U(1)_{\rm H}$ fugacity, $\vec \mu$ are the flavor fugacities associated with the fundamental hypers, and the integration is 
performed over a contour given by a union of unit circles. For an $SU(N)$ gauge theory with  $N_f$ flavors, the corresponding formula is 
given as
\begin{align}
\CI^H_{SU(N),N_f}(x, \vec \mu) =  \frac{(1-x)^{N-1}}{N!}\,\oint_{|\vec z|=1}\, \prod^{N-1}_{i=1}\frac{dz_i}{z_i}\,
\CI^{\rm int}_{\CT^N_{N_f,0}}(\vec z, x, \vec \mu)|_{z_N = \prod^{N-1}_{i=1}z^{-1}_i},
\end{align}
where $\CI^{\rm int}_{\CT^N_{N_f,0}}$ is the integrand defined above.\\

Finally, the HS for the theory $\CT^N_{N_f,P}$ -- a $U(N)$ gauge theory with $N_f$ fundamental hypers and $P$ Abelian 
hypers, is given as
\begin{align}
\CI^H_{\CT^N_{N_f,P}}(x, \vec \mu, \vec y) = \frac{(1-x)^N}{N!}\,\oint_{|\vec z|=1}\, \prod^N_{i=1}\frac{dz_i}{z_i}\,
& \prod_{i\neq j}\, (1-\frac{z_i}{z_j})(1-\frac{x\,z_i}{z_j})\,\prod^N_{i=1}\prod^{N_f}_{k=1}\prod_{s=\pm}\,\frac{1}{(1-x^{1/2}\,z^s_i\,\mu^{-s}_k)} \nn \\
& \times \prod^P_{l=1}\,\prod_{s=\pm}\,\frac{1}{(1-x^{1/2}\,(\prod^N_{i=1}\,z_i)^s\,y^{-s}_l)},
\end{align}
where $\vec y$ are the flavor fugacities associated with the Abelian hypers. 

%\section{Sphere partition function identity for an ugly theory}\label{PFId-ugly}
%
%In this section, we will derive the partition function identity \eref{ugly-id}, following the presentation in \cite{}. 
%The starting point is the partition function formula for a $U(N)$ theory with $N_f$ flavors:
%\be
%Z^{\CT^N_{N_f,0}}(\vec m; \eta) =  \int \, [d \vec \s] \, e^{2\pi i \eta\,\tr \vec \s}\, \frac{\prod_{j<k}\, \sinh^2{\pi(\s_j- \s_k)}}{\prod^{N}_{j=1} \prod^{N_f}_{i=1}\ch{(s_j- m_i)}}.
%\ee

\section{Construction of the 3d mirror by $S$-type operation}\label{PF-SOp}

In this section, we will discuss the construction of the 3d mirror of the theory $\CT^N_{N_f,P}$ by an $S$-type operation
as shown in \figref{SOp-IRdual-gen1} in terms of the sphere partition function. For $N_f > 2N$, the starting point is the pair of linear quivers 
$(X,Y)$ given in the first row of  \figref{SOp-IRdual-gen1}, i.e.

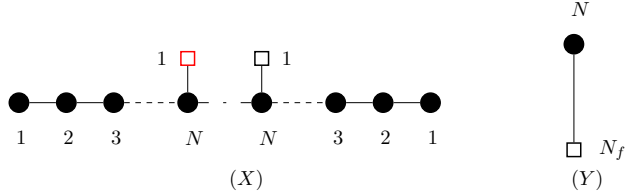
\begin{center}
\begin{tabular}{ccc}
 \scalebox{0.7}{\begin{tikzpicture}
\node[unode] (1) {};
\node[unode] (2) [right=.5cm  of 1]{};
\node[unode] (3) [right=.5cm of 2]{};
\node[unode] (4) [right=1cm of 3]{};
\node[] (5) [right=0.5cm of 4]{};
\node[unode] (6) [right=1 cm of 4]{};
\node[unode] (7) [right=1cm of 6]{};
\node[unode] (8) [right=0.5cm of 7]{};
\node[unode] (9) [right=0.5cm of 8]{};
\node[fnode, red] (13) [above=0.5cm of 4]{};
%\node[snode] (14) [left=0.5cm of 13]{1};
\node[fnode] (15) [above=0.5cm of 6]{};
%\node[snode] (11) [below=0.5cm of 9]{1};
\node[] (20) [right=0.2cm of 4]{};
\node[] (21) [left=0.2cm of 6]{};
\node[text width=0.1cm](41)[below=0.2 cm of 1]{1};
\node[text width=0.1cm](42)[below=0.2 cm of 2]{2};
\node[text width=0.1cm](43)[below=0.2 cm of 3]{3};
\node[text width=0.1cm](44)[below=0.2 cm of 4]{$N$};
\node[text width=0.1cm](45)[below=0.2 cm of 6]{$N$};
\node[text width=0.1cm](46)[below=0.2 cm of 7]{3};
\node[text width=0.1cm](47)[below=0.2 cm of 8]{2};
\node[text width=0.1cm](48)[below=0.2 cm of 9]{1};
\node[text width=0.1cm](49)[right=0.1 cm of 15]{1};
\node[text width=0.1cm](50)[left=0.2 cm of 13]{1};
\draw[-] (1) -- (2);
\draw[-] (2)-- (3);
\draw[dashed] (3) -- (4);
\draw[-] (4) --(20);
\draw[-] (21) --(6);
\draw[dashed] (20) -- (21);
\draw[dashed] (6) -- (7);
\draw[-] (7) -- (8);
\draw[-] (8) --(9);
%\draw[-] (5) -- (10);
\draw[-] (4) -- (13);
%\draw[-] (13) -- (14);
\draw[-] (6) -- (15);
\node[text width=0.1cm](30)[below=1 cm of 5]{$(X)$};
\end{tikzpicture}}
& \qquad   \qquad 
&\scalebox{.7}{\begin{tikzpicture}
\node[unode] (1) at (0,0){};
\node[fnode] (2) at (0,-2){};
\node[text width=0.1cm](41)[above=0.2 cm of 1]{$N$};
\node[text width=0.1cm](42)[right=0.2 cm of 2]{$N_f$};
%\node[snode,blue] (3) at (3,0){$1$};
\draw[-] (1) -- (2);
\node[text width=0.1cm](30)[below=0.1 cm of 2]{$(Y)$};
%\draw[-, thick, blue] (1)-- (3);
%\node[text width=1cm](10) at (1.2, 0.2){$N-1$};
\end{tikzpicture}}
\end{tabular}
\end{center}
where the number of $U(N)$ gauge nodes in $X$ is $N_f -2N +1$. 3d mirror symmetry for linear quivers implies that the partition 
functions of $X$ and $Y$ are related in the following fashion:
\be \label{3dMS-LQ-1}
Z^{(X)}(m_1, m_2; \vec t) = C_{XY}(\vec m, \vec t)\cdot Z^{(Y)}(\vec t; -m_1, -m_2),
\ee
where $C_{XY}(\vec m, \vec t) = e^{2\pi i \sum_{i,l}m_i\,t_l}$ with $b_{ij}$ being an integer-valued matrix. 
An $S$-type operation $\CO^1$ of the flavoring-gauging type, where one first attaches a single hypermultiplet to the flavor node in $X$ 
marked in red and then gauges the flavor node, is implemented as:
\be \label{3dMS-S}
Z^{\CO^1(X)}(m^1_f, m_2; \vec t, \eta_1) = \int\, du\, \frac{e^{2\pi i \eta_1 u}}{\ch{(u- m^1_f)}}\,Z^{(X)}(u, m_2; \vec t).
\ee
On the dual side, the resultant operation on the theory $Y$ can be read off from the dual partition function:
\begin{align}  \label{3dMS-Sd}
Z^{\wt{\CO}^1(Y)}= e^{2\pi i \sum_l (b_{2l}m_2  + b_{1l}m^1_f )t_l}\, e^{2\pi i m^1_f \eta_1}\, \int\,\Big[ d\vec \s\Big]\,
\frac{e^{-2\pi i \tr \vec \s(m^1_f-m_2)}}{\ch{(\tr \vec \s - \eta_1 - \sum_l b_{1l}t_l)}}\, Z^{(Y)}_{\rm 1-loop}(\vec \s, \vec t)
\end{align}
which is obtained by using the identity \eref{3dMS-LQ-1} on the RHS of \eref{3dMS-S} to substitute $Z^{(X)}$, 
exchanging the order of integration, and integrating over $u$. The dual partition function can then be identified
as the partition function of the theory $\CT^N_{N_f,1}$ up to an overall phase factor:
\be
Z^{\wt{\CO}^1(Y)}= C(m_2,m^1_f, \vec t, \eta_1)\, Z^{\CT^N_{N_f,1}}(\vec t, m_{\rm ab}=\eta_1 + \sum_lb_{1l}t_l; -m^1_f, -m_2),
\ee
which implies that $\CT^N_{N_f,1}$ is the 3d mirror of the quiver gauge theory $\CO^1(X)$. In the next step, we implement another 
flavoring-gauging operation $\CO^2$ of the same type as above at the $U(1)$ flavor node of the theory $\CO^1(X)$ associated with the mass parameter $m^1_f$. Proceeding as above, one can show that the 3d mirror of the quiver $\CO^2 \circ \CO^1(X)$ is $\CT^N_{N_f,2}$. 
Repeating the procedure $P$ times, we arrive at the conclusion that $\CT^N_{N_f,P}$ is the 3d mirror of the theory 
$\wt{\CT}^N_{N_f,P}= \CO^P \circ \ldots \circ \CO^2 \circ \CO^1(X)$, where $\wt{\CT}^N_{N_f,P}$ is the quiver shown in the bottom left 
corner of \figref{SOp-IRdual-gen1}. For $N_f=2N$ and $N_f=2N-1$, the exercise can be performed in a similar fashion.

\bibliography{cpn1-1}
\bibliographystyle{JHEP}

\end{document}